\documentclass[a4paper,11pt]{article}
\pdfoutput=1 % if your are submitting a pdflatex (i.e. if you have
\usepackage{jheppub} % for details on the use of the package, please
\usepackage{amssymb,amsmath}
\numberwithin{equation}{section}  % Number equations within sections
\usepackage[table]{xcolor}
\usepackage{graphicx}
\usepackage{hyperref}
\usepackage{float} 
\usepackage[T1]{fontenc}
\usepackage{braket}
\usepackage{subfigure}
\usepackage{enumitem}
\usepackage{braket}
\usepackage{multicol}

\usepackage{adjustbox}

\usepackage{placeins}

\usepackage{soul}

\definecolor{light-gray}{gray}{0.8}
\usepackage[normalem]{ulem}

\definecolor{mygreen}{rgb}{.4,.76,.26}
\definecolor{myblue}{rgb}{.4,.76,.84}
\definecolor{myred}{rgb}{.85,.34,.25}

\newsavebox\MBox
\newcommand\Cline[2][red]{{\sbox\MBox{$#2$}%
  \rlap{\usebox\MBox}\color{#1}\rule[-1.2\dp\MBox]{\wd\MBox}{1.5pt}}}

\usepackage[utf8]{inputenc}
\usepackage{lmodern}
\usepackage{textcomp}
\usepackage{amsfonts}
\usepackage{mathrsfs}
\usepackage{physics}

\usepackage{pifont} % xmark

\usepackage{mathtools} %write above arrow

\usepackage{amstext} % for \text macro
\usepackage{array}   % for \newcolumntype macro
\newcolumntype{L}{>{$}l<{$}} % math-mode version of "l" column type

\newcounter{defcount}
\begin{document}

\title{Systematics of U-Spin Sum Rules for Systems with Direct Sums}

\author[a]{Margarita Gavrilova}
\author[b]{Stefan Schacht}

\affiliation[a]{Department of Physics, LEPP, Cornell University, Ithaca, NY 14853, USA}
\affiliation[b]{Department of Physics and Astronomy, University of Manchester, Manchester M13 9PL, United Kingdom}

\emailAdd{mg2333@cornell.edu}
\emailAdd{stefan.schacht@manchester.ac.uk}

\abstract{
A rich mathematical structure underlying flavor sum rules has been discovered recently. In this work, we extend these findings to systems with a direct sum of representations. We prove several results for the general case. We derive an algorithm that enables the determination of all $U$-spin amplitude sum rules at arbitrary order of the symmetry breaking for any system containing a direct sum of the representations $0 \oplus 1$. Potential applications are numerous and include, for example, higher order sum rules for CP-violating charm decays with an arbitrary number of final states.
}

\maketitle
\flushbottom

\section{Introduction}

Sum rules derived from the approximate flavor symmetry of QCD are an essential tool for the theoretical description of hadronic decays. Flavor sum rules are crucial for testing the Standard Model (SM), overconstraining the unitarity triangle, and probing the non-perturbative regime of QCD. 

Recent experimental advances in the measurement of hadronic decays~\cite{LHCb:2019hro, LHCb:2022lry, LHCb:2023mwc, LHCb:2023rae, LHCb:2019oke, CMS:2024hsv, Belle-II:2023vra, Belle:2023bzn, Belle:2023str, LHCb:2024rkp, LHCb:2022lzp, LHCb:2024vhs, LHCb:2024hfo, Belle-II:2024vfw, Belle-II:2024xtf, Belle:2024dhj, LHCb:2023ngz, LHCb:2023tma}, in particular decays with multi-body final states, as well as future prospects~\cite{Belle-II:2018jsg, Cerri:2018ypt, LHCb:2018roe} and recent measurements indicating signs of large $U$-spin breaking~\cite{LHCb:2022lry, Schacht:2022kuj, Bhattacharya:2022akr} highlight the increasing need for a more comprehensive theoretical understanding of flavor sum rules. A theoretical framework that can describe current and future data on multi-body hadronic decays must be capable of generating predictions for processes with an arbitrary number of hadronic final states, while also minimizing theoretical uncertainties by accounting for arbitrarily high orders of symmetry breaking.

To that end, a formalism for deriving $U$-spin amplitude sum rules for an arbitrary system of decays at any order of symmetry breaking was developed in Ref.~\cite{Gavrilova:2022hbx}. Generally speaking, the results in Ref.~\cite{Gavrilova:2022hbx} rely solely on the assumption of an approximate $SU(2)$ symmetry and can, therefore, be applied to any subgroup of $SU(3)_F$, including isospin. However, applications of this formalism beyond $U$-spin, \emph{i.e.}~to $V$-spin and isospin, are limited to the case of semileptonic final states only. The conclusion of Ref.~\cite{Gavrilova:2022hbx} points out several ways in which the developed formalism could be extended. One of the discussed extensions is to move beyond the case where each particle in the initial and final states, as well as the Hamiltonian, is a component of a single $U$-spin multiplet with a fixed value of total $U$-spin — a scenario we refer to as ``pure multiplets.'' An important generalization would involve extending the formalism to cases where the group-theoretical description of the system includes a direct sum of irreducible representations (irreps) of $SU(2)$. This is the generalization that we focus on in the present work.

In this paper, we extend the formalism of Ref.~\cite{Gavrilova:2022hbx} to $U$-spin systems that include a direct sum of irreps. While our approach, like in Ref.~\cite{Gavrilova:2022hbx}, is based solely on the approximate $SU(2)$ symmetry and can also be applied to isospin or $V$-spin, it is particularly robust for $U$-spin. For this reason, we primarily focus on $U$-spin throughout this work. We derive several general results for systems with a direct sum of any two irreps and then focus on the simplest special case: a direct sum of a singlet and a triplet. For these systems, we prove a powerful theorem that enables the derivation of all linearly independent amplitude sum rules to any order of symmetry breaking without requiring an explicit Clebsch-Gordan decomposition. Similarly to systems with pure irreps, we find that the systems with a direct sum of a singlet and a triplet allow for an elegant geometric interpretation, where amplitudes related by flavor symmetry correspond to nodes of a multidimensional lattice, and the sum rules can be read off the lattice following a set of simple rules. Both the current work and the work in Ref.~\cite{Gavrilova:2022hbx} are necessary steps towards understanding all-order sum rules between observables,~\emph{i.e.},~decay rates, branching fractions, and CP asymmetries, which is a highly non-trivial step.

Nature provides numerous examples of physical systems to which the mathematical results derived in this work can be applied. A prominent application is the derivation of sum rules between amplitudes of CP-violating decays that are valid for both interfering amplitudes. For instance, in CP-violating charm decays, the Hamiltonian contains a singly Cabibbo-suppressed (SCS) triplet component and an $\mathcal{O}(\lambda^5)$ suppressed singlet component, see, \emph{e.g.},~Refs.~\cite{Brod:2012ud, Grossman:2019xcj}, where $\lambda$ is the expansion parameter of the Wolfenstein parametrization. Another example is the isospin structure of the Hamiltonian for $B \rightarrow K\pi$ decays, which is also given by a direct sum of a singlet and a triplet~\cite{Grinstein:2014aza}. Additionally, the results presented in this work are relevant to systems with direct sums in the final state, such as $\pi-\eta$ mixing.

The application of flavor-symmetry methods to hadronic decays has a long history. For charm decays, both meson~\cite{Kingsley:1975fe, Voloshin:1975yx, Barger:1979fu, Golden:1989qx, Buccella:1994nf, Pirtskhalava:2011va, Brod:2012ud, Bhattacharya:2012ah, Hiller:2012xm, Cheng:2012wr, Feldmann:2012js, Grossman:2012ry, Buccella:2013tya, Buccella:2019kpn, Nierste:2015zra, Muller:2015lua, Muller:2015rna, Nierste:2017cua, Grossman:2018ptn, Grossman:2019xcj, Cheng:2019ggx, Dery:2021mll, Gavrilova:2022hbx, Iguro:2024uuw} and baryon decays~\cite{Kingsley:1975fe, Savage:1989qr, Savage:1991wu, Sharma:1996sc, Wang:2017gxe, Zhong:2024qqs, Lu:2016ogy, Wang:2024ztg} have been extensively studied. Methodologies not based on flavor symmetries are explored \emph{e.g.},~in Refs.~\cite{Li:2012cfa, Li:2019hho, Lenz:2023rlq, Chala:2019fdb, Khodjamirian:2017zdu, Pich:2023kim, Bediaga:2022sxw, Schacht:2021jaz}. Recent works applying flavor-symmetry methods to $B$ decays can be found in Refs.~\cite{Dery:2020lbc, Hassan:2022ucn, Davies:2023arm, Davies:2024vmv, Amhis:2022hpm, Grossman:2024amc, Berthiaume:2023kmp, Bhattacharya:2022akr}; see Refs.~\cite{Zeppenfeld:1980ex, Chau:1982da,Gronau:1995hm, Gronau:1994rj, Buras:1998ra, Jung:2009pb, Grinstein:1996us} for works developing the topological amplitude and SU(3)$_F$ formalism for $B$ decays. The need for higher-order sum rules is less urgent in the case of isospin symmetry, where leading-order results already have the desired precision~\cite{Gronau:1990ka, Grossman:2012eb, Wang:2022nbm, Gavrilova:2023fzy, Luo:2023vbx, Wang:2023pnb}. Potential new physics scenarios are 
explored,~\emph{e.g.},~in Refs.~\cite{Grossman:2006jg, Altmannshofer:2012ur, Dery:2019ysp, Calibbi:2019bay, Bause:2020obd, Buras:2021rdg, Acaroglu:2021qae, Bause:2022jes}.

Our paper is structured as follows. In Sec.~\ref{sec:review}, we review the key results of Ref.~\cite{Gavrilova:2022hbx}. In Sec.~\ref{sec:gen-case}, we discuss systems with a direct sum of two arbitrary irreps. In Sec.~\ref{sec:0+1}, we present results for the special case of systems with a direct sum of a triplet and a singlet. We then apply our results to two physical systems in Sec.~\ref{sec:examples}, and conclude in Sec.~\ref{sec:conclusion}. Additional details and derivations are provided in Appendices~\ref{app:details-trivial}-\ref{sec:proof}.

\section{Review of $U$-spin amplitude sum rules for systems with pure multiplets \label{sec:review}}

We start with a brief review. Ref.~\cite{Gavrilova:2022hbx} performs a rigorous and exhaustive analysis of $U$-spin amplitude sum rules to an arbitrary order of the symmetry breaking in the case when the system of interest can be described by pure $U$-spin multiplets. Since the present work relies substantially on the results of Ref.~\cite{Gavrilova:2022hbx}, in this section, for completeness, we briefly summarize the definitions and the main results that are relevant to the current work.

The assumption of pure $U$-spin multiplets is crucial to the results of Ref.~\cite{Gavrilova:2022hbx}. In particular, this assumption implies that the Hamiltonian contains operators with only one fixed value of the total $U$-spin and there is no mixing of representations in the initial and final state. As a consequence of this assumption, one can define CKM-free amplitudes, by dividing out the relevant CKM factors for each amplitude. If $\mathcal{A}_i$ is a decay amplitude and $f_{u,m}$ is the relevant CKM factor then we define CKM-free amplitudes as
\begin{equation}\label{eq:A-CKM-free-def}
    A_i \equiv \frac{\mathcal{A}_i}{f_{u,m}}\,.
\end{equation}
The definition of CKM-free amplitudes allows to remove the CKM factors from the problem. By adopting this definition, we can focus on the group-theoretical structure of the system of interest, derive all the sum rules for CKM-free amplitudes and then restore the CKM factors at the very last step.

The two key insights of Ref.~\cite{Gavrilova:2022hbx} are that, first, from the group-theoretical point of view, the assignment of $U$-spin representations to the initial state, final state or the Hamiltonian is irrelevant for the structure of sum rules, and second, all higher representations can be built from doublets. Thus it is enough to study the system with the following schematic structure under $U$-spin
\begin{equation}\label{eq:system-doub}
    0 \rightarrow \left(\frac{1}{2}\right)^{\otimes n}\,,
\end{equation}
that is a system with $U$-spin singlets in the initial state and the Hamiltonian, and $n$ $U$-spin doublets in the final state. All results obtained for the system in Eq.~\eqref{eq:system-doub} can be straightforwardly generalized to the case of systems with higher representations and the systems with non-singlet initial state and the Hamiltonian. For simplicity, for the rest of this section we only consider systems of the form in Eq.~\eqref{eq:system-doub}.

$U$-spin is an approximate symmetry which is broken by the mass difference between $d$- and $s$-quarks. The breaking is of the order $\varepsilon\sim30\%$. We take the breaking of the $U$-spin symmetry into account by insertion of the spurion operator $H_\varepsilon$, which is given by the $m=0$ component of a triplet. We use $b$ to denote the order of the symmetry breaking, that is $b = 0$ corresponds to the symmetry limit, $b = 1$ to the leading order of the symmetry breaking and so on. Order $b$ of the symmetry breaking is realized via the insertion of $b$ copies of the spurion operator. Thus in the case of the system in Eq.~\eqref{eq:system-doub}, the full Hamiltonian including all the orders of the symmetry breaking is given by
\begin{equation}
    H = \sum_{b=0} H_0^0 \otimes H_\varepsilon^{\otimes b}\,,
\end{equation}
where $H_0^0$ denotes a $U$-spin singlet.

Our goal is to write the sum rules among CKM-free amplitudes $A_i$ of the system in Eq.~\eqref{eq:system-doub}. It is useful to represent the amplitudes $A_i$ as $n$-tuples. Each element of the $n$-tuple represents the $m$ quantum number (QN) of one of the $n$ doublets in the final state of the system in Eq.~\eqref{eq:system-doub}. The '$+$' in the $n$-tuple corresponds to $m = +1/2$ and the '$-$' to $m = -1/2$. Only the $n$-tuples with an equal number of plus and minus signs are allowed. Without loss of generality, we refer to the subset of amplitudes whose $n$-tuples start with a minus sign simply as \textit{amplitudes}, and to the corresponding amplitudes whose $n$-tuples are obtained by flipping all the signs as \textit{$U$-spin conjugate amplitudes}. We denote the amplitudes by $A_i$ and the $U$-spin conjugate amplitudes as $\overline{A}_i$, where $i = 1,\dots, n_A/2$ is an index that we use to label the amplitudes, and $n_A$ is the number of amplitudes in the system. The index $i$ runs up to $n_A/2$ since each amplitude of the system in Eq.~\eqref{eq:system-doub} has a  $U$-spin conjugate partner.

Amplitudes $A_i$ and $\overline{A}_i$ form a $U$-spin pair. Next we define the $a$- and $s$-type amplitudes as follows
\begin{equation}\label{eq:as-def-pure}
    a_i \equiv A_i - (-1)^{n/2} \overline{A}_i\,,\qquad s_i \equiv A_i + (-1)^{n/2} \overline{A}_i\,.
\end{equation}
These definitions prove to be particularly useful in the case of systems with pure $U$-spin representations. Defining $a$- and $s$-type amplitudes is simply a basis choice, thus all the sum rules can always be expressed in terms of $a$- and $s$-type amplitudes. This basis choice however is very illuminating and has a number of nice properties that allow us to systematically construct amplitude sum rules up to an arbitrary order of the symmetry breaking. Here we list some of these properties:
\begin{enumerate}
    \item $a$- and $s$-type amplitudes for any $U$-spin system built from pure $U$-spin representations fully decouple to any order of the symmetry breaking. That is all the sum rules can be written as sum rules among $a$- and $s$-type amplitudes separately.
    \item In the symmetry limit there are always $n_A/2$ trivial $a$-type sum rules given by
    \begin{equation}
        a_i = 0\,,
    \end{equation}
    where $n_A$ is the number of amplitudes in the $U$-spin system.
    \item For even orders of breaking $b$, the $s$-type sum rules that hold up to order $b$ are preserved at order $b+1$.
    \item For odd orders of breaking $b$, the $a$-type sum rules that hold up to order $b$ are preserved at order $b+1$.
    \item The highest order at which the system can still have sum rules is $b_\text{max} = n/2-1$.
    \item At order $b_\text{max}$ there is always only one sum rule. This sum rule is an $s$-type sum rule for odd $b_\text{max}$ (even $n$), and it is an $a$-type sum rule for even $b_\text{max}$ (odd $n$). The highest order sum rule is given by the sum of all $a$- or $s$-type amplitudes.
\end{enumerate}
Note that all of these results also hold in the most general case of systems with pure irreps, except that the definitions of $a$- and $s$-type amplitudes are slightly modified (see Eq.~(5.7) of Ref.~\cite{Gavrilova:2022hbx} and the discussion therein) and that the amplitudes in the sum rules are weighted by symmetry factors (see Eqs.~(5.8)--(5.11) of Ref.~\cite{Gavrilova:2022hbx}).

We conclude this brief review by some counting results. The number of amplitudes in a $U$-spin system of $n$ doublets is given by
    \begin{equation}\label{eq:nA}
    n_A^d(n) = \binom{n}{n/2} = \frac{n!}{\left(n/2\right)! \left(n/2\right)!}.
\end{equation}
The number of sum rules at the order of breaking $b$ is given by
\begin{equation}\label{eq:nSR_doublets}
    n_{SR}^{d}(n,b) = \frac{n!}{\left({n/2} + b + 1\right)!\left({n/2} - b - 1\right)!}\,.
\end{equation}

\section{General discussion of systems with a direct sum of representations}\label{sec:gen-case}

In the previous section, we reviewed the systematics of $U$-spin amplitude sum rules in the case when all the particles in the initial and final state as well as the Hamiltonian of a $U$-spin system are given by pure $U$-spin multiplets. This is a good assumption for a wide class of $U$-spin systems and in physical terms it means that (\emph{i})~there is a single dominant representation in the Hamiltonian, \emph{i.e.},~we neglect CKM-subleading operators, and (\emph{ii})~we neglect the mixing of particle multiplets, for example the mixing between $\pi^0$, $\eta_0$ and $\eta_8$. 
Both of these effects, however, can be taken into account if we allow the initial state, final state, or the Hamiltonian to be described by direct sums of different $U$-spin representations. We refer to this as \emph{representation mixing} as opposed to pure representations.
The case of representation mixing in the Hamiltonian is of particular interest to us as it allows to take into account CP violation.
In this work, we focus on systems with one direct sum only.

In this section, we define a class of systems that we would like to study, discuss some general properties of these systems and comment on the theoretical challenges that arise on the way towards the systematic description of these systems.

\subsection{Definitions and assumptions}

We would like to study $U$-spin systems with the following group-theoretical structure
\begin{equation}\label{eq:system-dir-sum}
    0 \rightarrow \left(\Sigma\, u_1 \oplus \Delta\, u_2\right) \otimes \left(\frac{1}{2}\right)^{\otimes k}\,,
\end{equation}
where $u_1$ and $u_2$ are the two $U$-spin representations that mix, and $\Sigma$ and $\Delta$ are the mixing coefficients given by:
\begin{equation}\label{eq:mixing_coeff}
    \Sigma = \left(\Sigma_{-u_1},\,\dots, \Sigma_{+u_1}\right),\qquad \Delta = \left(\Delta_{-u_2},\,\dots, \Delta_{+u_2}\right)\,,
\end{equation}
where the indices denote the $m$ QNs of the components of representations $u_1$ and $u_2$, respectively. Hereafter, we refer to this system as the \emph{$u_1 \oplus u_2$ system}.
We further assume, without loss of generality, that $u_1 > u_2$. The case of $u_1 = u_2$ is trivially covered by the results for the systems containing pure representations only through a redefinition of the representations; therefore, we do not consider it here. Note that, by construction, $u_1$ and $u_2$ are either both integers or both half-integers.

Our goal is to derive amplitude sum rules for the system with the group-theoretical structure given in Eq.~\eqref{eq:system-dir-sum}. This system is described by two representations $u_1$ and $u_2$ that mix and $k$ doublets, all in the final state. This choice of the system might seem somewhat specific, but it is not. As in the case of pure representations~\cite{Gavrilova:2022hbx}, crossing symmetry ensures that moving some of the $U$-spin representations and/or the direct sum into the Hamiltonian and/or initial state results in a system with the same counting and structure of sum rules as the system in Eq.~\eqref{eq:system-dir-sum}. Moreover, since any higher representation of $SU(2)$ can be constructed from doublets, the results for the system in Eq.~\eqref{eq:system-dir-sum} can be straightforwardly generalized to the case of a system described by a direct sum and any number of arbitrary $U$-spin representations. The generalization procedure is analogous to that described in Section IV of Ref.~\cite{Gavrilova:2022hbx}.

Next, we define $n$, the number of would-be doublets for the system in Eq.~\eqref{eq:system-dir-sum}, as the minimal number of doublets needed in order to construct all the representations of the system. In the case of the system in Eq.~(\ref{eq:system-dir-sum})
\begin{equation}
    n \equiv 2u_1 + k\,, \label{eq:number-of-would-be-doublets}
\end{equation}
and $n$ is required to be even in order to ensure non-zero amplitudes.
Recall that we have chosen $u_1 > u_2$. To ensure that the system under consideration is complete, meaning that all the components of all the irreps are allowed, we also require that $k \geq 2u_1$.

The amplitudes of systems with pure representations can always be labeled as follows~\cite{Gavrilova:2022hbx}. We choose some arbitrary but fixed ordering for the representations. For each amplitude of the system we list the corresponding $m$ QNs of the representations according to the chosen order. Similarly, the amplitudes of the systems with direct sums of representations can be labeled by listing the $m$ QN of all pure representations and all direct sums. This is possible since only the components of the multiplets in the direct sum that have the same $m$ QN can contribute to the same amplitude. Thus, each amplitude has a definite value of the $m$ QN for each direct sum in the system. 

All amplitudes of the $u_1 \oplus u_2$ system can thus be labeled via a set of $k + 1$ $m$ QNs, that is
\begin{equation}
    \mathcal{A}_i^{u_1\oplus u_2} \equiv \mathcal{A}^{u_1\oplus u_2}(\overline{m})\,,
\end{equation}
where $i$ is an index that enumerates the amplitudes and $\overline{m}$ is a set of $m$ QNs
\begin{equation}
    \overline{m} = \{m_1,\,\dots\,,m_{k},\,M\}\,,
\end{equation}
where we have chosen the first $k$ $m$ QNs to correspond to the $k$ doublets and the last $m$ QN to the direct sum $u_1 \oplus u_2$. We use $\mathcal{A}$ to denote the physical amplitudes, that is the amplitudes that depend on the mixing coefficients, similarly to Eq.~\eqref{eq:A-CKM-free-def}.

Without loss of generality, we choose to refer to the amplitudes for which $m_1 = -1/2$ as amplitudes and to the ones that have $m_1 = +1/2$ as $U$-spin conjugate amplitudes. We use the notation $\overline{\mathcal{A}}_i^{u_1\oplus u_2}$ or $\overline{\mathcal{A}}^{u_1\oplus u_2}(\overline{m})$ to distinguish the $U$-spin conjugate amplitudes when necessary.

Using the Wigner-Eckart theorem, one can perform the group-theoretical decomposition of the amplitudes of the $u_1 \oplus u_2$ system in terms of reduced matrix elements (RME). We write this decomposition as
\begin{equation}\label{eq:A-u1+u2-RME-decomp}
    \mathcal{A}_i^{u_1\oplus u_2} = \sum_\alpha c_{i\alpha} X_\alpha\,,
\end{equation}
where $X_\alpha$ are the RMEs and $c_{i\alpha}$ are coefficients of the decomposition given by products of Clebsch-Gordan (CG) coefficients and the CKM-factors in Eq.~\eqref{eq:mixing_coeff}. At orders of breaking less than some maximum order the number of linearly independent RMEs in the decomposition is less than the number of amplitudes and thus there exist sum rules between amplitudes. As it was demonstrated in Ref.~\cite{Gavrilova:2022hbx}, in the case of pure irreps, \emph{i.e.} without direct sums, the $U$-spin symmetry ensures an elegant mathematical structure of the decomposition analogous to Eq.~\eqref{eq:A-u1+u2-RME-decomp}. The understanding of this structure allows for the construction of an algorithm for deriving the sum rules that does not require one to explicitly evaluate the coefficients $c_{i\alpha}$. In this section we use our understanding of the RME decomposition in the case of pure irreps in order to infer some general properties of the decomposition for the $u_1\oplus u_2$ system and its sum rules.

\subsection{The case of trivial mixing coefficients \label{sec:case-no-coeff}  }

A lot of qualitative results about the sum rules for the systems with direct sums of representations can be understood by studying a special case of the $u_1\oplus u_2$ system with trivial mixing coefficients
\begin{equation}
    \Sigma_{m_1} = 1,\qquad \Delta_{m_2} = 1\,,\qquad \forall\, m_1\,,m_2\,:\quad \abs{m_1}\leq u_1\,,\, \abs{m_2} \leq u_2\,.
\end{equation}
This is the case that we study in this section.
It will also be helpful to consider two auxiliary systems:
\begin{equation}\label{eq:auxiliary}
    0 \rightarrow u_1 \oplus \left(\frac{1}{2}\right)^{\oplus k}\,, \qquad 0 \rightarrow u_2 \oplus \left(\frac{1}{2}\right)^{\oplus k}\,,
\end{equation}
which we refer to as the \emph{$u_1$ system} and \emph{$u_2$ system}, respectively. These two auxiliary systems can be considered subsystems of the $u_1 \oplus u_2$ system. They are examples of systems with pure representations and are covered in Ref.~\cite{Gavrilova:2022hbx}. Thus, the sum rules for the two auxiliary systems are known to an arbitrary order of symmetry breaking. 

Since the auxiliary systems are subsystems of the $u_1 \oplus u_2$ system, all the sum rules of the $u_1 \oplus u_2$ system should also hold for the $u_1$ and $u_2$ systems. Our goal in this section is to derive some general rules regarding which of the sum rules of the auxiliary systems carry over to the $u_1 \oplus u_2$ system and which do not.

In order to do this, it is useful to introduce the following definitions. We divide all the amplitudes $A^{u_1\oplus u_2}_i$ of the $u_1\oplus u_2$ system into two subsets. We denote the amplitudes of these subsets as $A_j^{u_1\oplus u_2}(u_1)$ and $A_\ell^{u_1\oplus u_2}(u_1,\,u_2)$. The amplitudes $A_j^{u_1\oplus u_2}(u_1)$ are the amplitudes of the $u_1\oplus u_2$ system for which $\abs{M}>u_2$, while amplitudes $A_\ell^{u_1\oplus u_2}(u_1,\,u_2)$ are the ones for which $\abs{M}\leq u_2$. The amplitudes $A_\ell^{u_1\oplus u_2}(u_1,\,u_2)$ get non-trivial contributions from both representations in the direct sum, $u_1$ and $u_2$, while the amplitudes $A_j^{u_1\oplus u_2}(u_1)$ only get contributions from $u_1$. Note that, in the case of the $u_1 \oplus u_2$ system with trivial coefficients, we use $A$ to denote the amplitudes to emphasize that they are independent of the mixing coefficients, similar to how we denote CKM-free amplitudes in the case of pure irreps (see Eq.~\eqref{eq:A-CKM-free-def}). Also note that we use different indices, $j$ and $\ell$, to label the amplitudes from the two subsets. This distinction emphasizes that the sets are different and do not overlap. Occasionally, we use the index $i$ as a generic index that runs over all amplitudes in the system. Similarly, we divide all the RMEs in the decomposition that enter at order of breaking $b$ into two subsets of RMEs, $X^{(b)}_\alpha(u_1)$ and $X^{(b)}_\beta(u_2)$. The RMEs $X^{(b)}_\alpha(u_1)$ are the RMEs that pick up the representation $u_1$ from the direct sum, while the RMEs $X^{(b)}_\beta(u_2)$ pick up the representation $u_2$. As for the amplitudes, we use indices $\alpha$ and $\beta$ to emphasize that the two subsets do not overlap.

Since $u_1 \neq u_2$, the $u_1$ and $u_2$ systems are described by different numbers of would-be doublets, $n_1 = 2u_1 +k$ and $n_2 = 2u_2 +k$ respectively. This leads to two possibilities. According to Eq.~\eqref{eq:as-def-pure}, in the case when $(-1)^{n_1/2} = (-1)^{n_2/2}$, the amplitudes and their $U$-spin conjugates enter the $a$($s$)-type amplitudes of the $u_1$ and $u_2$ systems with the same relative sign, while in the case when $(-1)^{n_1/2} \neq (-1)^{n_2/2}$, the definitions of $a$($s$)-type amplitudes are different for the $u_1$ and $u_2$ systems. We say that in the first case $u_1$ and $u_2$ systems have \emph{the same parity}, while in the second case they have \emph{different parity}. In the case of different parity, the $a$($s$)-type amplitudes of the $u_1$ system are defined in the same way as $s$($a$)-type amplitudes of the $u_2$ system. Note that this notion of parity is unrelated to the one of spacetime.

We choose a convention in which the $a$($s$)-type amplitudes of the $u_1 \oplus u_2$ system are defined in the same way as in Eq.~\eqref{eq:as-def-pure} with $n = n_1$, where $n$ represents the number of would-be doublets in the $u_1 \oplus u_2$ system, or equivalently, the number of would-be doublets in the $u_1$ system:
\begin{equation}\label{eq:as-def-u1+u2}
    a_i = {A}^{u_1\oplus u_2}_i - (-1)^{n/2} \overline{A}^{u_1\oplus u_2}_i\,,\qquad s_i = {A}^{u_1\oplus u_2}_i + (-1)^{n/2} \overline{A}^{u_1\oplus u_2}_i\,.
\end{equation}
In this convention, the $a$- and $s$-type amplitudes of the $u_1\oplus u_2$ system are always defined in the same way as the $a$- and $s$-type amplitudes of the $u_1$ system, while the definitions for the $u_2$ system might be switched depending on the relative parity of $u_1$ and $u_2$. In particular, when the two systems have the same parity, and thus the definitions of $a$- and $s$-type amplitudes for the $u_1$ and $u_2$ systems match, we find that
\begin{equation}\label{eq:as-same-parity}
    a_i^{u_1\oplus u_2} = a_i^{u_1} + a_i^{u_2}, \qquad s_i^{u_1\oplus u_2} = s_i^{u_1} + s_i^{u_2}\,,\qquad \text{(same parity)}\,,
\end{equation}
where $a_i^{u}$ and $s_i^u$ are the $a$- and $s$-type amplitudes of the system with irrep $u = u_1$ or $u = u_2$. If $i$ is such that $\abs{M} > u_2$, then the amplitudes of the $u_2$ system are set to zero. In the case of different relative parity between $u_1$ and $u_2$, we have
\begin{equation}\label{eq:as-diferent-parity}
    a_i^{u_1\oplus u_2} = a_i^{u_1} + s_i^{u_2}, \qquad s_i^{u_1\oplus u_2} = s_i^{u_1} + a_i^{u_2}\,,\qquad \text{(different parity)}\,.
\end{equation}
This shows that in the same parity case, any $a$($s$)-type sum rule that holds for the $u_1\oplus u_2$ system also holds for the $a$($s$)-type amplitudes of both auxiliary systems. In the different parity case, any $a$($s$)-type sum rule that holds for the $u_1\oplus u_2$ system also holds for the $a$($s$)-type amplitudes of the $u_1$ system and for the $s$($a$)-type amplitudes of the $u_2$ system.

Analogous to how we divide all the amplitudes $A_i^{u_1\oplus u_2}$ of the $u_1\oplus u_2$ system into two subsets, we also divide the $a$- and $s$-type amplitudes of the $u_1\oplus u_2$ system, as defined in Eq.~\eqref{eq:as-def-u1+u2}, into two subsets. We use $a_j(u_1)$ and $s_j(u_1)$ to denote the $a$- and $s$-type amplitudes that are constructed from the amplitudes $A_j^{u_1\oplus u_2}(u_1)$, and $a_\ell(u_1, u_2)$ and $s_\ell(u_1, u_2)$ to denote the $a$- and $s$-amplitudes that are constructed from the amplitudes $A_\ell^{u_1\oplus u_2}(u_1, u_2)$. That is, in the chosen convention, we have
\begin{align}
    a_j(u_1) &= A_j^{u_1\oplus u_2}(u_1) - (-1)^{n/2} \overline{A}_j^{u_1\oplus u_2}(u_1),\nonumber\\
      s_j(u_1) &= A_j^{u_1\oplus u_2}(u_1) + (-1)^{n/2} \overline{A}_j^{u_1\oplus u_2}(u_1),\label{eq:as-u1-def}
\end{align}
and
\begin{align}
    a_\ell(u_1, u_2) &= A_\ell^{u_1\oplus u_2}(u_1, u_2) - (-1)^{n/2} \overline{A}^{u_1\oplus u_2}_\ell(u_1, u_2),\nonumber\\
    s_\ell(u_1, u_2) &= A_\ell^{u_1\oplus u_2}(u_1, u_2) + (-1)^{n/2} \overline{A}^{u_1\oplus u_2}_\ell(u_1, u_2)\,.\label{eq:as-u1u2-def}
\end{align}
For brevity, on the LHS of Eqs.~(\ref{eq:as-u1-def}), (\ref{eq:as-u1u2-def}) and in the following we omit the superscripts $u_1 \oplus u_2$ where the meaning is clear from the context.
Using the results derived for systems of pure representations (see Section III.B of Ref.~\cite{Gavrilova:2022hbx}), we can write the RME decompositions of the $a$- and $s$-type amplitudes of the $u_1\oplus u_2$ system. First, for amplitudes that only get non-trivial contributions from $u_1$, we can directly use the result obtained in Eq.(3.12) of Ref.~\cite{Gavrilova:2022hbx}, we have
\begin{equation}\label{eq:as-u1}
    a_j (u_1) = \sum_{\alpha} \left(1 - (-1)^b\right) C_{j\alpha} X^{(b)}_\alpha (u_1)\,,\qquad 
    s_j (u_1) = \sum_{\alpha} \left(1 + (-1)^b\right) C_{j\alpha} X^{(b)}_\alpha (u_1)\,,
\end{equation}
that is, the $a_j(u_1)$ amplitudes only get contributions from odd orders of breaking, $b$, while the $s_j(u_1)$ amplitudes only get contributions from even orders of breaking. Next, for the amplitudes that get non-trivial contributions from both $u_1$ and $u_2$, the form of the decomposition is different depending on $u_1$ and $u_2$ either having the same or different parity. We have for these amplitudes
\begin{align}
    a_\ell(u_1, u_2) &=  \sum_{\alpha} \left(1 - (-1)^b\right) C_{\ell\alpha} X^{(b)}_\alpha (u_1) + \sum_\beta \left(1 - (-1)^{u_1-u_2+b}\right) C_{\ell\beta} X^{(b)}_\beta(u_2)\,,\nonumber\\
    s_\ell(u_1, u_2) &=  \sum_{\alpha} \left(1 + (-1)^b\right) C_{\ell\alpha} X^{(b)}_\alpha (u_1) + \sum_\beta \left(1 + (-1)^{u_1-u_2+b}\right) C_{\ell\beta} X^{(b)}_\beta(u_2)\,,  \label{eq:as-u1-u2}
\end{align}
In order to derive Eq.~(\ref{eq:as-u1-u2}), we use the CG decomposition for the amplitudes in the form
\begin{align}\label{eq:A-decomp}
 A_\ell^{u_1\oplus u_2}(u_1, u_2) &=  \sum_{\alpha} C_{\ell\alpha} X^{(b)}_\alpha (u_1) + \sum_{\beta} C_{\ell\beta} X^{(b)}_\beta (u_2)\,,
\end{align}
and the decomposition for the corresponding $U$-spin conjugate amplitudes
\begin{align}\label{eq:A-conj-decomp}
(-1)^{\frac{n_1}{2}} \overline{A}_\ell(u_1, u_2) &= (-1)^{\frac{n_1}{2}} \left((-1)^{\frac{n_1}{2}}\sum_{\alpha} (-1)^b C_{\ell\alpha} X^{(b)}_\alpha (u_1) + (-1)^{\frac{n_2}{2}}\sum_{\beta} (-1)^b C_{\ell\beta} X^{(b)}_\beta (u_2)\right) \nonumber\\
%%%
&= \sum_{\alpha} (-1)^b C_{\ell\alpha} X^{(b)}_\alpha (u_1) + \sum_{\beta} (-1)^{\frac{n_1}{2}+\frac{n_2}{2}+b} C_{\ell\beta} X^{(b)}_\beta (u_2)\nonumber\\
%%%%
&= \sum_{\alpha} (-1)^b C_{\ell\alpha} X^{(b)}_\alpha (u_1) + \sum_{\beta} (-1)^{u_1-u_2+b} C_{\ell\beta} X^{(b)}_\beta (u_2)\,.
\end{align}
The first equality in Eq.~\eqref{eq:A-conj-decomp} is due to the general property of the CG decomposition of $U$-spin conjugate amplitudes in the case of pure irreps (see Eq.~(3.8) of Ref.~\cite{Gavrilova:2022hbx}), and in the last line we used 
\begin{align}
(-1)^{\frac{n_1}{2} + \frac{n_2}{2} +b} = 
(-1)^{n_2}(-1)^{\frac{n_1}{2}-\frac{n_2}{2} + b} = 
(-1)^{n_2}(-1)^{\frac{2 u_1+k}{2}-\frac{2 u_2+k}{2} + b} =
(-1)^{u_1 - u_2+b}\,,
\end{align}
where to write the last equality we used the fact that $n_2$ is even (as is $n_1$). We then plug Eqs.~\eqref{eq:A-decomp} and~\eqref{eq:A-conj-decomp} into the definitions of the $a$- and $s$-type amplitudes in Eq.~\eqref{eq:as-u1u2-def} (with $n = n_1$) and arrive at Eq.~\eqref{eq:as-u1-u2}.

The factor $(-1)^{u_1 - u_2}$ in Eq.~\eqref{eq:as-u1-u2} accounts for the relative parity of $u_1$ and $u_2$. The case $(-1)^{u_1 - u_2} = 1$ corresponds to $u_1$ and $u_2$ having the same parity, while the case $(-1)^{u_1 - u_2} = -1$ corresponds to $u_1$ and $u_2$ having different parity.

\begin{table}[t]
\subfigure[Same parity]{
\begin{tabular}{c|c|c||c|c||c|c||c}
& $X^{(0)}(u_1)$ & $X^{(0)}(u_2)$ & $X^{(1)}(u_1)$ & $X^{(1)}(u_2)$ & $X^{(2)}(u_1)$ & $X^{(2)}(u_2)$ &...\\
\hline
$a(u_1)$ & 0 & 0 & \cellcolor{black!25} & 0 & 0 & 0 &\\
\hline
$a(u_1,u_2)$ & 0 & $0$ &\cellcolor{black!25}  &\cellcolor{black!25} & 0 & 0 &\\
\hline\hline
$s(u_1)$ & \cellcolor{black!25} & 0 & 0 & 0& \cellcolor{black!25} &0 &\\
\hline
$s(u_1,u_2)$ & \cellcolor{black!25} & \cellcolor{black!25} & 0 & $0$ & \cellcolor{black!25} & \cellcolor{black!25} &
\end{tabular}}
\subfigure[Different parity]{
\begin{tabular}{c|c|c||c|c||c|c||c}
& $X^{(0)}(u_1)$ & $X^{(0)}(u_2)$ & $X^{(1)}(u_1)$ & $X^{(1)}(u_2)$ & $X^{(2)}(u_1)$ & $X^{(2)}(u_2)$ &...\\
\hline
$a(u_1)$ & 0 & 0 &  \cellcolor{black!25}& 0 & 0 & 0 & \\
\hline
$a(u_1,u_2)$ & 0 & \cellcolor{black!25} & \cellcolor{black!25} & 0 & 0 & \cellcolor{black!25} & \\
\hline\hline
$s(u_1)$ & \cellcolor{black!25} & 0 & 0 & 0& \cellcolor{black!25} & 0 &\\
\hline
$s(u_1,u_2)$ & \cellcolor{black!25} & 0 & 0 & \cellcolor{black!25} & \cellcolor{black!25} & 0 & 
\end{tabular}
}
\caption{Schematic representation of the CG decomposition for the $u_1\oplus u_2$ system, for $u_1$ and $u_2$ having the same (a) and different (b) parity. The shaded cells represent contributions to the decomposition that can contain non-vanishing coefficients. The pattern keeps repeating itself as we go to higher orders in the symmetry breaking.} \label{fig:CG-u1u2}
\end{table}

We schematically demonstrate the structure of the resulting decompositions for the $a$- and $s$-type amplitudes of the $u_1\oplus u_2$ system in Table~\ref{fig:CG-u1u2}.
Tables~\ref{fig:CG-u1u2}(a) and~\ref{fig:CG-u1u2}(b) illustrate the RME decompositions in the case when the $u_1$ and $u_2$ systems have the same and different parity,~respectively. The zeroes in the table indicate that the corresponding RMEs do not contribute to a given $a$($s$)-type amplitude, while the shaded cells represent coefficients in the decomposition that, in principle, can be non-zero. We use the rules in Table~\ref{tab:parity-cases} that follow from Eqs.~\eqref{eq:as-u1} and~\eqref{eq:as-u1-u2} to construct the tables in Table~\ref{fig:CG-u1u2}.
\begin{table}[h!]
    \centering
    \setlength{\leftmargini}{0.4cm}
    \begin{tabular}{p{7cm}p{7cm}}
         \begin{center}\textbf{Same parity case}\end{center} & \begin{center}\textbf{Different parity case}\end{center} \\
         \begin{itemize}
    \item RMEs $X^{(b)}_\beta(u_2)$ only contribute to amplitudes $a_\ell(u_1, u_2)$ and $s_\ell(u_1, u_2)$.
    \item Only RMEs with odd(even) orders $b$ contribute to $a$($s$)-type amplitudes.
\end{itemize} & \begin{itemize}
    \item RMEs $X^{(b)}_\beta(u_2)$ only contribute to amplitudes $a_\ell(u_1, u_2)$ and $s_\ell(u_1, u_2)$.
    \item Among RMEs $X^{(b)}_\alpha(u_1)$, only the ones with odd(even) orders $b$ contribute to $a$($s$)-type amplitudes.
    \item Among RMEs $X^{(b)}_\beta(u_2)$, only the ones with even(odd) orders $b$ contribute to $a$($s$)-type amplitudes.
\end{itemize}
    \end{tabular}
    \caption{Rules for constructing the tables in Table~\ref{fig:CG-u1u2}.} \label{tab:parity-cases}
\end{table}
\FloatBarrier

The key insight that we learn from the structure of the RME decompositions, as shown in Eqs.~\eqref{eq:as-u1} and~\eqref{eq:as-u1-u2} and illustrated in Table~\ref{fig:CG-u1u2}, is that, as in the case of systems with pure irreps~\cite{Gavrilova:2022hbx}, $a$- and $s$-type amplitudes decouple from each other. This is because the sets of RMEs that enter the decompositions of $a$- and $s$-type amplitudes do not overlap. Thus, as in the case of pure irreps, all sum rules of the $u_1\oplus u_2$ system with trivial coefficients can be written as sum rules among $a$-type and $s$-type amplitudes separately.

Moreover, by studying the general structure of the decompositions and assuming that the sum rules for the auxiliary systems in Eq.~\eqref{eq:auxiliary} are known, we can infer general rules regarding the relation between the sum rules of the $u_1 \oplus u_2$ system with trivial coefficients and the sum rules of the $u_1$ and $u_2$ systems, which we summarize below.

\paragraph{Same parity case}

In the case when $u_1$ and $u_2$ have the same parity, we make the following observations.
\begin{enumerate}
\item At LO, there are $n_A/2$ trivial sum rules given by $a_i = 0$, $i = 1,\,\dots,n_A/2$, where $n_A$ is the number of amplitudes in the system. These sum rules hold because none of the $a$-type amplitudes receive contributions from any of the LO RMEs. However, they are broken by $b = 1$ corrections, as none of these sum rules hold for either the $u_1$ or $u_2$ system (see Ref.~\cite{Gavrilova:2022hbx} and the review in Section~\ref{sec:review}).

\item All $a$($s$)-type sum rules that involve only the amplitudes $a_j(u_1)$ or $s_j(u_1)$ and hold for the $u_1$ system at order $b$ also hold for the $u_1 \oplus u_2$ system at order $b$. This is because the $a_j(u_1)$ and $s_j(u_1)$ amplitudes receive contributions only from the RMEs $X^{(b)}_\alpha(u_1)$.
\end{enumerate}

\paragraph{Different parity case}
In the case when $u_1$ and $u_2$ have different parity, we make the following observations. 
\begin{enumerate}
    \item LO trivial $a$-type sum rules that only involve the amplitudes $a_j(u_1)$ also hold for the $u_1\oplus u_2$ system, while the trivial $a$-type sum rules involving the amplitudes $a_j(u_1, u_2)$ are broken.
    \item Additional LO $a$-type sum rules are obtained from the LO $s$-type sum rules of the $u_2$ system via the substitution $s^{u_2}_\ell \rightarrow a^{u_1\oplus u_2}_\ell(u_1, u_2)$. This also follows from Eq.~\eqref{eq:as-diferent-parity}, as at the LO $a^{u_1}_i = 0$, $\forall\, i$. The $a$-type sum rules in this and the previous point form a full set of LO $a$-type sum rules of the $0\oplus 1$ system.
    \item All LO $s$-type sum rules that hold for the $u_1$ system also hold for the $u_1 \oplus u_2$ system. They form the full set of LO $s$-type sum rules for the $u_1\oplus u_2$ system.
    \item All $a$($s$)-type sum rules that only involve amplitudes $a_j(u_1)$ or $s_j(u_1)$ and hold for the $u_1$ system at order $b$ also hold for the $u_1 \oplus u_2$ system at order $b$.
\end{enumerate}

The rules summarized above provide guidance on how to construct some of the sum rules for the $u_1\oplus u_2$ system with trivial coefficients, but they do not guarantee the construction of all linearly independent sum rules. As we will see in later sections, there are additional sum rules that are more subtle and do not follow from the observations made above.

\subsection{The case of non-trivial mixing coefficients}\label{sec:case-with-coeff}

Next we move to the case of $u_1\oplus u_2$ systems with arbitrary non-zero mixing coefficients $\Sigma$ and $\Delta$. We would like to understand how the structure of the RME decomposition in Table~\ref{fig:CG-u1u2} changes when we allow for non-trivial coefficients.

In the case of systems with pure representations in order to ensure the decoupling of $a$- and $s$-type amplitudes we have introduced the CKM-free amplitudes, see Eq.~\eqref{eq:A-CKM-free-def}.  This was possible because in the case of pure irreps each amplitude has exactly one CKM matrix element associated with it. When the CKM factors are divided out and $a$- and $s$-type amplitudes are defined using CKM-free amplitudes via Eq.~\eqref{eq:as-def-pure}, we obtain the decoupling of $a$- and $s$-type amplitudes and all the useful properties that we use to derive the sum rules.

Unfortunately, this basis choice does not translate exactly to the case of $u_1\oplus u_2$ systems, as each amplitude $A_\ell^{u_1 \oplus u_2}(u_1, u_2)$ comes with two different mixing coefficients, one associated with a component of $u_1$ and one with a component of $u_2$. This poses a challenge that comes with non-trivial coefficients $\Sigma$ and $\Delta$. In order to address this challenge, we need to find a basis in which the $a$- and $s$-type amplitudes decouple and recover the structure of the decomposition in Table~\ref{fig:CG-u1u2}. We could not identify such a basis in the most general case of arbitrary irreps $u_1$ and $u_2$ and for all orders of symmetry breaking.

However, there is one special case when the basis of interest is easy to find, this is the case when $u_2 = 0$ and $u_1\equiv u$ is an arbitrary irrep. We discuss this special case in the following section.

\subsection{The special case $u \oplus 0$}\label{sec:u0-coeff}

Let us consider a special case of the $u_1\oplus u_2$ system where $u_1 \equiv u$ and $u_2 = 0$. We write the RME decomposition of amplitudes of this system as follows
\begin{equation}\label{eq:A10-decomp}
    \mathcal{A}_i^{u \oplus 0} = \sum_\alpha C_{i\alpha}\, \Sigma_{m} X^{(b)}_\alpha(u) + \sum_\beta  C_{i\beta}\, \Delta_0 X^{(b)}_\beta(0)\,,
\end{equation}
where we emphasize that we are discussing the system with non-trivial mixing coefficients by using $\mathcal{A}_i^{u \oplus 0}$ to denote the amplitudes. We separate the contributions from the RMEs $X^{(b)}_\alpha(u)$ and $X^{(b)}_\beta(0)$ and explicitly write down the coefficients in the decomposition as products of CG coefficients $C_{i\alpha}$ or $C_{i\beta}$ and mixing coefficients $\Sigma_{m}$ or $\Delta_0$. The RME decompositions of amplitudes $\mathcal{A}_j^{u \oplus 0}(u)$ and $\mathcal{A}_\ell^{u \oplus 0}(u,0)$ are given by
\begin{align}
        \mathcal{A}_j^{u \oplus 0}(u) &= \sum_{\alpha} C_{j\alpha}\,\Sigma_{m} X^{(b)}_\alpha(u)\,,\\
        \mathcal{A}^{u \oplus 0}_\ell(u, 0) &= \sum_{\alpha} C_{\ell\alpha}\,\Sigma_{0} X^{(b)}_\alpha(u) + \sum_\beta  C_{\ell\beta}\,\Delta_0 X^{(0)}_\beta\,.
\end{align}

For the amplitudes $\mathcal{A}_j^{u \oplus 0}(u)$, it is straightforward to define amplitudes free of mixing coefficients by dividing them by the corresponding mixing coefficients, as follows:
\begin{equation}
    A_j^{u \oplus 0}(u) \equiv \frac{\mathcal{A}^{u \oplus 0}_j(u)}{\Sigma_m}\,.
\end{equation}
This definition ensures that when $a_j(u)$ and $s_j(u)$ amplitudes are defined using the convention of Eq.~\eqref{eq:as-u1-def}, the resulting RME decomposition of these $a$- and $s$-type amplitudes has the same structure as in Eq.~\eqref{eq:as-u1} and as illustrated in Table~\ref{fig:CG-u1u2}.

Next, let us consider the RME decompositions of the amplitudes $\mathcal{A}^{u \oplus 0}_\ell(u,0)$ divided by $\Sigma_0$. We have
\begin{equation}
    A_\ell^{u \oplus 0}(u,0) \equiv \frac{\mathcal{A}^{u \oplus 0}_\ell(u, 0)}{\Sigma_0} = \sum_{\alpha} C_{\ell\alpha} X^{(b)}_\alpha + \sum_\beta  C_{\ell\beta}\,\frac{\Delta_0}{\Sigma_0} X^{(0)}_\beta\,.
\end{equation}
From this equation, it becomes clear why the $u \oplus 0$ system we consider here is special. The first sum on the RHS is free of mixing coefficients. The second sum is not free of mixing coefficients; however, the factor $\Delta_0/\Sigma_0$ is the same for all $\ell$ and thus is the same for all amplitudes on the LHS. This is a consequence of the fact that the singlet has only one component. Because of this, even though the amplitudes $A^{u \oplus 0}_\ell(u,0)$, strictly speaking, depend on the mixing coefficients, they still ensure that, given the definitions of $a$- and $s$-type amplitudes in Eq.~\eqref{eq:as-u1u2-def}, the structure of Eq.~\eqref{eq:as-u1-u2} and Table~\ref{fig:CG-u1u2} is recovered.

Another way to think about this is that the ratio of mixing coefficients $\Delta_0/\Sigma_0$ can be absorbed by redefinitions of RMEs. We can define RMEs $\widetilde{X}_{\beta}^{(b)}(0)$ as follows
\begin{equation}
    \widetilde{X}_{\beta}^{(b)}(0) = \frac{\Delta_0}{\Sigma_0}X^{(b)}_\beta(0)\,.
\end{equation}
We have then for the amplitudes $A_\ell^{u \oplus 0}(u,0)$, 
\begin{equation}
    A_\ell^{u \oplus 0}(u,0) \equiv \frac{\mathcal{A}^{u \oplus 0}_j(u, 0)}{\Sigma_0} = \sum_{\alpha} C_{\ell\alpha} X^{(b)}_\alpha(u) + \sum_\beta  C_{\ell\beta} \widetilde{X}^{(b)}_\beta(0)\,,
\end{equation}
and it is straightforward to see that the structure of the decomposition in the case with trivial coefficients is recovered. Thus in the special case of the $u \oplus 0$ systems a convenient basis choice for the amplitudes is
\begin{equation}\label{eq:A_u+0-def}
    A_i^{u \oplus 0} \equiv \frac{\mathcal{A}^{u \oplus 0}_i}{\Sigma_m}\,,
\end{equation}
where $\Sigma_m$ is the relevant mixing coefficient. This definition of the amplitudes $A_i^{u \oplus 0}$ ensures that the RME decomposition of the $a$- and $s$-type amplitudes defined using Eqs.~\eqref{eq:as-u1-def} and~(\ref{eq:as-u1u2-def}) has the same structure as the decomposition of the $u \oplus 0$ system with trivial coefficients. Thus as in the case of systems with pure representations, we can first find the sum rules between amplitudes $A^{u \oplus 0}_i$ and then use the definition in Eq.~\eqref{eq:A_u+0-def} to rewrite them in terms of physical amplitudes $\mathcal{A}_i^{u \oplus 0}$. From this we also learn that, in the case of $u \oplus 0$ systems, the sum rules between physical amplitudes do not depend on the coefficient $\Delta$ that accompanies the singlet in the direct sum.

The special case of the $u \oplus 0$ system with $u = 1$ is the simplest mathematically and is of particular physical interest. The direct sum $0\oplus 1$ describes the group-theoretical structure of several physical systems, including the $U$-spin structure of the effective Hamiltonian for charm decays and the isospin structure of $\pi-\eta$ mixing. Due to its mathematical simplicity and physical significance, we will focus, for the remainder of this paper, on studying systems whose $SU(2)$ flavor symmetry structure includes a direct sum of a singlet and a triplet, $0\oplus 1$.

\section{$0 \oplus 1$ Systems \label{sec:0+1}}

Our goal is to derive all linearly independent sum rules to an arbitrary order of symmetry breaking for a system with the following group-theoretical structure
\begin{equation}\label{eq:system-0}
    0 \rightarrow \left(\Delta\, 0 \oplus \Sigma\, 1\right)\otimes\left(\frac{1}{2}\right)^{\otimes (n - 2)}\,, 
\end{equation}
where, in accordance with Eq.~(\ref{eq:number-of-would-be-doublets}), the number of would-be doublets is $n$. Here, $\Sigma$ and $\Delta$ represent the triplet and singlet mixing coefficients, respectively
\begin{equation}
    \Sigma = \left(\Sigma_{-1},\, \Sigma_{0},\, \Sigma_{+1}\right)\,, \qquad \Delta = \Delta_0\,.
\end{equation}
We denote the amplitudes of this system as $\mathcal{A}^{0\oplus 1}_i$. As shown in Section~\ref{sec:u0-coeff}, for systems of the form given in Eq.~\eqref{eq:system-0}, it is convenient to introduce the amplitudes $A_i^{0\oplus 1}$ as
\begin{equation}\label{eq:A-0+1-def}
    A^{0\oplus 1}_i = \frac{\mathcal{A}_i^{0\oplus 1}}{\Sigma_m}\,,
\end{equation}
where the mixing coefficient $\Sigma_m$ corresponds to the amplitude $\mathcal{A}^{0\oplus 1}_i$. As explained in Section~\ref{sec:u0-coeff}, the sum rules between the amplitudes $A_i^{0\oplus 1}$ are identical to those for a $0\oplus 1$ system with trivial coefficients, that is the system given by
\begin{equation}\label{eq:system-0+1}
    0 \rightarrow \left(0 \oplus 1\right)\otimes\left(\frac{1}{2}\right)^{\otimes (n - 2)}\,. 
\end{equation}
Thus, in this section, we focus on deriving the sum rules for the system in Eq.~\eqref{eq:system-0+1}. When considering the physical system, the sum rules between the physical amplitudes $\mathcal{A}_i^{0\oplus 1}$ are obtained from the sum rules for the amplitudes $A_i^{0\oplus 1}$ using Eq.~\eqref{eq:A-0+1-def}. For simplicity, throughout the remainder of this section, we will refer to this system as the \emph{$0\oplus 1$ system}. At the very end of this section, we also briefly discuss the generalization of the results to the case where, in addition to the direct sum, the system includes arbitrary irreps beyond doublets, and where the initial state and the Hamiltonian are non-singlets.

\subsection{Amplitudes of the $0\oplus 1$ system}\label{sec:amps-0+1}

The amplitudes of the system in Eq.~\eqref{eq:system-0+1} can be labeled by listing $n-1$ $m$ QNs. To label the amplitudes we choose to assign the first $n-2$ $m$ QNs to the $n-2$ doublets and the last $m$ QN to the direct sum $0\oplus 1$. We thus denote the amplitudes of this system as $A^{0\oplus 1}(m_1,\,\dots,\,m_{n-2},\,M)$, where $m_i = \pm 1/2$ for $i = 1,\,\dots,\,n-2$ and we use $M$ for the last $m$ QN to emphasize that it corresponds to the direct sum, $M = \pm 1,\,0$. As always the complete set of amplitudes of the system is given by all the sets $\{m_1,\,\dots,\,m_{n-2},\,M\}$ that fulfill
\begin{equation}\label{eq:sum-m-0}
    \sum_{i=1}^{n-2} m_i + M = 0\,.
\end{equation}

For example, consider a system of the form in Eq.~\eqref{eq:system-0+1} with $n = 4$. There are 4 possible sets $\{m_1, m_2, M\}$ that satisfy the condition in Eq.~\eqref{eq:sum-m-0}, thus the system with $n = 4$ has 4 amplitudes
\begin{align}\label{eq:amps-0+1-mQN-not}
    &A^{0\oplus 1}(-1/2,\,-1/2,\,+1)\,, &&A^{0\oplus 1}(+1/2,\,+1/2,\,-1)\,,\nonumber\\
    &A^{0\oplus 1}(-1/2,\,+1/2,\,0)\,, &&A^{0\oplus 1}(+1/2,\,-1/2,\,0)\,.
\end{align}

Since the amplitudes of the system with a direct sum can be labeled by the $m$ QNs of irreps in exactly the same way the amplitudes of the systems with pure representations are, we can use the generalized $n$-tuple notation introduced in Section IV.A of Ref.~\cite{Gavrilova:2022hbx} for the system with direct sums as well. A generalized $n$-tuple in the case of the system in Eq.~\eqref{eq:system-0+1} is a string of $n$ plus and minus signs that is built according to the following rules:
\begin{enumerate}
    \item The $n$-tuple has $n-1$ positions separated by commas. Without loss of generality, we choose the first $n-2$ positions to correspond to the $n-2$ doublets and the last position to correspond to the direct sum.
    \item The first $n-2$ positions of the $n$-tuple are assigned signs based on the $m$ QNs $m_1$ through $m_{n-2}$. If $m_i = -1/2$ the corresponding position of the $n$-tuple has a ``$-$'' sign, if $m_i = +1/2$, the corresponding position of the $n$-tuple has a ``$+$'' sign, where $i = 1,\,\dots,\,n-2$.
    \item The last position of the $n$-tuple has two signs. For $m_{n-1} = -1$ the signs are ``$--$'', for $m_{n-1} = +1$ the signs are ``$++$'', for $m_{n-1} = 0$ the signs are ``$-+$''.
\end{enumerate}

For example, for $n = 4$ the amplitudes in Eq.~\eqref{eq:amps-0+1-mQN-not} can be mapped into $n$-tuples as follows
\begin{align}\label{eq:amps-0+1-n-tup-not}
    &A^{0\oplus 1}(-1/2,\,-1/2,\, + 1) \equiv (-,\,-,\,++)\,,&&A^{0\oplus 1}(+1/2,\,+1/2,\, -1)\equiv (+,\,+,\,--)\,,\nonumber\\
    &A^{0\oplus 1}(-1/2,\,+1/2,\, 0)\equiv (-,\,+,\,-+)\,,&&A^{0\oplus 1}(+1/2,\,-1/2,\, 0)\equiv (+,\,-,\,-+)\,.
\end{align}

For an arbitrary $n$ the number of amplitudes in the system described by $n-2$ doublets and a direct sum $0\oplus1$ is found as the number of all possible sets $\{m_1,\,\dots,\,m_{n-2},\,M\}$ such that the condition in Eq.~\eqref{eq:sum-m-0} is satisfied. We denote this number by $n_A^{0\oplus 1} (n)$, where $n$ in parentheses indicates the number of would-be doublets of the system. We derive $n_A^{0\oplus 1} (n)$ in Appendix~\ref{app:counting-0+1} (Eq.~(\ref{eq:nA-0+1-final})) to be 
\begin{equation}\label{eq:nA-0+1}
    n_A^{0\oplus 1} (n) = 
    \frac{3n-4}{n}\binom{n-2}{n/2-1}\,,
\end{equation}
which can be expressed in terms of the number of amplitudes in the system of doublets as 
\begin{equation}
    n_A^{0\oplus 1} (n) = n_A^d(n) - n_A^d(n-2)\,,
\end{equation}
where $n_A^d(n)$ is given in Eq.~\eqref{eq:nA}. The values of this expression for several values of $n$ are shown in Table~\ref{tab:nA-nRME-0+1}. The number of sum rules for the $0\oplus 1$ system, $n_{SR}^{0 \oplus 1}(n,\,b)$, also given in Table~\ref{tab:nA-nRME-0+1}, is further discussed after Eq.~(\ref{eq:nSR-0+1}) below.

\begin{table}[t]
\centering
\begin{tabular}{|c|c|c|c|c|c|}
\hline
~$n$~ & ~$n_A^{0 \oplus 1}(n)$~ & ~$n_{SR}^{0 \oplus 1}(n,\,0)$~ & ~$n_{SR}^{0 \oplus 1}(n,\,1)$~ & ~$n_{SR}^{0 \oplus 1}(n,\,2)$~ & ~$n_{SR}^{0 \oplus 1}(n,\,3)$~\\
\hline
~~$4$~~ & $4$ & $2$ & $0$ & $0$ & $0$ \\
~~$6$~~ & $14$ & $9$ & $2$ & $0$ & $0$ \\
~~$8$~~ & $50$ & $36$ & $13$ & $2$ & $0$ \\
~~$10$~~ & $182$ & $140$ & $64$ & $17$ & $2$ \\
\hline
\end{tabular}
\caption{Counting of the numbers of amplitudes $n_A^{0 \oplus 1}$ and numbers of sum rules $n_{SR}^{0 \oplus 1}(n,\,b)$ at different orders of breaking $b$ for systems of the form Eq.~\eqref{eq:system-0+1}. \label{tab:nA-nRME-0+1}}
\end{table}

\subsection{The emergence of the $0 \oplus 1$ system from the system of doublets}\label{sec:0+1-from-n-doublets}

The system of $n$ doublets contains in itself the $0\oplus 1$ system. In this section, we discuss how the $0\oplus 1$ system emerges from the system of $n$-doublets.
Let us start by considering a system of $n$ doublets in the final state
\begin{equation}\label{eq:system-n-doub}
    0 \rightarrow \left(\frac{1}{2}\right)^{\otimes n}\,.
\end{equation}
We denote the amplitudes of this system as $A^d(m_1,\,\dots,\,m_n)$, where the amplitudes are labeled by listing the $m$ QNs of the $n$ doublets, $m_i = \pm1/2$ for all $i = 1,\,\dots,\,n$, and the sum of the $m$ QN used to label the amplitudes is equal to zero.

As in Section~\ref{sec:gen-case}, we consider two auxiliary systems. First, an auxiliary system build from the system of $n$ doublets in Eq.~\eqref{eq:system-n-doub} by symmetrizing two of the doublets. This results in the emergence of a system with the following symmetry structure
\begin{equation}\label{eq:system-n-sym}
    0\rightarrow 1\otimes\left(\frac{1}{2}\right)^{\otimes (n-2)}\,.
\end{equation}
We refer to this system as \emph{$u = 1$ system}. The symmetrization procedure is discussed in detail in Section IV.B of Ref.~\cite{Gavrilova:2022hbx}. To construct the $u = 1$ system from the system in Eq.~\eqref{eq:system-n-doub}, we choose to symmetrize the last two doublets of the system of doublets only. We denote the amplitudes of the system with one symmetrization by $A^{(1)}(m_1,\,\dots,\,m_{n-2},\,M)$, where $m_i = \pm 1/2$ for $i = 1,\,\dots,\,n-2$ and we use $M$ for the last $m$~QN to emphasize that it corresponds to the triplet, $M = \pm 1,\,0$. Next, we perform the symmetrization explicitly. The amplitudes of the system of $n$ doublets for which $m_{n-1} = m_{n}$ are already symmetrized and thus they map trivially onto the amplitudes of the system with one symmetrization
\begin{equation}\label{eq:symmetrization-1}
    A^{(1)}(m_1,\,\dots,m_{n-2},\,\pm 1) = A^d(m_1,\,\dots,\,m_{n-2},\,\pm 1/2,\,\pm 1/2)\,.
\end{equation}
The amplitudes for which $m_{n-1} \neq m_n$ require symmetrization, which we write explicitly as
\begin{align}\label{eq:symmetrization-2}
    A^{(1)}(m_1,\,\dots,m_{n-2},\, 0) = \frac{1}{\sqrt{2}}\left(\right. &A(m_1,\,\dots,m_{n-2},\, +1/2,\,-1/2) +\nonumber\\
    &\left. A(m_1,\,\dots,m_{n-2},\, -1/2,\,+1/2)\right)\,.
\end{align}

The second auxiliary system that we consider is a system of $n$ doublets with an anti-symmetrization of two doublets. The anti-symmetrization of two doublets in the system in Eq.~\eqref{eq:system-n-doub} gives a system with the following group-theoretical structure
\begin{equation}\label{eq:system-n-antisym}
    0\rightarrow 0\otimes\left(\frac{1}{2}\right)^{\otimes (n-2)}\,,
\end{equation}
which is equivalent to a system of just $n-2$ doublets in the final state. We refer to this system as \emph{$u = 0$ system}. We denote the amplitudes of the system with an anti-symmetrization as $A^{(0)}(m_1,\,\dots,\,m_{n-2},\,M)$, where $m_i = \pm 1/2$ for $i = 1,\,\dots,\,n-2$ and $M = 0$ for all amplitudes. The only amplitudes of the system of $n$ doublets that allow for anti-symmetrization are the ones for which $m_{n-1}\neq m_{n}$. For these amplitudes the anti-symmetrization takes the following form
\begin{align}\label{eq:symmetrization-3}
    A^{(0)}(m_1,\,\dots,m_{n-2},\, 0) = \frac{1}{\sqrt{2}}\left(\right. &A(m_1,\,\dots,m_{n-2},\, +1/2,\,-1/2) -\nonumber\\
    &\left. A(m_1,\,\dots,m_{n-2},\, -1/2,\,+1/2)\right)\,.
\end{align}

Our goal is to see how the amplitudes $A^d(m_1,\,\dots,\,m_{n-1},\,m_n)$ of the system of $n$~doublets map into the amplitudes $A^{0 \oplus 1}(m_1,\,\dots,\,m_{n-2}, M)$ of the $0\oplus 1$ system. The amplitudes of the $0\oplus1$ system are obtained by simply adding together the amplitudes of the two auxiliary systems. For the amplitudes of the $0\oplus 1$ system with $M = \pm1$, the only contribution is coming from the system with one symmetrization, that is we have
\begin{equation}\label{eq:A-0+1-def-1}
    A^{0\oplus 1}(m_1,\,\dots,m_{n-2},\,\pm 1) \equiv A^{(1)}(m_1,\,\dots,m_{n-2},\,\pm 1) = A^d(m_1,\,\dots,\,m_{n-2},\,\pm 1/2,\,\pm 1/2)\,.
\end{equation}
For the amplitudes of the $0\oplus 1$ system with $M = 0$, there are contributions from both auxiliary systems
\begin{align}\label{eq:A-0+1-def-2}
    A^{0\oplus 1}(m_1,\,\dots,m_{n-2},\,0) &\equiv A^{(1)}(m_1,\,\dots,m_{n-2},\,0) + A^{(0)}(m_1,\,\dots,m_{n-2},\,0)\nonumber\\
    &= \sqrt{2}A^d(m_1,\,\dots,\,m_{n-2},\,+ 1/2,\,- 1/2)\,.
\end{align}

We emphasize the following points about this construction:
\begin{enumerate}
    \item The system of $n$ doublets contains the $0\oplus 1$ system and each amplitude of the $0\oplus 1$ system maps into exactly one amplitude of the system of doublets.
    \item There are less amplitudes in the $0\oplus 1$ system than in the system of $n$ doublets, see also the explicit expressions for the numbers of amplitudes in the two cases in Eqs.~\eqref{eq:nA-0+1} and~\eqref{eq:nA}, respectively.
    \item The number of amplitudes is still even and each amplitude has a corresponding $U$-spin partner\footnote{Note that this feature is specific to systems with at least one half-integer representation. In integer-only systems, some of the amplitudes are their own $U$-spin conjugates, see the discussion in Sec.~IV.A of Ref.~\cite{Gavrilova:2022hbx}.}.
\end{enumerate}

The first point is to be contrasted with the case of the system with one symmetrization when some of the amplitudes map into a linear combination of the amplitudes of the system of doublets, and the case of the systems with one anti-symmetrization when all the amplitudes map into linear combinations, see Eqs.~\eqref{eq:symmetrization-1}--\eqref{eq:symmetrization-3}. The fact that each amplitude of the $0\oplus 1$ system maps into exactly one amplitude of the system of doublets allows us to think about the relation between the system of $n$ doublets and the system with the direct sum $0\oplus 1$ in the following way. Up to overall symmetry factors, the system with the direct sum $0 \oplus 1$ is obtained from the system of $n$ doublets by disregarding some of the amplitudes. To be more concrete, we first choose which two doublets of the system of $n$ doublets to use to construct the direct sum $0 \oplus 1$. Then, among the amplitudes that have different $m$ QNs corresponding to these two doublets, we only keep the amplitudes with a fixed ordering of these two QNs, and disregard the rest. All the amplitudes of the system of doublets for which the chosen two doublets have the same $m$ QN are kept. In the language of $n$-tuples, this means that we keep all the amplitudes where the $n$-tuples have the same signs at the two chosen positions. Among the $n$-tuples that have different signs at the chosen positions, we only keep one of the two possible orderings. As Eq.~\eqref{eq:A-0+1-def-2} shows, with the chosen phase convention for the CG coefficients in Eqs.~\eqref{eq:symmetrization-2} and~\eqref{eq:symmetrization-3}, the ordering ``$+-$'' is the one that is kept and the ordering ``$-+$'' is disregarded. Essentially what this means is that when we construct a system with a direct sum from a system of doublets some of the amplitudes of the system of doublets are identified. The identified amplitudes are the ones that have ``$+-$'' and ``$-+$'' at the position of $n$-tuples that correspond to the direct sum $0\oplus 1$, but otherwise are identical. From there, the specific ordering of the two signs is a convention that does not affect the sum rules.

\subsubsection{Example $n = 4$}

To illustrate the emergence of the $0\oplus1$ system from the system of $n$ doublets, we consider the simplest example of $n = 4$. This $U$-spin system has $4$ amplitudes that are listed in Eq.~\eqref{eq:amps-0+1-n-tup-not} with their corresponding $n$-tuples.

We would like to express the amplitudes in Eq.~\eqref{eq:amps-0+1-n-tup-not} in terms of the amplitudes of the system of $n = 4$ doublets. According to Eq.~\eqref{eq:nA}, the system of $4$ doublets has $n_A = 6$ amplitudes. The $n$-tuples for this system are given by
\begin{align}
    A_{1}^d & = \left(-,-,+,+\right), & \overline{A}_{1}^d &= \left(+,+,-,-\right),\\
    A_{2}^d& = \left(-,+,-,+\right), & \overline{A}_{2}^d& = \left(+,-,+,-\right),\\
    A_{3}^d & = \left(-,+,+,-\right),& \overline{A}_{3}^d & = \left(+,-,-,+\right)\,,
\end{align}
where we choose an arbitrary labeling of the amplitudes via $i = 1,\,2,\,3$ for brevity.

As above to express the amplitudes of the system with the direct sum in terms of amplitudes of the system of doublets, we choose to (anti-)symmetrize the last two doublets in the $n$-tuple. In this convention and using Eqs.~\eqref{eq:A-0+1-def-1} and~\eqref{eq:A-0+1-def-2}, we write for the amplitudes of the $0\oplus 1$ system
\begin{align}
    &A^{0 \oplus 1}(-1/2,\,-1/2,\, + 1) = A_1^d,&& {A}^{0 \oplus 1}(+1/2,\,+1/2,\, -1) = \overline{A}_1^d\nonumber\\
   &A^{0 \oplus 1}(-1/2,\,+1/2,\, 0) =\sqrt{2} A_3^d,&& {A}^{0 \oplus 1}(+1/2,\,-1/2,\, 0) = \sqrt{2} \overline{A}_2^d\,.
\end{align}
Two amplitudes in each line above form a $U$-spin pair of the $0\oplus 1$ system with $n = 4$.

\subsection{Sum rules of the $0\oplus 1$ system}\label{sec:sum-rules}

The notion of the emergence of the $0\oplus 1$ system from the system of doublets is crucial for our understanding of the relation between the sum rules of the auxiliary systems in Eqs.~\eqref{eq:system-n-sym} and~\eqref{eq:system-n-antisym} and the sum rules of the $0\oplus 1$ system. Here we discuss the important consequences of this construction.

First of all, the fact that the $0\oplus 1$ system can be constructed from the system of doublets by removing some of the amplitudes teaches us that in principle all RMEs that are present in the decomposition of the amplitudes for the system of doublets are also present for the system with the direct sum.  
Naively, it might seem as if this means that the rank of the coefficient matrix (and thus the number of sum rules) can be found by simply taking the total number of amplitudes $n_A^{0\oplus 1}(n)$ and subtracting from it the number of linearly independent RMEs for the system of $n$ doublets. This is, however, incorrect, and something more interesting takes place. As we show in detail in Appendix~\ref{app:counting-0+1}, the removal of amplitudes from the system of doublets not only reduces the number of amplitudes but also introduces new linear dependencies between RMEs that are not present for the system of doublets.

In Appendix~\ref{app:counting-0+1}, we find that the linear dependencies have the following structure. For any order of breaking $b$, all RMEs of the $u = 0$ system at order $b$, $X_\beta^{(b)}(0)$, are linearly dependent with certain RMEs of the $u = 1$ system $X_\alpha^{(b-1)}(1)$ at one order lower and RMEs $X_{\alpha}^{(b + 1)}(1)$ at one order higher. Here we use the notation for RMEs introduced in Section~\ref{sec:case-no-coeff}.
This leads to the following important consequences that are essential to our derivation of the sum rules below.
\begin{enumerate}
    \item All the sum rules of the $0\oplus 1$ system can be constructed as sum rules of the $u = 1$ system.
    \item For any $b$, all the sum rules of the $u = 1$ system that hold at order $b + 1$ also hold for the $0\oplus 1$ system at order $b$.
    \item Accounting for the linear dependencies described above allows us to find $n_{SR}^{0\oplus 1}(n,\,b)$, the number of sum rules for the $0\oplus 1$ system at order $b$, with the number of doublets given by $n$ (see Appendix~\ref{app:counting-0+1} for details). We find that the final answer can be written in terms of $n_{SR}^d(n,\,b)$, the number of sum rules for the system of doublets given in Eq.~\eqref{eq:nSR_doublets}, in the following compact form:
    \begin{equation}\label{eq:nSR-0+1}
    n_{SR}^{(0\oplus1)}(n,b) = n^{d}_{SR}(n, b) - n^{d}_{SR}(n-2, b-1)\,.
    \end{equation}
     We note in passing that Eq.~(\ref{eq:nSR-0+1}) is remarkably simple. We suppose therefore that a simpler derivation that does not require an explicit study of the linear dependencies of the CG decomposition might also exist.
\end{enumerate}

For the reader's reference, we show the values of $n_{SR}^{(0\oplus1)}(n,b)$ for a few values of $n$ in Table~\ref{tab:nA-nRME-0+1}. Using Eqs.~\eqref{eq:nSR_doublets} and~\eqref{eq:nSR-0+1}, one can show that
\begin{equation}
    n_{SR}^{0\oplus 1}(n,\, b) > 0, \qquad \forall b \leq n/2-2\,,
\end{equation}
and
\begin{equation}
    n_{SR}^{0\oplus 1}(n,\, b) = 0, \qquad b = n/2-1\,,
\end{equation}
\emph{i.e.},~the system does not have any sum rules for $b\geq n/2 - 1$.
This allows us to find $b_\text{max}^{0\oplus 1}$, the maximum order at which the $0\oplus 1$ system has sum rules, as
\begin{equation}\label{eq:b_max}
    b_\text{max}^{0\oplus 1} = n/2 - 2\,.
\end{equation}

In what follows, we present a simple algorithm for the derivation of sum rules for the $0\oplus 1$ system. In the algorithm, we consider the $u = 1$ system and construct the sum rules that carry over to the $0\oplus 1$ system. We present the algorithm using the lattice approach introduced in Sections III.F and IV.C of Ref~\cite{Gavrilova:2022hbx}. Below, we first review the construction of the lattice for the system with one symmetrization, and then we formulate the algorithm that allows us to read off the sum rules from the lattice.

\subsubsection{Lattice for the $0\oplus 1$ system}\label{sec:lattice-review}

Since, as explained above, we are constructing the sum rules for the $0 \oplus 1$ system from the sum rules for the $u = 1$ system, the lattice for the $0\oplus 1$ system, in our approach, is identical to that for the $u = 1$ system. Let us review therefore the lattice construction and the harvesting of sum rules in the case of a system of $n$ doublets with one symmetrization.

\paragraph{Coordinate notation.} In the geometrical picture, amplitude pairs of the $0 \oplus 1$ system are mapped onto a $d = n/2 - 1$ dimensional lattice, where each node represents a single amplitude pair. In Section~\ref{sec:amps-0+1}, we discuss the $m$ QN and $n$-tuple notation for the $0\oplus 1$ system, which is identical to that for a system with one symmetrization. Here, we adopt the same convention for $n$-tuples as in Section~\ref{sec:amps-0+1} (see Eq.~\eqref{eq:amps-0+1-n-tup-not} and the discussion above it). Specifically, we choose the first $n-2$ positions of the $n$-tuple to represent $n-2$ doublets, while the last position to represent the triplet. Without loss of generality, we choose to represent $U$-spin pairs by $n$-tuples that start with a minus sign.

We use $r$ to denote the position in the $n$-tuple corresponding to the triplet. In accordance with our convention, for the $0 \oplus 1$ system in Eq.~\eqref{eq:system-0+1}, we have $r = n - 2$. To map an amplitude pair onto a lattice node, we enumerate the positions of the $n$-tuple starting from $0$ for the first minus sign and up to $r$ for the position that corresponds to the triplet. For each minus sign (excluding the first one), we list its position, resulting in a string of $d = n/2 - 1$ numbers that constitute the coordinate notation. For example, for $n = 6$, we have:
\begin{align}
    (\underset{0}{-},\underset{1}{-},\underset{2}{+},\underset{3}{+}, \underset{4}{-+}) &= (1,\, 4)\,,\\
    (\underset{0}{-},\underset{1}{+},\underset{2}{+},\underset{3}{+}, \underset{4}{--}) &= (4,\, 4)\,,\\
    (\underset{0}{-},\underset{1}{-},\underset{2}{-},\underset{3}{+}, \underset{4}{++}) &= (1,\, 2)\,,
\end{align}
where on the LHS are the $n$-tuples and on the RHS are the corresponding lattice nodes. Note that, by construction, out of the $d$ coordinates, no more than one can be equal to $1,\,\dots,\,n-3$, and no more than two can be equal to $r$. Additionally, each lattice node corresponds to a unique amplitude pair;  however, the reverse is not true. In general, the lattice contains multiple nodes that map onto the same amplitude. These nodes are related by the permutation of coordinates. For example, $(1, 2, 3) = (2, 1, 3) = (3, 1, 2)$ are all identical nodes, i.e., they all map onto a single amplitude pair.

\paragraph{Symmetry factors.}  Every lattice node is accompanied by a $\mu$-factor, which arises from the symmetrization of the two doublets. For a general discussion of $\mu$-factors, see Section IV.C.2 and 
Eq.~(4.34) of Ref.~\cite{Gavrilova:2022hbx}. In the case of the lattice of the $u = 1$ system (and thus the $0\oplus 1$ system), the $\mu$-factor for a node with the coordinate notation $(x_1,\,\dots,\, x_d)$ is given by
\begin{equation}\label{eq:mu-u1-def-main}
    \mu(x_1,\,\dots,\,x_{d}) = \begin{cases}
        1, \quad \text{if } x_i \neq r\text{ for all } i,\\
        \sqrt{2}, \quad \text{if only one of the } x_i \text{ is equal to } r,\\
        2, \quad \text{if two of the } x_i \text{ are equal to } r.
    \end{cases}
\end{equation}

\begin{figure}[t]
\centering
\includegraphics[width=0.4\textwidth]{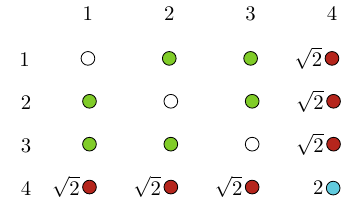}
\caption{Lattice for a $0\oplus 1$ system with $n=6$. Figure adapted from Ref.~\cite{Gavrilova:2022hbx}.}
\label{fig:lattice-0+1-n=6}
\end{figure}

\paragraph{Harvesting the sum rules.} Once the lattice is constructed and the $\mu$-factors are assigned to each node, we can harvest the sum rules. For any even (odd) $b$, the sum rules are read off the lattice as follows:
\begin{itemize}
    \item The $a$($s$)-type sum rules are obtained by summing the $a$($s$)-type amplitudes over all $b$-dimensional subspaces of the lattice, with each amplitude weighted by its corresponding $\mu$-factor. These sum rules are broken by corrections at order $b + 1$.
    \item The $s$($a$)-type sum rules are obtained by summing the $s$($a$)-type amplitudes over all $(b + 1)$-dimensional subspaces of the lattice, weighted by their corresponding $\mu$-factors. These sum rules carry to order $b + 1$.
\end{itemize}
A $b$-dimensional subspace of the lattice is defined by fixing the values of any $d - b$ coordinates in the coordinate notation.

As an example, in Fig.~\ref{fig:lattice-0+1-n=6}, we show the lattice for the $u = 1$ system in the case where the number of would-be doublets is $n = 6$. All the solid nodes represent ``allowed'' nodes, meaning nodes that map to an amplitude pair, while the empty (white) nodes are not allowed. In the context of the $u = 1$ lattice being used for the construction of sum rules for the $0\oplus 1$ system, it is useful to distinguish three types of lattice nodes:
\begin{itemize}
    \item[(\emph{i})] Nodes for which none of the coordinates are equal to $r$. In the case of the lattice in Fig.~\ref{fig:lattice-0+1-n=6}, these nodes are shown in green.
    \item[(\emph{ii})] Nodes for which exactly one coordinate is equal to $r$. These nodes are shown in red.
    \item[(\emph{iii})] Nodes for which two coordinates are equal to $r$. These nodes are shown in blue.
\end{itemize}
The green nodes of type (\emph{i}) represent amplitude pairs with the $m$ QN of the direct sum equal to $M = \pm 1$. The blue nodes of type (\emph{iii}) represent amplitude pairs with $M = \mp 1$, and the red nodes of type (\emph{ii}) correspond to $M = 0$. Thus the green and blue nodes represent amplitude pairs that only get contributions from the triplet in the direct sum, while the red nodes get contributions from both the triplet and the singlet.

\subsubsection{Systematic picture of sum rules for the $0\oplus 1$ system \label{sec:algorithm}}

We discuss the sum rules of the $0\oplus 1$ system separately for the case of $b = 0$ and for $1 \leq b \leq b_\text{max}$, where $b_\text{max}$ is given in Eq.~\eqref{eq:b_max}.

First, we consider the $b = 0$ case. In Section~\ref{sec:case-no-coeff}, by studying the general structure of the CG decomposition for the different parity $u_1 \oplus u_2$ system with trivial coefficients, we found the full set of LO sum rules for this system. The $0\oplus 1$ system is a special case of the different parity $u_1\oplus u_2$ system. Additionally, as we show in Section~\ref{sec:u0-coeff}, for the systems with $u_2 = 0$, the structure of the CG decomposition with non-trivial coefficients is equivalent to that for the systems with trivial coefficients. Therefore, the LO sum rules obtained in Section~\ref{sec:case-no-coeff} for the different parity $u_1 \oplus u_2$ system with trivial coefficients also form a full set of LO sum rules for the $0\oplus 1$ system.

Thus, based on the discussion at the end of Section~\ref{sec:case-no-coeff}, the $b = 0$ sum rules for the $0\oplus 1$ system can be read off the lattice as follows, where everywhere we implicitly assume that all amplitudes in the sums are weighted by the corresponding $\mu$-factors.\\

\underline{$b = 0$:}
\begin{enumerate}
    \item[(A)] Trivial $a$-type sum rules are obtained from the lattice nodes for which none of the coordinates are equal to $r$ or for which two of the coordinates are equal to $r$. These correspond to the green and blue nodes in the example in Fig.~\ref{fig:lattice-0+1-n=6}.
    \item[(B)] $a$-type sum rules are obtained by summing over the amplitudes in $1$-dimensional subspaces, obtained by fixing $d - 1$ coordinates, such that exactly one of the fixed coordinates is equal to $r$. The sum goes only over the nodes for which the free coordinate takes values $1,\,\dots,\, n-3$. These correspond to the sums over the red nodes in the example in Fig.~\ref{fig:lattice-0+1-n=6}. 
    \item[(C)] $s$-type sum rules are obtained by summing over all nodes in all $1$-dimensional subspaces of the lattice, weighted by the corresponding $\mu$-factors.
\end{enumerate}

To provide a bit of additional intuition, let us briefly discuss how these sum rules are consistent with the general points 1--2 about the sum rules of the $0\oplus 1$ system summarized in Section~\ref{sec:sum-rules} before Eq.~(\ref{eq:nSR-0+1}). First, by construction, all the sum rules described by (A), (B), and (C) are sum rules of the $u = 1$ system at order $b = 0$. This statement is straightforward for sum rules in (A) and (C), as they are covered exactly in our discussion of harvesting of sum rules that follows Eq.~\eqref{eq:mu-u1-def-main}. For (B), note that at order $b = 0$, each node of the lattice corresponds to a trivial $a$-type sum rule, and thus any linear combination of $a$-type amplitudes is also a sum rule of the $u = 1$ system at $b = 0$. Finally, note that the sum rules in (C) are $b = 1$ sum rules of the $u = 1$ system, which is consistent with point 2.

Next, we discuss the case $b \geq 1$. The sum rules in this case can be found following the theorem in Appendix~\ref{sec:proof}. Here, for convenience, we reproduce the relevant statements from Appendix~\ref{sec:proof}. The theorem states that the full set of amplitude sum rules for the $0 \oplus 1$ system at even (odd) order of breaking $b \geq 1$  can be read off the lattice as follows\\

\underline{even (odd) $b \geq 1$:}
\begin{enumerate}
    \item[(1)] $s$($a$)-type sum rules, given by sums of all $s$($a$)-type amplitudes corresponding to the nodes in all $(b + 1)$-dimensional subspaces of the lattice.
    
    \item[(2)] $a$($s$)-type sum rules, given by:
    \begin{enumerate}
        \item[(2a)] The sum of all $a$($s$)-type amplitudes corresponding to the nodes in all $b$-dimensional subspaces obtained by fixing $d - b$ coordinates, such that two of the $d - b$ fixed coordinates are equal to $r$. These correspond to the sum rules that only involve blue nodes in the example in Fig.~\ref{fig:lattice-0+1-n=6}.

        \item[(2b)] For all $(b + 1)$-dimensional subspaces obtained by fixing $d - b - 1$ coordinates such that none of the fixed coordinates is equal to $r$, the sum of all nodes in the subspace without any coordinates being equal to $r$ minus the sum of all nodes in the subspace for which two of the coordinates are equal to $r$. These correspond to the sum rules that are given by the difference of all green nodes minus blue nodes in the example in Fig.~\ref{fig:lattice-0+1-n=6}.
        \item[(2c)] The sum of all nodes in all $(b + 2)$-dimensional subspaces obtained by fixing $d - (b + 2)$ coordinates such that none of the fixed coordinates is equal to $r$. 
    \end{enumerate}
\end{enumerate}

Again, we immediately see that all the sum rules in (1), (2a), and (2c) are sum rules of the $u = 1$ system, as they are covered directly by our discussion of harvesting the sum rules. It is a bit more subtle to see this for (2b). In Appendix~\ref{app:2b-proof}, we show that the sum rules described by (2b) are linear combinations of order $b$ sum rules of the $u = 1$ system and thus are themselves order $b$ sum rules for this system. 

Furthermore, by construction, the sum rules in (1) and (2c) are sum rules of the $u = 1$ system that hold at order $b + 1$, and thus, they also hold for the $0\oplus 1$ system in accordance with point 2 in our discussion of general results for the sum rules of the $0\oplus 1$ system in Section~\ref{sec:sum-rules} before Eq.~(\ref{eq:nSR-0+1}). 

Finally, it is also easy to understand why the sum rules in (2a) and (2b) are sum rules of the $0\oplus 1$ system. In both cases, the sum rules only involve the amplitudes with $M = \pm 1$, and thus if they hold for the $u = 1$ system, they also hold for the $0\oplus 1$ system. This is consistent with point 4 in our discussion of the general properties of sum rules for odd $u_1 \oplus u_2$ systems at the end of Section~\ref{sec:case-no-coeff}.

Before we proceed to give an example, we would like to point out one more general and elegant result. At the order of breaking $b^{0\oplus 1}_\text{max} = n/2 - 2$ (see Eq.~\eqref{eq:b_max}), there are always exactly two sum rules in the $0\oplus 1$ system. This can be verified explicitly using Eqs.~\eqref{eq:nSR_doublets} and~\eqref{eq:nSR-0+1} for $b = n/2 - 2$, which gives
\begin{equation}
    n_{SR}^{0\oplus 1} (n, n/2 - 2) = 2\,, \qquad \forall\,n\geq 4\,.
\end{equation}

According to the theorem quoted above, one of the two sum rules at order $b_{\text{max}}^{0\oplus 1}$ is an $a$-type sum rule and the other is an $s$-type sum rule. This can be seen as follows: for even (odd) $b_{\text{max}}^{0\oplus 1}$, there is an $s$($(a)$)-type sum rule given by the sum of all lattice nodes (corresponding to (1)), and an $a$($s$)-type sum rule given by the difference between the sum of all nodes for which none of the coordinates are equal to $r$ and the sum of all nodes for which two of the coordinates are equal to $r$ (corresponding to (2b)). 
Note that the sum rules in (2a) and (2c) do not appear at order $b_\text{max}^{0\oplus 1}$, as the subspaces described by (2a) and (2c) do not exist for the $d = n/2 - 1$ dimensional lattice that we consider.
For $b = b_\text{max}^{0\oplus 1}$, in (2c) we would have to fix $d - (b_{\text{max}}^{0\oplus 1} + 2) = -1$ coordinates which is impossible. In (2a) we would have to fix $d - b_{\text{max}}^{0\oplus 1} = 1$ coordinate, \emph{i.e.},~less than the required minimum amount of two which should be fixed to $r$.

\subsubsection{Example: $n = 6$}\label{sec:sum-rules-ex}

We consider an example of the $0\oplus 1$ system with the number of would-be doublets $n = 6$. The lattice for this system is shown in Fig.~\ref{fig:lattice-0+1-n=6}. For this system, we have $r = n - 2 = 4$, and we use $a_{(x_1, x_2)}$ and $s_{(x_1, x_2)}$ to denote the $a$- and $s$-type amplitudes that correspond to the lattice node with coordinates $(x_1, x_2)$. Since all the nodes that are related by permutation of coordinates represent a single amplitude pair, without loss of generality, we label the amplitudes by pairs $(x_1, x_2)$ such that $x_1 \leq x_2$.

We start by listing the sum rules that hold at order $b = 0$. Following points (A)--(C) above, we have:
\begin{align}
    a_{(1, 2)} = a_{(1, 3)} = a_{(2, 3)} = a_{(4, 4)} & = 0\,,\label{eq:SR-n6-A}\\
    a_{(1, 4)} + a_{(2, 4)} + a_{(3, 4)} & = 0\,,\label{eq:SR-n6-B}\\
    {s_{(1,2)}} + {s_{(1,3)}} + \sqrt{2} s_{(1,4)}& = 0\,,\label{eq:SR-n6-C-1}\\
    {s_{(1,2)}} + {s_{(2,3)}} + \sqrt{2} s_{(2,4)} & = 0\,,\\
    {s_{(1,3)}} + {s_{(2,3)}} + \sqrt{2} s_{(3,4)} & = 0\,,\\
    s_{(1,4)} + s_{(2,4)} + s_{(3,4)} + \sqrt{2} s_{(4,4)} & = 0\,,\label{eq:SR-n6-C-4}
\end{align}
where the four sum rules in Eq.~\eqref{eq:SR-n6-A} follow from (A), the sum rule in Eq.~\eqref{eq:SR-n6-B} follows from (B), and the four sum rules in Eqs.~\eqref{eq:SR-n6-C-1}--\eqref{eq:SR-n6-C-4} follow from (C).

Next, we list the sum rules that hold at order $b = 1$:
\begin{align}
    {a_{(1,2)}} + {a_{(1,3)}} + {a_{(2,3)}} + {a_{(4,4)}} + \sqrt{2} \left(a_{(1,4)} + a_{(2,4)} + a_{(3,4)}\right) & = 0\,, \label{eq:SR-n6-b1-1}\\
    {s_{(1,2)}} + {s_{(1,3)}} + {s_{(2,3)}} - {s_{(4,4)}} & = 0\,,\label{eq:SR-n6-2b}
\end{align}
where the $a$-type sum rule follows from (1), and the $s$-type sum rule follows from (2b). There are no sum rules that follow from (2a) and (2c) since the subspaces described by these points do not exist for the 2-dimensional lattice. According to Eq.~\eqref{eq:b_max}, the maximum order at which the $0\oplus 1$ system with $n = 6$ has sum rules is $b^{0\oplus 1}_\text{max} = 1$, thus there are no sum rules for this system for $b \geq 2$.

For the readers' reference, in Table~\ref{tab:1plus0times4_summary}, we summarize the sum rules side by side for the $u = 1$ system, the $u = 0$ system, and the $0\oplus 1$ system in the case when $n = 6$. We write the amplitudes of the $u = 0$ system using the notation of the lattice for the $u = 1$ system. Since all amplitudes of the $u = 0$ system have $M = 0$, they correspond to the nodes for which exactly one of the coordinates is equal to $r = n - 2 = 4$. Note how the role of $a$- and $s$-type amplitudes is switched in the $u=0$ system, see also the discussion in Sec.~\ref{sec:case-no-coeff}. Also, note that any sum rule that holds at order $b > 0$ also holds at any order $b^\prime < b$. In the table, we underline some of the amplitudes in green, blue, or red to show the correspondence of these amplitudes to the nodes in the lattice in Fig.~\ref{fig:lattice-0+1-n=6}, where important. Note how the green and blue nodes give rise to trivial $a$-type sum rules at $b = 0$ in accordance with (A), a sum rule that only involves $a$-type amplitudes corresponding to red nodes holds for the $0\oplus 1$ system in accordance with (B), and how the $b = 1$ $s$-type sum rule has a structure of a difference between green and blue nodes in accordance with (2b).

\begin{table}[t]
\centering
% Increase space between rows
\renewcommand{\arraystretch}{1.4} 
% Adjust column separation
\setlength{\tabcolsep}{10pt}
\begin{adjustbox}{max width=\textwidth}
\begin{tabular}{|c|c|c|c|}
\hline
~~$b$~~ & ~~$0 \rightarrow 1\otimes \left(\frac{1}{2}\right)^{\otimes 4}$~~ & ~~$0 \rightarrow 0\otimes \left(\frac{1}{2}\right)^{\otimes 4}$~~  & ~~$0 \rightarrow \left(1\oplus 0\right)\otimes \left(\frac{1}{2}\right)^{\otimes 4}$~~\\
\hline
0 &~~\(\displaystyle  {a_{(1,2)}} = {a_{(1,3)}} = a_{(1,4)} = {a_{(2,3)}}\) ~~& ~~\(\displaystyle {s_{(1,4)}} = {s_{(2,4)}} = {s_{(3,4)}} = 0 \)~~ &~~\(\displaystyle  \Cline[mygreen]{a_{(1,2)}} = \Cline[mygreen]{a_{(1,3)}} = \Cline[mygreen]{a_{(2,3)}} = \Cline[myblue]{a_{(4,4)}} = 0 \) ~~ \\
& ~~\(\displaystyle = {a_{(2,4)}} = {a_{(3,4)}} = {{a_{(4,4)}}} = 0\)~~ & & \(\displaystyle \Cline[myred]{a_{(1,4)}} + \Cline[myred]{a_{(2,4)}} + \Cline[myred]{a_{(3,4)}} = 0\) \\
 & & & ~~\(\displaystyle {s_{(1,2)}} +  {s_{(1,3)}} + \sqrt{2} {s_{(1,4)}} = 0, \)~~ \\
 & & & ~~\(\displaystyle  {s_{(1,2)}} +  {s_{(2,3)}} + \sqrt{2} {s_{(2,4)}} = 0, \)~~ \\
 & & & ~~\(\displaystyle {s_{(1,3)}} +  {s_{(2,3)}} + \sqrt{2} {s_{(3,4)}} = 0, \)~~ \\
 & & & ~~\(\displaystyle {s_{(1,4)}} + {s_{(2,4)}} + {s_{(3,4)}} + \sqrt{2} {s_{(4,4)}} = 0 \)~~ \\
\hline
1 & ~~\(\displaystyle  {s_{(1,2)}} + {s_{(1,3)}} + \sqrt{2} s_{(1,4)} = 0,\)~~  &~~\(\displaystyle a_{(1,4)} + a_{(2,4)} + a_{(3,4)} = 0\) ~~& ~~\(\displaystyle \Cline[mygreen]{s_{(1,2)}} + \Cline[mygreen]{s_{(1,3)}} + \Cline[mygreen]{s_{(2,3)}} - \Cline[myblue]{s_{(4,4)}} = 0,\)~~ \\
 & ~~\(\displaystyle  {s_{(1,2)}} + {s_{(2,3)}} + \sqrt{2} s_{(2,4)} = 0,\)~~  & & \({a_{(1,2)}} + {a_{(1,3)}} + {a_{(2,3)}} + {a_{(4,4)}}\displaystyle \) \\
  & ~~\(\displaystyle {s_{(1,3)}} + {s_{(2,3)}} + \sqrt{2} s_{(3,4)} = 0, \)~~  & & \(\displaystyle + \sqrt{2} \left(a_{(1,4)} + a_{(2,4)} + a_{(3,4)}\right) = 0 \) \\
    & ~~\(\displaystyle s_{(1,4)} + s_{(2,4)} + s_{(3,4)} + \sqrt{2} s_{(4,4)} = 0 \)~~  & & \\
\hline
2 &\({a_{(1,2)}} + {a_{(1,3)}} + {a_{(2,3)}} + {a_{(4,4)}}\displaystyle \) & & \\
 &\(\displaystyle + \sqrt{2} \left(a_{(1,4)} + a_{(2,4)} + a_{(3,4)}\right) = 0 \) & & \\
\hline
\end{tabular}
\end{adjustbox}
\caption{Comparison of sum rules for $u = 1$, $u = 0$ and $0\oplus 1$ systems with $n = 4$. \label{tab:1plus0times4_summary}}
\end{table}

\subsubsection{Generalization}\label{sec:gen-alg}

There are several generalizations of the results for the $0\oplus 1$ system discussed above that are called for. First, in our discussion of systems with a group-theoretical description involving a direct sum $0\oplus 1$, we only considered the case when the remaining irreps are doublets. We would like to generalize this to cases where, in addition to the direct sum, the system contains arbitrary irreps. Second, we would like to allow for non-singlet irreps not only in the final state but also in the initial state and the Hamiltonian. Since the lattice that we construct to derive the sum rules for the $0\oplus 1$ system is a lattice for the system with pure irreps, both of these generalizations are straightforward and are completely the same as what is discussed in Sections III.G and IV of Ref.~\cite{Gavrilova:2022hbx} for systems with pure irreps. Here, we only briefly outline these generalizations and refer the reader to Ref.~\cite{Gavrilova:2022hbx} for additional details and derivations. Let us start by commenting on systems with the following group-theoretical structure: 
\begin{equation}\label{eq:gen-system}
    0 \rightarrow (0 \oplus 1) \otimes u_0\otimes \dots \otimes u_{r - 1}\,,
\end{equation}
where $u_0 = 1/2$ and $u_i$, $i = 1,\,\dots,\, r - 1$, are some arbitrary irreps. The case when the system does not contain any doublets is somewhat special and is discussed in Section IV.A and Step 2.3 of the algorithm in Section V.B of Ref.~\cite{Gavrilova:2022hbx}. We refer the reader to the above-mentioned discussion and here only focus on the case when at least one of the irreps is a doublet. For the system in Eq.~\eqref{eq:gen-system}, we define the number of would-be doublets as
\begin{equation}\label{eq:n-gen}
    n = 2 + \sum_{i=0}^{r-1} 2 u_i\,.
\end{equation}
To generalize the results of Section~\ref{sec:sum-rules} to the system in Eq.~\eqref{eq:gen-system}, we need to build a generalized lattice. The generalization of the lattice construction is discussed in Sections~IV.C.1 and~IV.C.2 of Ref.~\cite{Gavrilova:2022hbx} and consists of two steps. First, we introduce the generalized coordinate notation, and second, we assign $\mu$-factors to all the nodes. In the generalized coordinate notation, the dimension of the lattice is still determined by the number of would-be doublets and is given by $d = n/2 -1$. Each coordinate, however, now takes values from $1$ to $r$. As before, without loss of generality, we choose the last position of the $n$-tuple, and thus, the coordinate value $r$ to correspond to the direct sum. The $\mu$-factors for the case of arbitrary irreps are discussed in Section IV.C.2 of Ref.~\cite{Gavrilova:2022hbx} and are given in Eqs.~(4.33)--(4.34) therein.

Once the generalized lattice is constructed and the $\mu$-factors are assigned, the sum rules can be derived from the lattice following the algorithm steps (A)--(C) and (1), (2a)--(2c) in Sec.~\ref{sec:algorithm}. The only modification is that the position of the $n$-tuple that represents the direct sum $r$ is no longer equal to $n - 2$; instead, it is determined by the number of irreps in Eq.~\eqref{eq:gen-system}.
The doublet-only case is readily recovered from Eq.~\eqref{eq:gen-system} setting $u_i=1/2$ and $r=n-2$. The indices of the irreps in Eq.~\eqref{eq:gen-system} go then from $0$ to $r-1=n-3$, corresponding to $n-2$ doublets.

Finally, let us comment on the case when non-singlet irreps also appear in the initial state and the Hamiltonian. We account for this in exactly the same way as in Sec.~III.G and Sec.~IV.D of Ref.~\cite{Gavrilova:2022hbx}, which involves the following two steps.  

The first step relates to the construction of $n$-tuples. As in Ref.~\cite{Gavrilova:2022hbx}, we adopt the convention of inverting the $m$ QNs of the multiplets in the initial state and the Hamiltonian when constructing the $n$-tuples. This convention ensures that each $n$-tuple representing an amplitude of the system has the same number of plus and minus signs as in the case when all irreps are in the final state. 

The second step involves modifying the definitions of $a$- and $s$-type amplitudes from their form in Eq.~\eqref{eq:as-def-u1+u2}. In complete analogy with Eq.~(3.31) of Ref.~\cite{Gavrilova:2022hbx}, the modified definitions are
\begin{equation}\label{eq:as-def-u1+u2-gen}
        a_i \equiv (-1)^{q_i}( {A}^{u_1\oplus u_2}_i - (-1)^{p} \overline{A}^{u_1\oplus u_2}_i)\,,\qquad s_i \equiv (-1)^{q_i}({A}^{u_1\oplus u_2}_i + (-1)^{p} \overline{A}^{u_1\oplus u_2}_i)\,,
\end{equation}
where the $p$-factor is given by (see Eq.~(C9) in Ref.~\cite{Gavrilova:2022hbx})
\begin{equation}
    p = 2 \sum_{i} u_i^F - \frac{n}{2}\,, \label{eq:p-def}
\end{equation}
with the sum running over the irreps $u_i^F$ in the final state . For cases where the direct sum $0\oplus 1$ is in the final state, we use $u_r = 1$ for the direct sum in our calculation of $p$. As described in detail in Sec.~IV.D of Ref.~\cite{Gavrilova:2022hbx}, the $q_i$-factors are determined by the product of minus signs at the positions of $n$-tuples that correspond to the final state: if the $n$-tuple corresponding to the amplitude has an even number of minus signs in the final state, then $(-1)^{q_i} = 1$; if it has an odd number of minus signs, then $(-1)^{q_i} = -1$. 

Now, equipped with these generalizations and the rules in (A)--(C) and (1), (2a)--(2c), we are ready to derive the sum rules for physical systems whose group-theoretical description involves a direct sum of a singlet and a triplet.

\section{Physical examples \label{sec:examples}}

In this section we apply the mathematical results derived above to two $U$-spin systems of charm decays whose group-theoretical description contains a direct sum of a singlet and a triplet.
The singlet contribution to the charm decay amplitude is heavily CKM-suppressed compared to the triplet one, such that for the calculation of branching ratios the former is completely negligible. However, in order to derive relations between CP violating observables it is essential to take into account both interfering amplitudes, \emph{i.e.},~the contributions from both singlet and triplet. Therefore, below we construct amplitude sum rules for physical systems taking into account both contributions.

\subsection{$D^0\rightarrow P^+P^-$}

We start by considering as an illustration the $D^0\rightarrow P^+P^-$ system of charm decays,
which has been studied extensively~\cite{Gavrilova:2022hbx, Brod:2012ud, Grossman:2019xcj, Muller:2015lua, Muller:2015rna, Grossman:2006jg, Hiller:2012xm, Pirtskhalava:2011va, Grossman:2012ry, Grossman:2013lya}. 
Our notation, that we briefly review below, follows Ref.~\cite{Gavrilova:2022hbx}. 
$D^0$ is a neutral $D$-meson, which is a singlet under $U$-spin, and $P^\pm$ are positively and negatively charged pseudoscalar mesons given by
\begin{equation} \label{eq:Pp-Pm-def}
P^+ = \begin{bmatrix}
K^+\\
\pi^+
\end{bmatrix}, \hspace{25pt}
P^- = \begin{bmatrix}
\pi^-\\
K^-
\end{bmatrix}\,.
\end{equation}
The effective Hamiltonian is given by a sum of a singlet and a triplet that we write as~\cite{Gavrilova:2022hbx}
\begin{equation} \label{eq:D-Ham}
    \mathcal{H}_\text{eff}^{(0)} = f_{0,0} H^0_0 + \sum_{m = -1}^{1} f_{1,m} H^1_m,
\end{equation}
with the operators of the effective Hamiltonian $H^u_m$ and the corresponding CKM-factors $f_{u, m}$. The components of the singlet and the triplet, $H_0^0$ and $H^1_m$ respectively, are given by~\cite{Gavrilova:2022hbx}
\begin{align} 
	H^0_0 &=  {(\bar{u} s) (\bar{s} c)+(\bar{u} d) (\bar d c)\over
  \sqrt{2}}, \label{H0-charm} \\
H^1_{1} =  (\bar{u} s) (\bar d c),\qquad
H^1_{-1} &= -(\bar{u} d) (\bar s c), \qquad
H^1_0 =  {(\bar{u} s) (\bar s c)-(\bar{u} d) ( \bar d c)\over \sqrt{2}}\,, \label{H1-charm}
\end{align}
and the corresponding CKM-factors are~\cite{Gavrilova:2022hbx}
\begin{equation}\label{eq:charmCKM-f00}
    f_{0,0} = \frac{V_{cs}^* V_{us} + V_{cd}^* V_{ud}}{2}\,,
\end{equation}
\begin{align}
f_{1,1} = V_{cd}^* V_{us}, \qquad 
f_{1,-1} &= -V_{cs}^* V_{ud}, \qquad 
f_{1,0} = \frac{V_{cs}^* V_{us} - V_{cd}^* V_{ud}}{\sqrt{2}}\,.\label{eq:charmCKM-f1m}
\end{align}

\begin{figure}[t]
\centering
\includegraphics[width=0.2\textwidth]{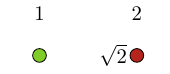}
\caption{Lattice for a $0\oplus 1$ system with $n=4$.}
\label{fig:lattice-0+1-n=4}
\end{figure}

The $D^0 \rightarrow P^+ P^-$ system contains four decays related by $U$-spin that we show
in Table~\ref{tab:map-c-mesons} along with the corresponding $n$-tuples, lattice coordinates, $\mu$-factors and $(-1)^{q_i}$ for each amplitude pair. When writing the $n$-tuples we use the convention that the first component of the $n$-tuple represents $P^+$, the second component $P^-$ and the last component the Hamiltonian. The number of would-be doublets for this system is $n = 4$ and thus the lattice is $1$-dimensional and it is given in Fig.~\ref{fig:lattice-0+1-n=4}. The maximum order at which this system has sum rules is $b_\text{max}^{0\oplus 1} = 0$, \emph{i.e.} the system only has LO sum rules. As we discussed at the end of Sec.~\ref{sec:algorithm} for any $0\oplus 1$ system there are two sum rules at order $b^{0\oplus 1}_\text{max}$, one $s$-type sum rule and one $a$-type sum rule. The $a$-type sum rule is the trivial one and it follows from (A)
\begin{equation}
    a_1 = 0\,,
\end{equation}
the $s$-type sum rule follows from (C) and is given by
\begin{equation}
    s_1 + \sqrt{2} s_2 = 0\,,
\end{equation}
where we use $a_{x}$ and $s_{x}$ to denote the $a$- and $s$-type amplitudes corresponding to the node with coordinate $x$. Accounting for the CKM, $p$-factor (here, $p=0$ from Eq.~(\ref{eq:p-def})) and $(-1)^{q_i}$ factors given in Table~\ref{tab:map-c-mesons}, we obtain the following sum rules in terms of physical amplitudes:
\begin{equation}\label{eq:DPP-aSR}
    \frac{\mathcal{A}(D^0 \rightarrow \pi^+ K^-)}{ -V_{cs}^* V_{ud} } = \frac{\mathcal{A}(D^0 \rightarrow K^+\pi^-)}{ V_{cd}^* V_{us} }\,,
\end{equation}
\begin{equation}\label{eq:DPP-sSR}
    \frac{\mathcal{A}(D^0 \rightarrow \pi^+ K^-)}{ -V_{cs}^* V_{ud} } + \frac{\mathcal{A}(D^0 \rightarrow K^+\pi^-)}{ V_{cd}^* V_{us} }-
\sqrt{2}\frac{\mathcal{A}(D^0 \rightarrow \pi^+\pi^-)}{\sqrt{2} V_{cd}^* V_{ud} } - \sqrt{2}\frac{\mathcal{A}(D^0 \rightarrow K^+ K^-)}{\sqrt{2} V_{cs}^* V_{us} }= 0\,.
\end{equation}
These two sum rules are among three amplitude sum rules found for the $D \rightarrow P^+ P^-$ system at the LO when only the triplet contribution to the Hamiltonian is taken into account, see for example Eqs.~(5.24)--(5.25) of Ref.~\cite{Gavrilova:2022hbx}. Here we emphasize the result that the two sum rules in Eqs.~\eqref{eq:DPP-aSR} and~\eqref{eq:DPP-sSR} also hold when the singlet component of the Hamiltonian is taken into account in addition to the triplet component.

\begin{table}[t]
\centering
\begin{tabular}{|c|c|c|c|c|c|}
\hline
Decay &  $U$-spin conjugate & $n$-tuple & Node & ~~$\mu$-factor~~ &~~$(-1)^{q_i}$~~\\
\hline
~~~$D^0 \rightarrow \pi^+ K^-$~~~ & 
~~~$D^0 \rightarrow K^+\pi^-$~~~& ~~$(-,-,++)$~~ &~~$(1)$~~ & 1 & $+1$ \\
~~~$D^0 \rightarrow \pi^+ \pi^- $~~~ & 
~~~$D^0 \rightarrow K^+ K^-$~~~& ~~$(-,+,-+)$~~ &~~$(2)$~~ & $\sqrt{2}$ & $-1$ \\
\hline
\end{tabular}
\caption{Amplitude pairs of $D^0\to P^+ P^-$ decays and the corresponding $n$-tuples, nodes in the coordinate notation, $\mu$-factors, and $(-1)^{q_i}$ factors. The table is adapted from Ref.~\cite{Gavrilova:2022hbx}.\label{tab:map-c-mesons}}
\end{table}

\subsection{Three body charm baryon decays $C_b\rightarrow L_b P^+ P^-$}

Next, we consider the system of three-body charm baryon decays, ${C_b} \to {L_b} P^- P^+$, where $C_b$ is a doublet of charmed baryons, $L_b$ is a doublet of light baryons, and $P^\pm$ are the doublets of pseudoscalar mesons. Following the convention of Ref.~\cite{Gavrilova:2022hbx}, we have
\begin{equation}
C_b = \begin{bmatrix}
\Lambda_c^+ \\ \Xi_c^+
    \end{bmatrix} =
    \begin{bmatrix}
    \ket{cud}\\
    \ket{cus}
    \end{bmatrix} =
    \begin{bmatrix}
    \ket{\frac{1}{2}, +\frac{1}{2}}\\
    \ket{\frac{1}{2}, -\frac{1}{2}}
    \end{bmatrix}\,, 
    \qquad
L_b = \begin{bmatrix}
    p \\
    \Sigma^+
    \end{bmatrix} =
    \begin{bmatrix}
    \ket{uud}\\
    \ket{uus}
    \end{bmatrix} =
    \begin{bmatrix}
    \ket{\frac{1}{2}, +\frac{1}{2}}\\
    \ket{\frac{1}{2}, -\frac{1}{2}}
    \end{bmatrix}\,,
\end{equation}
and the $P^\pm$ are defined in Eq.~\eqref{eq:Pp-Pm-def}. The Hamiltonian for this system is defined by Eqs.~\eqref{eq:D-Ham}--\eqref{eq:charmCKM-f1m}. The CKM-leading contribution to the amplitudes of this system has been analyzed previously in Refs.~\cite{Gavrilova:2022hbx, Savage:1989qr}, and LO sum rules between CP asymmetries have been derived in Ref.~\cite{Grossman:2018ptn}. Here, we derive the complete set of amplitude sum rules taking into account both CKM-leading and -subleading contributions. 

\begin{table}[t]
\centering
\resizebox{\textwidth}{!}{
\begin{tabular}{|c|c|c|c|c|c|}
\hline
Decay &  $U$-spin conjugate & $n$-tuple & Node &~~ $\mu$-factor~~&~~$(-1)^{q_i}$~~ \\
\hline
~~~$\Lambda^+_c \to \Sigma^+ K^- K^+$~~~ & ~~~$\Xi_c^+ \to p\pi^- \pi^+$~~~ & ~~$(-, -, -, +, + +)$~~ & ~~$(1,2)$~~ & 1 & $+1$ \\
$\Lambda_c^+ \to \Sigma^+ \pi^- \pi^+$ & $\Xi_c^+ \to p K^- K^+$ & $(-,-,+,-,++)$ & $(1,3)$ & 1 & $+1$ \\
$\Lambda^+_c \to \Sigma^+ \pi^- K^+$ & $\Xi_c^+ \to p K^- \pi^+$ & $(-,-,+,+,-+)$ & $(1,4)$ & $\sqrt{2}$ & $-1$ \\
$\Lambda_c^+ \to p K^- \pi^+$ & $\Xi_c^+ \to \Sigma^+ \pi^- K^+$ & $(-,+,-,-,++)$ & $(2,3)$ & $1$ & $+1$ \\
$\Lambda_c^+ \to p K^- K^+$ & $\Xi_c^+ \to \Sigma^+ \pi^- \pi^+$ & $(-,+,-,+,-+)$ & $(2,4)$ & $\sqrt{2}$ & $-1$ \\
$\Lambda_c^+ \to p \pi^- \pi^+$ & $\Xi_c^+ \to \Sigma^+ K^- K^+$ & $(-,+,+,-,-+)$ & $(3,4)$ & $\sqrt{2}$ & $-1$ \\
$\Lambda_c^+ \to p \pi^- K^+$ & $\Xi_c^+ \to \Sigma^+ K^- \pi^+$ & $(-,+,+,+,--)$ & $(4,4)$ & 2 & $+1$ \\
\hline
\end{tabular}
}
\caption{Amplitude pairs of $C_b \to L_b P^- P^+$ decays and the corresponding $n$-tuples, nodes in the coordinate notation, $\mu$-factors, and $(-1)^{q_i}$ factors. The table is adapted from Ref.~\cite{Gavrilova:2022hbx}.\label{tab:map-c-baryons} }
\end{table}

The $C_b\rightarrow L_b P^+ P^-$ system contains 14 decays listed in Table~\ref{tab:map-c-baryons}, which also shows the corresponding $n$-tuples for each amplitude pair, the coordinates of the corresponding nodes in the lattice, the $\mu$-factors, and the $(-1)^{q_i}$ factors. In the $n$-tuples, we use the first position to represent the component of $C_b$, the second position to represent the component of $L_b$, the third and fourth positions to represent the components of $P^-$ and $P^+$, respectively, and the last position to represent the Hamiltonian. The number of would-be doublets for this system is $n = 6$ and thus the lattice has the dimension $d = n/2 - 1 = 2$. The lattice is shown in Fig~\ref{fig:lattice-0+1-n=6}. We list the sum rules for this system at order $b = 0$ in Eqs.~\eqref{eq:SR-n6-A}--\eqref{eq:SR-n6-C-4}, and the sum rules at order $b = 1$ in Eqs.~\eqref{eq:SR-n6-b1-1}--\eqref{eq:SR-n6-2b}. The order $b = 1$ is the highest order at which this system still has sum rules. Accounting for the $p$-factor (here $p=0$) and the factors of $(-1)^{q_i}$, we obtain the following $b = 0$ sum rules in terms of coefficient-free amplitudes 
\begin{align}
A(\Lambda^+_c \to \Sigma^+ K^- K^+) &= A(\Xi_c^+ \to p\pi^- \pi^+),
\end{align}
\begin{align}
A(\Lambda_c^+ \to \Sigma^+ \pi^- \pi^+) &= A(\Xi_c^+ \to p K^- K^+),
\end{align}
\begin{align}
A(\Lambda_c^+ \to p K^- \pi^+) &= A(\Xi_c^+ \to \Sigma^+ \pi^- K^+),
\end{align}
\begin{align}
A(\Lambda_c^+ \to p \pi^- K^+)& = A(\Xi_c^+ \to \Sigma^+ K^- \pi^+).
\end{align}
%%%%%%%%
\begin{align}
& A(\Lambda_c^+\rightarrow \Sigma^+ \pi^- K^+) - A(\Xi_c^+\rightarrow p K^-\pi^+) \nonumber\\
& A(\Lambda_c^+\rightarrow pK^-K^+) - A(\Xi_c^+\rightarrow \Sigma^+\pi^-\pi^+ ) \nonumber\\
& A(\Lambda_c^+ \rightarrow p \pi^-\pi^+) - A(\Xi_c^+\rightarrow \Sigma^+ K^- K^+) = 0
\end{align}
\begin{align}
A(\Lambda^+_c \to \Sigma^+ K^- K^+)+A(\Xi_c^+ \to p\pi^- \pi^+) + A(\Lambda_c^+ \to \Sigma^+ \pi^- \pi^+) & \nonumber \\
+A(\Xi_c^+ \to p K^- K^+) - \sqrt{2}A(\Lambda^+_c \to \Sigma^+ \pi^- K^+) - \sqrt{2} A(\Xi_c^+ \to p K^- \pi^+) & = 0,
\end{align}
%%%%%%%
\begin{align}
A(\Lambda^+_c \to \Sigma^+ K^- K^+)+A(\Xi_c^+ \to p\pi^- \pi^+) + A(\Lambda_c^+ \to p K^- \pi^+) & \nonumber \\ 
+ A(\Xi_c^+ \to \Sigma^+ \pi^- K^+) - \sqrt{2} A(\Lambda_c^+ \to p K^- K^+) - \sqrt{2} A(\Xi_c^+ \to \Sigma^+ \pi^- \pi^+) & = 0,
\end{align}
\begin{align}
A(\Lambda_c^+ \to \Sigma^+ \pi^- \pi^+) + A(\Xi_c^+ \to p K^- K^+) + A(\Lambda_c^+ \to p K^- \pi^+) & \nonumber \\ 
+ A(\Xi_c^+ \to \Sigma^+ \pi^- K^+) - \sqrt{2}A(\Lambda_c^+ \to p \pi^- \pi^+) - \sqrt{2} A(\Xi_c^+ \to \Sigma^+ K^- K^+) & = 0,
\end{align}
\begin{align}
- A(\Lambda^+_c \to \Sigma^+ \pi^- K^+) - A(\Xi_c^+ \to p K^- \pi^+) - A(\Lambda_c^+ \to p K^- K^+) \nonumber \\ 
- A(\Xi_c^+ \to \Sigma^+ \pi^- \pi^+) - A(\Lambda_c^+ \to p \pi^- \pi^+) - A(\Xi_c^+ \to \Sigma^+ K^- K^+) \nonumber \\
+\sqrt{2} A(\Lambda_c^+ \to p \pi^- K^+) + \sqrt{2} A(\Xi_c^+ \to \Sigma^+ K^- \pi^+) & = 0.
\end{align}

For the $b = 1$ sum rules we have
\begin{align}
&+\left[A(\Lambda^+_c \to \Sigma^+ K^- K^+) - A(\Xi_c^+ \to p\pi^- \pi^+)\right] \nonumber\\&
+ \left[A(\Lambda_c^+ \to \Sigma^+ \pi^- \pi^+) - A(\Xi_c^+ \to p K^- K^+)\right]\nonumber\\&
+\left[A(\Lambda_c^+ \to p K^- \pi^+) - A(\Xi_c^+ \to \Sigma^+ \pi^- K^+)\right]
\nonumber\\&
+\left[A(\Lambda_c^+ \to p \pi^- K^+) - A(\Xi_c^+ \to \Sigma^+ K^- \pi^+)\right]
\nonumber\\&
-\sqrt{2}\left[A(\Lambda^+_c \to \Sigma^+ \pi^- K^+)  -A(\Xi_c^+ \to p K^- \pi^+)\right]
\nonumber\\ &-
\sqrt{2}\left[A(\Lambda_c^+ \to p K^- K^+) - A(\Xi_c^+ \to \Sigma^+ \pi^- \pi^+)\right]
\nonumber\\&
-\sqrt{2}\left[A(\Lambda_c^+ \to p \pi^- \pi^+)- A(\Xi_c^+ \to \Sigma^+ K^- K^+)\right] =0,
\end{align}
\begin{align}
    A(\Lambda^+_c \to \Sigma^+ K^- K^+)+A(\Xi_c^+ \to p\pi^- \pi^+) + A(\Lambda_c^+ \to \Sigma^+ \pi^- \pi^+) & \nonumber \\
+A(\Xi_c^+ \to p K^- K^+) + A(\Lambda_c^+ \to p K^- \pi^+) + A(\Xi_c^+ \to \Sigma^+ \pi^- K^+) &\nonumber\\
- A(\Lambda_c^+ \to p \pi^- K^+) - A(\Xi_c^+ \to \Sigma^+ K^- \pi^+) & = 0\,.
\end{align}

\section{Conclusion \label{sec:conclusion}}

We have studied $U$-spin systems with a direct sum of irreducible representations. We have obtained several general results that hold for a direct sum of two arbitrary irreps, and extended the formalism developed in Ref.~\cite{Gavrilova:2022hbx} to $U$-spin systems that involve a direct sum of a singlet and a triplet. 
We found a modified algorithm that enables the systematic determination of all linearly independent amplitude sum rules up to an arbitrary order of symmetry breaking without the need for the calculation of explicit Clebsch-Gordan decompositions. This generalization allows for the study of physical systems with two different $U$-spin multiplets in the Hamiltonian, which is relevant for the discussion of CP violation, and systems that have representation mixing in the final state. We present the algorithm in Sec.~\ref{sec:algorithm}.

We identify several directions for future work, which we list below. 
\begin{itemize}
\item[(\emph{i})] The formalism may be extended to include further physically relevant cases like for example the direct sum $1/2 \oplus 3/2$, and by further exploring its application to isospin with fully hadronic final states. 
\item[(\emph{ii})] The algorithm developed in Ref.~\cite{Gavrilova:2022hbx} and extended in this work is ideally suited to be implemented as a computer program.
\item[(\emph{iii})] Finally, and most importantly, future work should focus on developing sum rules at arbitrary order also at the observable level, \emph{i.e.}~for branching ratios and CP asymmetries. This step is highly non-trivial and requires not only the formulation of relations that are non-linear in amplitudes but also the examination of phase-space effects and the momentum dependence of the amplitudes.
\end{itemize}
Given the abundance of measurements of multi-body decays at LHCb and Belle~II as well as future prospects, we anticipate that the methodology of higher order flavor-symmetry sum rules will be turned from theory into practice in the near future.

\acknowledgments

We would like to thank Yuval Grossman for stimulating discussions and valuable comments on the manuscript. S.S.~is supported by a Stephen Hawking Fellowship from UKRI under reference EP/T01623X/1 and the STFC research grant ST/X00077X/1.

\appendix

\section{Details for the general system with trivial coefficients \label{app:details-trivial}}

In this Appendix, we provide additional details about the $u_1 \oplus u_2$ systems with trivial coefficients discussed in Sec.~\ref{sec:case-no-coeff}. Here, we explicitly present the order-by-order structure of the $a$- and $s$-type amplitudes of the $u_1 \oplus u_2$ system, as defined in Eqs.~\eqref{eq:as-u1-def}--\eqref{eq:as-u1u2-def}, in terms of contributions to the $a$- and $s$-type amplitudes of the $u_1$ and $u_2$ systems (see Eq.~\eqref{eq:auxiliary}). First, we write a general expansion for the $a$- and $s$-type amplitudes of the $u_1 \oplus u_2$ system in terms of contributions that enter at different orders of breaking $b$:
\begin{align}
a_j^{u_1\oplus u_2}(u_{1}) &= \sum_b a_j^{u_1\oplus u_2, (b)}(u_{1})\,, & 
s_j^{u_1\oplus u_2}(u_{1}) &= \sum_b s_j^{u_1\oplus u_2, (b)}(u_{1})\,,  \label{eq:order-by-order-1}\\
a_{\ell}^{u_1\oplus u_2}(u_1,u_2) &= \sum_b a_{\ell}^{u_1\oplus u_2, (b)}(u_1, u_2)\,, & 
s_{\ell}^{u_1\oplus u_2}(u_1,u_2) &= \sum_b s_{\ell}^{u_1\oplus u_2, (b)}(u_1, u_2)\,, \label{eq:order-by-order-3}
\end{align}
where we follow the index convention from Sec.~\ref{sec:case-no-coeff}: the index $j$ runs over the amplitudes with $\abs{M} > u_2$ (these amplitudes receive contributions only from $u_1$ in the direct sum), the index $\ell$ runs over the amplitudes with $\abs{M} \leq u_2$ (these amplitudes receive contributions from both $u_1$ and $u_2$).

Similarly, for the amplitudes of the $u_1$ and $u_2$ systems, we write the order-by-order expansions as follows:
\begin{align}
a^{u_1} &= \sum_{\text{odd } b} a^{u_1, (b)}\,, &
s^{u_1} &= \sum_{\text{even } b} s^{u_1, (b)}\,, \label{eq:order-by-order-u1}\\
%%%
a^{u_2} &= \sum_{\text{odd } b} a^{u_2, (b)}\,, &
s^{u_2} &= \sum_{\text{even } b} s^{u_2, (b)}\,, \label{eq:order-by-order-u1u2}
\end{align}
where the $a$-type amplitudes get non-zero contributions at odd orders of breaking $b$, while $s$-type amplitudes at even orders $b$. This is in accordance with general results for $a$- and $s$-type amplitudes for systems with pure irreps derived in Section III.C of Ref.~\cite{Gavrilova:2022hbx}. We emphasize the distinction between the $a$- and $s$-type amplitudes of the system with the direct sum $u_1\oplus u_2$ in Eqs.~\eqref{eq:order-by-order-1}--\eqref{eq:order-by-order-3}, and the amplitudes of the systems with pure irreps ($u_1$ and $u_2$ systems) in Eqs.~\eqref{eq:order-by-order-u1}--\eqref{eq:order-by-order-u1u2}.

The amplitudes $a^{u_1 \oplus u_2}_j(u_1)$ and $s^{u_1 \oplus u_2}_j(u_1)$, when written in terms of amplitudes of the auxiliary systems only get contributions from the $u_1$ system. Moreover, based on our convention in Eq.~\eqref{eq:as-def-u1+u2}, the definitions of $a$- and $s$-type amplitudes of the $u_1 \oplus u_2$ and the $u_1$ system match. Thus, we have the following equalities for the contributions at order $b$
\begin{align}
    a_j^{u_1 \oplus u_2, (b)}(u_1) & = a_j^{u_1, (b)}\,,\qquad \text{odd $b$,}\nonumber\\
    s_j^{u_1 \oplus u_2, (b)}(u_1) & = s_j^{u_1, (b)}\,, \qquad\text{even $b$\,,}
\end{align}
while
\begin{align}
    a_j^{u_1 \oplus u_2, (b)}(u_1) & = a_j^{u_1, (b)} = 0\,,\qquad \text{even $b$,}\nonumber\\
    s_j^{u_1 \oplus u_2, (b)}(u_1) & = s_j^{u_1, (b)} = 0\,, \qquad\text{odd $b$\,.}
\end{align}

Next, the amplitudes $a^{u_1 \oplus u_2}_j(u_1, u_2)$ and $s^{u_1 \oplus u_2}_j(u_1, u_2)$ get contributions from both $u_1$ and $u_2$ amplitudes and the exact form of these contributions depends on the relative parity of $u_1$ and $u_2$ (see the discussion above Eq.~\eqref{eq:as-def-u1+u2} in Section~\ref{sec:case-no-coeff}). We distinguish four cases of contributions to amplitudes in Eqs.~(\ref{eq:order-by-order-1})--(\ref{eq:order-by-order-3}) at order $b$.
\begin{itemize}
%%%
\item \underline{Same parity, $b$ even:}
\begin{align}
a_{\ell}^{u_1\oplus u_2, (b)}(u_1, u_2) &= 0\,, \\
s_{\ell}^{u_1\oplus u_2, (b)}(u_1, u_2) &= s_{\ell}^{u_1, (b)} + s_{\ell}^{u_2, (b)} \,.
\end{align}
%%%
\item \underline{Same parity, $b$ odd:}
\begin{align}
a_{\ell}^{u_1\oplus u_2, (b)}(u_1, u_2) &= a_{\ell}^{u_1, (b)} + a_{\ell}^{u_2, (b)}\,,\\
s_{\ell}^{u_1\oplus u_2, (b)}(u_1, u_2) &= s_{\ell}^{u_2, (b)}\,.
\end{align}
%%%
\item \underline{Different parity, $b$ even:}\\
\begin{align}
a_{\ell}^{u_1\oplus u_2, (b)}(u_1,u_2) &= s_{\ell}^{u_2, (b)}\,, \label{eq:s-turned-a}\\
s_{\ell}^{u_1\oplus u_2, (b)}(u_1,u_2) &= s_{\ell}^{u_1, (b)}\,.
\end{align}
%%%%
\item \underline{Different parity, $b$ odd:}
\begin{align}
a_{\ell}^{u_1\oplus u_2, (b)}(u_1, u_2) &= a_{\ell}^{u_1, (b)}\,, \\
s_{\ell}^{u_1\oplus u_2, (b)}(u_1, u_2) &= a_{\ell}^{u_2, (b)}\,.
\end{align}
%%%
\end{itemize}
We illustrate the order by order structure of amplitudes of the $u_1 \oplus u_2$ system that follows from these equations in  Table~\ref{fig:amplitude-structure}, which is equivalent to the following equations: 
\begin{itemize}
\item Same parity:
\begin{align}
a_j^{u_1\oplus u_2}(u_1) &= a_j^{u_1,(1)} + \mathcal{O}(\varepsilon^3)\,, \label{eq:order-by-order-first}\\
a_{\ell}^{u_1\oplus u_2}(u_1, u_2) &= a_{\ell}^{u_1,(1)} + a_{\ell}^{u_2,(1)} + \mathcal{O}(\varepsilon^3)\,,\\
%%%
s_{j}^{u_1\oplus u_2}(u_1) &= s_j^{u_1,(0)} + s_j^{u_1,(2)} + \mathcal{O}(\varepsilon^4)\,,\\
%%%
s_{\ell}^{u_1\oplus u_2}(u_1,u_2) &= s_{\ell}^{u_1,(0)} +s_{\ell}^{u_2,(0)} + s_{\ell}^{u_1,(2)}+s_{\ell}^{u_2,(2)} + \mathcal{O}(\varepsilon^4)\,.
\end{align}
\item Different parity:
\begin{align}
a_j^{u_1\oplus u_2}(u_1) &= a_j^{u_1, (1)} + \mathcal{O}(\varepsilon^3)\,,\\
%%%%
a_{\ell}^{u_1\oplus u_2}(u_1,u_2) &= s_{\ell}^{u_2, (0)} + a_{\ell}^{u_1, (1)} + s_{\ell}^{u_2, (2)} + \mathcal{O}(\varepsilon^3)\,,\label{eq:order-by-order-au1u2}\\
%%%
s_{j}^{u_1\oplus u_2}(u_1) &= s_j^{u_1, (0)} + s_j^{u_1,(2)} + \mathcal{O}(\varepsilon^4)\,,\\
%%%
s_{\ell}^{u_1\oplus u_2}(u_1, u_2) &= s_{\ell}^{u_1, (0)} + a_{\ell}^{u_2, (1)} + s_{\ell}^{u_1, (2)} + \mathcal{O}(\varepsilon^3)\,.\label{eq:order-by-order-last}
\end{align}
\end{itemize}

\begin{table}[t]
\subfigure[Same parity]{
\begin{tabular}{c|c|c||c|c||c|c||c}
& $u_1$, $b=0$ & $u_2$, $b=0$ & $u_1$, $b=1$ & $u_2$, $b=1$ & $u_1$, $b=2$ &  
$u_2$, $b=2$ &...\\
\hline
$a_j^{u_1\oplus u_2}(u_1)$ & 0 & 0 & $a_j^{u_1,(1)}$  & 0 & 0 & 0 &\\
\hline
$a_{\ell}^{u_1\oplus u_2}(u_1,u_2)$ & 0 & $0$ & $a_{\ell}^{u_1, (1)}$  & $a_{\ell}^{u_2, (1)}$ & 0 & 0 &\\
\hline\hline
$s_j^{u_1\oplus u_2}(u_1)$ & $s_j^{u_1, (0)}$ & 0 & 0 & 0& $s_j^{u_1, (2)}$ &0 &\\
\hline
$s_{\ell}^{u_1\oplus u_2}(u_1,u_2)$ & $s_{\ell}^{u_1, (0)}$ & $s_{\ell}^{u_2, (0)}$ & 0 & $0$ & $s_{\ell}^{u_1,(2)}$ & $s_{\ell}^{u_2, (2)}$  &
\end{tabular}}
\subfigure[Different parity]{
\begin{tabular}{c|c|c||c|c||c|c||c}
& $u_1$, $b=0$ & $u_2$, $b=0$ & $u_1$, $b=1$ & $u_2$, $b=1$ & $u_1$, $b=2$ &  
$u_2$, $b=2$ &...\\
\hline
$a_j^{u_1\oplus u_2}(u_1)$ & 0 & 0 &  $a_j^{u_1, (1)}$ & 0 & 0 & 0 & \\
\hline
$a_{\ell}^{u_1\oplus u_2}(u_1,u_2)$ & 0 & $s_{\ell}^{u_2, (0)}$ & $a_{\ell}^{u_1, (1)}$  & 0 & 0 & $s_{\ell}^{u_2, (2)}$ & \\
\hline\hline
$s_j^{u_1\oplus u_2}(u_1)$ & $s_j^{u_1, (0)}$ & 0 & 0 & 0&  $s_j^{u_1, (2)}$ & 0 &\\
\hline
$s_{\ell}^{u_1\oplus u_2}(u_1,u_2)$ & $s_{\ell}^{u_1, (0)}$ & 0 & 0 & $a_{\ell}^{u_2, (1)}$  & $s_{\ell}^{u_1,(2)}$  & 0 & 
\end{tabular}
}
\caption{Amplitude structure for even (a) and odd (b) relative parity of $u_1$ and $u_2$.
In order to construct a complete amplitude from the table, one has to sum the different contributions, see 
Eqs.~(\ref{eq:order-by-order-first})--(\ref{eq:order-by-order-last}). The pattern keeps repeating itself as we go to higher orders in the symmetry breaking as in Table~\ref{fig:CG-u1u2}.
} \label{fig:amplitude-structure}
\end{table}

\section{Number of sum rules for the $0\oplus 1$ system \label{app:counting-0+1}  }

\subsection{Generalities}

In this appendix we derive the closed form formula for the number of amplitude sum rules for a system described by $n$ would-be doublets that contains a direct sum of a singlet and a triplet and the rest of representations are $n-2$ doublets. As discussed in Ref.~\cite{Gavrilova:2022hbx} and reviewed in Section~\ref{sec:review}, the ordering of the representations and their assignment to the initial final state and the Hamiltonian are irrelevant to the counting of sum rules. Thus, without loss of generality we consider a system with the following group theoretical structure
\begin{equation}\label{eq:system-0+1-app}
    0 \rightarrow \left(0 \oplus 1\right)\otimes \left(\frac{1}{2}\right)^{\otimes (n-2)}\,,
\end{equation}
where the Hamiltonian is a singlet.
That is a system where all the $U$-spin representations are in the final state. We refer to this system as \emph{$0\oplus1$ system} for short.

In the standard approach to deriving $U$-spin amplitude sum rules, we express the amplitudes of the $U$-spin system in terms of reduced matrix elements (RME). $U$-spin symmetry, which is technically realized via the application of the Wigner-Eckart theorem, ensures that at orders of breaking $b$ that are less or equal than some finite $b_\text{max}$, the number of the linearly independent combinations of the RMEs is less than the number of amplitudes in the system and thus there are sum rules between the amplitudes. The order of breaking $b_\text{max}$ is defined as the highest order at which the system still has sum rules and is determined by the group theoretical structure of the system. The number of sum rules at a given order of breaking $b$ is found as the difference between the number of amplitudes in the system and the number of linearly independent reduced matrix elements at order $b$. For detailed discussion of the physical and $U$-spin bases and the decomposition of amplitudes in terms of RMEs we refer the reader to Section II.D and Appendices A and B of Ref.~\cite{Gavrilova:2022hbx}.

In Section~\ref{sec:0+1-from-n-doublets} we argue that the $0\oplus1$ system can be constructed from the system of $n$ doublets. In particular, this means that the group-theoretical decomposition of the $0\oplus1$ system can be obtained from the group-theoretical decomposition of the system of $n$-doublets.

First, let's consider the system of $n$ doublets in the final state. The amplitudes $A_j^d$ of this system can be written as follows
\begin{equation}\label{eq:A-decomps-n-doublets}
    A_j^d = \sum_\alpha C_{j\alpha} X_\alpha\,,
\end{equation}
where $X_\alpha$ are RMEs and $C_{j\alpha}$ are coefficients given by products of CG coefficients. The number of sum rules of the system of $n$ doublets that hold up to the order of breaking $b$ is then given by the difference between the number of amplitudes $n_A^d(n)$ and the number of linearly independent RMEs up to order $b$. The latter can be found as the rank of the matrix of coefficients $\left[C_{j\alpha}\right]$ in Eq.~\eqref{eq:A-decomps-n-doublets} at order $b$. The number of sum rules in the system that hold up to the order of breaking $b$ is then found as (see Eq.~(E10) in Ref.~\cite{Gavrilova:2022hbx})
\begin{equation}
    n_{SR}^{d}(n,\,b) = n_A^d(n) - \text{rank}\,\left[C_{j\alpha}\right]\,,
\end{equation}
where the RHS has an implicit dependence on $b$ via the index $\alpha$.

Similarly, we write the decomposition for the amplitudes $A_j^{(0\oplus1)}$ of the $0\oplus1$ system as follows
\begin{equation}\label{eq:A-decomp-0+1}
    A_j^{0 \oplus 1} = \sum_\alpha \widetilde{C}_{j\alpha} X_\alpha\,,
\end{equation}
where the RMEs $X_\alpha$ are the same as in the case of the system of doublets, but there is an important difference between the decompositions in Eq.~\eqref{eq:A-decomps-n-doublets} and Eq.~\eqref{eq:A-decomp-0+1}. The amplitudes $A_j^{(0\oplus1)}$ are given by a subset of amplitudes $A_j$, but otherwise are identical up to possible factors of $\sqrt{2}$ that are irrelevant for the counting of sum rules. Ignoring these irrelevant symmetry factors, the matrix $\left[\widetilde{C}_{j\alpha}\right]$ is obtained from the matrix $\left[C_{j\alpha}\right]$ by removing some of the rows. This means that even though the RMEs in the two decompositions are identical, the ranks of the two matrices can be different. In particular, the following inequality is satisfied
\begin{equation}\label{eq:rank-ineq}
    \text{rank}\,\left[\widetilde{C}_{j\alpha}\right] \leq \text{rank}\,\left[C_{j\alpha}\right],\qquad \forall \,b\,,
\end{equation}
where again the dependence on the order of breaking $b$ is implicit in the index $\alpha$.

The number of sum rules in the $0\oplus 1$ system at order $b$ is given by
\begin{equation}\label{eq:nSR-0+1-def}
    n_{SR}^{0\oplus 1}(n,\,b) = n_A^{(0\oplus1)}(n) - \text{rank}\,\left[\widetilde{C}_{j\alpha}\right]\,.
\end{equation}
Thus in order to find the number of sum rules $n_{SR}^{0\oplus 1}(n,\,b)$, we need to find the number of amplitudes in the $0\oplus 1$ system $n_A^{(0\oplus1)}(n)$ and the $\text{rank}\,\left[\widetilde{C}_{j\alpha}\right]$.

\subsection{The number of amplitudes, $n_A^{(0\oplus1)}(n)$}

There are many ways to derive $n_A^{0 \oplus 1}(n)$, here we choose to follow the derivation based on the construction of the $0\oplus1$ system from the system of $n$ doublets described in Section~\ref{sec:0+1-from-n-doublets}.

In Section~\ref{sec:0+1-from-n-doublets}, we start by considering $n_A^d(n)$ amplitudes of the system of $n$ doublets, and then we proceed to showing that the $0\oplus1$ system is obtained from the system of doublets by discarding some of the amplitudes. Our approach here is to count the number of amplitudes that are discarded $n_\text{disc}(n)$. Then the number of amplitudes in the $0 \oplus 1$ system is given by
\begin{equation}\label{eq:nA-0+1-def}
    n^{0\oplus 1}_A(n) = n_A^d(n) - n_\text{disc}(n)\,.
\end{equation}
The number of amplitudes of the system of $n$ doublets is derived in Appendix E of Ref.~\cite{Gavrilova:2022hbx} and is given in Eq.~\eqref{eq:nA} which we duplicate here for convenience
\begin{equation}\label{eq:nA-n-doublets-app}
    n_A^d(n) = \binom{n}{n/2}\,.
\end{equation}

We represent amplitudes of the system of $n$ doublets via $n$-tuples. According to the construction in Section~\ref{sec:0+1-from-n-doublets}, without loss of generality we choose the last two positions of the $n$-tuples to be symmetrized or anti-symmetrized in order to build a triplet and a singlet representation, respectively. We discard all the amplitudes of the system of doublets for which the $n$-tuples have ``$-,+$'' as the last two entries and keep the rest of the amplitudes. Thus in order to count $n_\text{disc}(n)$, we need to count the number of $n$-tuples that end with ``$-,+$''.

Since an $n$-tuple always has $n/2$ plus signs and $n/2$ minus signs, the first $n-2$ positions of the $n$-tuples with ``$-,+$'' at the end have $(n/2-1)$ plus and $(n/2-1)$ minus signs and any ordering of the $(n-2)$ signs between $(n-2)$ positions is allowed. This means that the first $(n-2)$ positions of the $n$-tuples that we discard also form a tuple. Thus the number of $n$-tuples that end with ``$-,+$'' is found as the number of all $(n-2)$-tuples. By definition, the number of $(n-2)$-tuples is equal to the number of amplitudes in the system of $n-2$ doublets, and we have the following for the number of discarded amplitudes
\begin{equation}\label{eq:n-disc}
    n_\text{disc}(n) = n_A^d(n-2) = \binom{n-2}{n/2-1}\,.
\end{equation}
Using Eqs.~\eqref{eq:nA-0+1-def},~\eqref{eq:nA-n-doublets-app} and~\eqref{eq:n-disc}, the number of amplitudes of the $0\oplus1$ system is given by
\begin{align}\label{eq:nA-0+1-final}
    n_A^{0 \oplus 1} &=  n_A^d(n) - n_A^d(n-2)\nonumber\\
    &= \frac{3n-4}{n}\binom{n-2}{n/2-1}\,.
\end{align}

\subsection{The rank of the matrix $\left[\widetilde{C}_{j\alpha}\right]$}

As we established in Eq.~\eqref{eq:rank-ineq}, the rank of $\left[\widetilde{C}_{j\alpha}\right]$ is less or equal to the rank of $\left[C_{j\alpha}\right]$. In order to find the rank of $\left[\widetilde{C}_{j\alpha}\right]$, we first review the calculation of the rank of $\left[C_{j\alpha}\right]$ discussed in detail in Appendix E of Ref.~\cite{Gavrilova:2022hbx}. We then study the linear dependencies that arise in the matrix $\left[\widetilde{C}_{j\alpha}\right]$ and calculate its rank.

\subsubsection{The system of $n$ doublets}

Here we briefly review some of the results related to the decomposition of amplitudes of a system of $n$ doublets in terms of RMEs, see for derivations and details Section II.D and Appendices A and B of Ref.~\cite{Gavrilova:2022hbx}

We start by considering the $U$-spin system with $n$ doublets in the final state. All amplitudes of this system can be labeled by sets $\overline{m}$ of $m$-QNs of the doublets in the final state
\begin{equation}
    \overline{m}^d = \{m_1,\,\dots,\,m_n\}\,,\qquad m_i = \pm 1/2\,,
\end{equation}
such that
\begin{equation}
    \sum_i^{n} m_i = 0\,.
\end{equation}
We denote the amplitudes of the system of $n$ doublets as $A^d(\overline{m}^d)$. The index $j$ in Eq.~\eqref{eq:A-decomps-n-doublets} enumerates different orderings in the sets $\overline{m}$ and thus
\begin{equation}
    A_j^d \equiv A^d(\overline{m}^d)\,.
\end{equation}
In order to write the amplitudes in the form of a group-theoretical decomposition in Eq.~\eqref{eq:A-decomps-n-doublets}, we need to perform the basis rotation from the physical basis to the $U$-spin basis. To perform the basis rotation, we evaluate the tensor product of the doublets in the final state, and the tensor product of spurion operators in the Hamiltonian up to the desired order of the symmetry breaking $b$. A specific choice of the order in which the tensor products are evaluated constitutes a basis choice. We evaluate the tensor products and label the terms in the resulting decompositions by the values of the total $U$-spin for each intermediate tensor product. We use sets $\overline{U}$ to label the terms in the decomposition of the tensor product of $n$ doublets in the final state and sets $\overline{U}^\text{br}$ to label the terms in the tensor products of spurions in the Hamiltionian,
\begin{equation}
    \overline{U} = \{U_1,\,\dots,\,U_{n-1}\},\qquad \overline{U}^\text{br} = \{U^\text{br}_1,\,\dots,\,U^\text{br}_{b^\prime - 1}\}, \qquad b^\prime \leq b\,.
\end{equation}
Here, we use $b^\prime$ as a generic index and $b$ as the order of breaking up to which we want to write the decomposition in Eq.~\eqref{eq:A-decomps-n-doublets}. Employing this notation, the RMEs that enter the decomposition in Eq.~\eqref{eq:A-decomps-n-doublets} can be written as (see Eq.~(E8) in Ref.~\cite{Gavrilova:2022hbx}) 
\begin{equation}
    X_\alpha \equiv \mel{\overline{U}}{H(\overline{U}^\text{br},b^\prime)}{0}\,,
\end{equation}
where $\alpha$ is a multi-index given by 
\begin{equation}
    \alpha \equiv \{\overline{U}, \overline{U}^\text{br}, b^\prime\}\,,
\end{equation}
and $U_{n-1} = U^\text{br}_{b^\prime -1}$. This is necessary in order to construct a singlet initial state, following from the general rule 
\begin{align}
|j_1 - j_2| \leq J \leq |j_1 + j_2|
\end{align}
for adding two angular momenta $j_1$ and $j_2$.

The rank of the matrix $\left[C_{j\alpha}\right]$ at order $b$ is given by (see Eq.~(E20) in Ref.~\cite{Gavrilova:2022hbx})
\begin{equation}\label{eq:rank-C}
    \text{rank\,} \left[C_{j\alpha}\right] = \sum_{U = 0}^{b} N_U^n\,,
\end{equation}
where (see Eq.~(E16) in Ref.~\cite{Gavrilova:2022hbx})
\begin{equation}\label{eq:N^n_U}
    N_U^n = \begin{cases}\frac{n!\left(2U+1\right)}{\left(n/2 - U\right)!\left(n/2 + U + 1\right)!}\,,&U\geq 0\\
    0\,,&U < 0\,\end{cases}\,,
\end{equation}
is the multiplicity of irrep $U$ in the tensor product of $n$ doublets, which gives the number of linearly independent RMEs that enter the decomposition in Eq.~\eqref{eq:A-decomps-n-doublets} at order $b^\prime = U$\,. Note that in Eq.~(\ref{eq:N^n_U}) we generalize the formula for $N_U^n$ to formally cover also the case $U<0$ needed below.

Before we proceed, we would like to explain the following point about the decomposition of amplitudes $A_j^d$ of the system of $n$ doublets in terms of RMEs. At the order of breaking $b = 0$, the number of RMEs in the decomposition in Eq.~\eqref{eq:A-decomps-n-doublets} is given by the multiplicity of the singlet representation in the tensor product of $n$ doublets $N_0^n$. All of them are linearly independent. When we move from $b = 0$ to $b = 1$, new RMEs of order $b = 1$ are added to the decomposition. There are $N^n_1$ of these RMEs and all of them are linearly independent with each other and with RMEs at order $b = 0$. However, for $b \geq 2$, the number of added RMEs at order $b$ and the number of linearly independent RMEs at order $b$ are not equal to each other anymore. For $b \geq 2$, all of the RMEs of order $b$ are linearly independent with each other, however only $N^n_b$ of them are linearly independent with all the RMEs that enter at lower orders. The RMEs that enter at order $b$ and that are linearly independent from the lower order RMEs can be chosen as the ones for which $U_{n-1} = U^\text{br}_{b-1} = b$. Since only the linearly independent combinations of RMEs are relevant for finding the rank of $\left[C_{j\alpha}\right]$ and deriving the sum rules, in our discussion we often ignore the linearly dependent RMEs. In particular, in what follows, when we discuss the RMEs that enter at order $b$ we only refer to the $N^n_b$ linearly independent RMEs which are chosen without loss of generality to be the ones with $U_{n-1} = U^\text{br}_{b-1} = b$.

\subsubsection{The basis choice}\label{app:bais-choice}

Elements of the matrix $\left[C_{j\alpha}\right]$ are given by (see Eq.~(E12) of Ref.~\cite{Gavrilova:2022hbx})
\begin{equation}\label{eq:C_jalpha-def}
    C_{j\alpha} = C^*(\overline{m}^d,\,\overline{U}) \times C^*(\overline{U}^\text{br})\,,
\end{equation}
where $C^*(\overline{m}^d,\,\overline{U})$ and $C^*(\overline{U}^\text{br})$ are defined as the coefficients of rotation from the physical to the $U$-spin basis. The rank of $\left[C_{j\alpha}\right]$ is fully determined by $C^*(\overline{m}^d,\,\overline{U})$. The $C^*(\overline{U}^\text{br})$ part does not affect the rank as it is independent on $\overline{m}^d$ and thus can be absorbed via redefinition of RMEs. The coefficients $C^*(\overline{m}^d,\,\overline{U})$ are defined via
\begin{equation}
    \ket{\overline{m}} = \sum_{\overline{U}} C^*(\overline{m}^d,\,\overline{U}) \ket{\overline{U},0}\,.
\end{equation}
The exact values of the coefficients $C^*(\overline{m}^d,\,\overline{U})$ are basis dependent. That is they depend on the specific choice of the order in which the tensor products of the doublets are evaluated.

As we use the system of doublets to construct the $0\oplus 1$ system, the following basis choice is convenient for $\overline{U}$. We first take the tensor product of the last two doublets and use $U_{1}$ to represent one of the two possible values of total $U$-spin of this tensor product in the set $\overline{U}$. We then evaluate the tensor product of the remaining $n-2$ doublets and use 
\begin{equation}
    \overline{U}_{n-2} = \{U_2,\,\dots,\,U_{n-2}\}
\end{equation}
to label the terms in the product of $n-2$ doublets. Finally $U_{n-1}$ is the total $U$-spin of the term in the product of $n$ doublets. With this we write the set $\overline{U}$ as
\begin{equation}\label{eq:set-U-basis}
    \overline{U} = \{U_1,\,\overline{U}_{n-2},\,U_{n-1}\}\,.
\end{equation}
We further define the set $\overline{m}_{n-2}$ that contains the $m$-QN of the $n-2$ doublets
\begin{equation}\label{eq:set-m_n-2-def}
    \overline{m}_{n-2} = \{m_1,\,\dots,\,m_{n-2}\}\,,
\end{equation}
as well as
\begin{equation}
    M_1 = m_{n-1} + m_n\,,
\end{equation}
and
\begin{equation}
    M_{n-2} = \sum_{i = 1}^{n-2} m_i\,.
\end{equation}
In the chosen basis, and using the above definitions, $C^*(\overline{m},\,\overline{U})$ becomes
\begin{equation}\label{eq:C*-def}
    C^*(\overline{m}^d,\,\overline{U}) = \mathop{C_{1/2, m_{n-1}}}_{\hspace{0pt} 1/2, m_{n}}^{\hspace{0pt} U_1, M_1} \times  C^*(\overline{m}_{n-2},\,\overline{U}_{n-2})\times \hspace{-10pt}\hspace{-12pt}\mathop{C_{U_{1}, M_{1}}}_{\hspace{28pt} U_{n-2}, M_{n-2}}^{\hspace{12pt} U_{n-1}, 0}\,.
\end{equation}

\subsubsection{The $0\oplus1$ system}

Next we move to considering the $0 \oplus 1$ system. As in the case of doublets, all amplitudes of the $0\oplus 1$ system can be labeled by the sets of $m$-QNs of irreps in the system. We denote these sets by $\overline{m}$ and
\begin{equation}\label{eq:m-set-0+1}
    \overline{m} = \{m_1,\,\dots,\,m_{n-2},\,M\}\,,
\end{equation}
where
\begin{align}
    m_i &= \pm 1/2,\qquad i = 1,\,\dots,\, n-2\,,\\
    M & = \pm 1\,,0\,,
\end{align}
and
\begin{equation}
    \sum_{i=1}^{n-2} m_i + M = 0\,.
\end{equation}
In Eq.~\eqref{eq:m-set-0+1}, we have chosen to use the first $n-2$ entries of the set $\overline{m}$ to denote the $m$-QN of the $n-2$ doublets in the $0\oplus 1$ system, and the last entry, $m_{n-1}$, to represent the $m$-QN of the direct sum of a singlet and a triplet, $0\oplus1$. Using the definition of $\overline{m}_{n-2}$ in Eq.~\eqref{eq:set-m_n-2-def}, the set $\overline{m}$ in Eq.~\eqref{eq:m-set-0+1} can be written as
\begin{equation}
    \overline{m} = \{\overline{m}_{n-2},\,m_{n-1}\}\,.
\end{equation}

We construct the $0\oplus 1$ system from the system of $n$ doublets by (anti-)symmetrizing the last two doublets of the system of doublets as explained in Section~\ref{sec:0+1-from-n-doublets}. Since singlet and a triplet form a complete basis for the product of two doublets, that is
\begin{equation}
    \frac{1}{2}\otimes \frac{1}{2} = 0\oplus 1,
\end{equation}
all the RMEs $X_\alpha$ that enter the decomposition for the system of $n$-doublets also enter the decomposition for the $0\oplus 1$ system. In the language of the $0\oplus 1$ system the RMEs with $U_1 = 0$ are contributions from the singlet in the direct sum and the RMEs with $U_1 = 1$ are contributions from the triplet.

\subsubsection{Three types of possible linear dependencies in $\left[\widetilde{C}_{j\alpha}\right]$}

We think about the matrix $\left[\widetilde{C}_{j\alpha}\right]$ as matrix $\left[C_{j\alpha}\right]$ with the rows corresponding to the amplitudes $A^d(\overline{m}_{n-2},\,-1/2,\,+1/2)$ being removed. We start with the decomposition of the form in Eq.~\eqref{eq:A-decomps-n-doublets} for the system of doublets, in which we only keep the linearly independent RMEs, see the discussion after Eq.~\eqref{eq:N^n_U}. As we remove rows from $\left[C_{j\alpha}\right]$ some of the RMEs can become linearly dependent. We organize all the possibilities for linear dependencies in $\left[\widetilde{C}_{j\alpha}\right]$ into three groups that we refer to as \emph{types of linear dependencies}.
\begin{itemize}[leftmargin=2.2cm]
    \item[\emph{Type I.}] Linear dependencies that only involve RMEs with the same $U_1$ and any order of breaking $b$.
    \item[\emph{Type II.}] Linear dependencies that only involve RMEs with different $U_1$, but the same order of breaking $b$.
    \item [\emph{Type III.}] Linear dependencies that involve RMEs with different $U_1$ and different order of breaking $b$.
\end{itemize}

In what follows, we show that linear dependencies of Type I and Type II can not be realized in case of the $0\oplus 1$ system and the only possibility are the linear dependencies of Type III. We formulate these results in the form of two lemmas.\\

\textbf{Lemma I:} There are no Type I linear dependencies in $[\widetilde{C}_{j\alpha}]$.

\textbf{Proof:}
As shown in Section~\ref{sec:0+1-from-n-doublets}, the amplitudes $A^{0 \oplus 1}_j$ can be written as a sum of amplitudes of two auxiliary systems:
\begin{equation}\label{eq:A0+A1}
    A_j^{0\oplus1} = A_j^{(0)} + A^{(1)}_j\,,
\end{equation}
where $A^{(0)}_j$ and $A^{(1)}_j$ are amplitudes of the systems with one anti-symmetrization and one symmetrization, respectively. Performing an (anti-)symmetrization in a system of doublets, in the language of the RME decomposition in Eq.~\eqref{eq:A-decomps-n-doublets}, is equivalent to (\emph{i}) reducing the total number of amplitudes in the system by constructing linear combinations of amplitudes of the system of doublets, and (\emph{ii}) removing the RMEs with $U_1 = 1$ in the case of anti-symmetrization and $U_1 = 0$ in the case of symmetrization.

Writing the RME decompositions for the two auxiliary systems, we have:
\begin{align}
    A_j^{(0)} & = \sum_{\alpha} C_{j\alpha}^{(0)}X_\alpha^{(0)}\,,\label{eq:A0-decomp}\\
    A_j^{(1)} & = \sum_{\alpha} C_{j\alpha}^{(1)}X_\alpha^{(1)}\,,\label{eq:A1-decomp}
\end{align}
where $X^{(U_1)}_{\alpha}$ and $C^{(U_1)}_{j\alpha}$ denote the RMEs with a given $U_1$ and the corresponding coefficients in the decomposition. The RMEs and the coefficients in these equations are subsets of those for the system of doublets. These two systems are complete systems, meaning that if the decomposition is written up to $b = n/2$, the maximal order at which the system of doublets has no more sum rules, the number of amplitudes in each of these systems is equal to the number of RMEs.

By construction in Section~\ref{sec:0+1-from-n-doublets} and as follows from Eq.~\eqref{eq:A0+A1}, the RMEs and the coefficients in Eqs.~\eqref{eq:A0-decomp} and~\eqref{eq:A1-decomp} are the same as those in Eq.~\eqref{eq:A-decomp-0+1}. That means the question about the linear dependencies between RMEs with the same value of $U_1$ in Eq.~\eqref{eq:A-decomp-0+1} is equivalent to the question of linear dependencies between RMEs in the two auxiliary systems. If there are linear dependencies between $U_1 = 0$ RMEs in the $0 \oplus 1$ system, for example, this would imply that the number of linearly independent RMEs in Eq.~\eqref{eq:A0-decomp} is less than the number of amplitudes, which cannot be the case as Eq.~\eqref{eq:A0-decomp} is a basis rotation. The same reasoning applies for $U_1 = 1$. We conclude that there are no linear dependencies of Type I in the $0 \oplus 1$ system. \textbf{End of proof.}\\

\textbf{Lemma II:} There are no Type II linear dependencies in $[\widetilde{C}_{j\alpha}]$.

\textbf{Proof:} Using the notation of Lemma I, the decompositions of the amplitudes $A^{0 \oplus 1}_j$ in terms of RMEs can be written as
\begin{equation}\label{eq:A-0-decomp}
    A^{0 \oplus 1}_j = \sum_{\alpha} C_{j\alpha}^{(0)} X_\alpha^{(0)} + \sum_{\alpha} C_{j\alpha}^{(1)}X_\alpha^{(1)}\,,
\end{equation}
while the decompositions of the corresponding $U$-spin conjugate amplitudes can be written as~(see Eqs.~(3.8) and (3.9) in Ref.~\cite{Gavrilova:2022hbx})
\begin{equation}\label{eq:A-bar-0-decomp}
    \overline{A}^{0 \oplus 1}_j = (-1)^{n/2-1}\sum_{\alpha}(-1)^b C_{j\alpha}^{(0)} X_\alpha^{(0)} + (-1)^{n/2}\sum_{\alpha} (-1)^b C_{j\alpha}^{(1)}X_\alpha^{(1)}\,.
\end{equation}

Let's assume that there exists a linear dependence between some RMEs with different $U_1$ and the same order $b$. In particular, this means that there exist sets of numbers $\{\omega_\alpha^{(0)}\}$ and $\{\omega_\alpha^{(1)}\}$ such that at least one number in each set is non-zero and
\begin{align}
    \sum_\alpha \omega_\alpha^{(0)} C_{j\alpha}^{(0)} + \sum_\alpha \omega_\alpha^{(1)} C_{j\alpha}^{(1)} &= 0\,,\nonumber\\
    -\sum_\alpha \omega_\alpha^{(0)} C_{j\alpha}^{(0)} + \sum_\alpha \omega_\alpha^{(1)} C_{j\alpha}^{(1)} &= 0\,, \label{eq:linear-dependencies}
\end{align}
for all $j$, and where the first line in Eq.~(\ref{eq:linear-dependencies}) is due to Eq.~(\ref{eq:A-0-decomp}), and the second line is due to Eq.~(\ref{eq:A-bar-0-decomp}). This implies that
\begin{equation}
    \sum_\alpha \omega_\alpha^{(0)} C_{j\alpha}^{(0)} = \sum_\alpha \omega_\alpha^{(1)} C_{j\alpha}^{(1)} = 0\,,
\end{equation}
or in other words, there are linear dependencies between RMEs with $U_1 = 0$ and $U_1 = 1$ separately. This is in contradiction with Lemma I. We thus conclude that there are no linear dependencies of Type II in the $0\oplus 1$ system. \textbf{End of proof.}\\

Note that lemmas I and II allow linear dependencies that involve any of the RMEs as long as they also involve at least one of each RME $X^{(0)}_\alpha$ and $X^{(1)}_\alpha$ taken at different orders of symmetry breaking. These are linear dependencies of Type III, which we study in what follows.

\subsubsection{Rank of the matrix $\left[\widetilde{C}_{j\alpha}\right]$}\label{app:rank-theorem}

In this section we formulate and prove the theorem that gives a closed-form formula for the rank of the coefficient matrix $\left[\widetilde{C}_{j\alpha}\right]$ at different orders of the symmetry breaking.\\

\textbf{Theorem.} The rank of the coefficient matrix $\left[\widetilde{C}_{j\alpha}\right]$ at order of breaking $b$ is given by
\begin{equation}\label{eq:rank-C-tilde}
    \text{rank\,} \left[\widetilde{C}_{j\alpha}\right] = \sum_{b^\prime = 0}^{b} \left(N^n_{b^\prime} - N^{n-2}_{b^\prime-1}\right)\,.
\end{equation}

\textbf{Outline of the proof.} From Lemmas I and II there are no Type I and Type II linear dependencies in $\left[\widetilde{C}_{j\alpha}\right]$. The only possibility are Type III linear dependencies, that is the dependencies that involve RMEs with different $U_1$ and different $b$. In our proof of Eq.~\eqref{eq:rank-C-tilde}, we start with the matrix $\left[\widetilde{C}_{j\alpha}\right]$ written at order $b = 0$ and then order by order add RMEs of the system of $n$ doublets with $b > 0$.
At each order of breaking $b$ we explicitly find $N_{b-1}^{n-2}$ RMEs that are linearly dependent with the lower order RMEs. We conclude by proving that these are the only linear dependencies and thus the rank of the coefficient matrix is given by Eq.~\eqref{eq:rank-C-tilde}.

\textbf{Proof.} We consider the cases $b = 0$, $b = 1$ and $b\geq 2$ separately. We start by studying the case $b = 0$. The only types of linear dependencies that are relevant for $b = 0$ are the Type I and Type II linear dependencies, which are prohibited by Lemmas I and II. We thus conclude that the removal of the amplitudes with $m_{n-1}\neq m_n$ does not introduce any new linear dependencies at $b = 0$ and
\begin{equation}
    \text{rank\,} \left[\widetilde{C}_{j\alpha}^{(b = 0)}\right] = \text{rank\,}\left[ {C}_{j\alpha}^{(b = 0)}\right]  = N^n_0\,,
\end{equation}
where we emphasize the order of breaking $b = 0$ in the LHS, for details see Eqs.~\eqref{eq:rank-C}--\eqref{eq:N^n_U} and the discussion following~\eqref{eq:N^n_U}.

Next, we are adding to the decomposition RMEs that correspond to $b = 1$.
Consider the following $b = 1$ RMEs with $U_1 = 1$
\begin{equation}\label{eq:XI-b1}
    X_\alpha^{I} = \mel{\overline{U}^\prime_{I}}{1}{0}\,, \qquad \overline{U}_{I} = \{1,\,\overline{U}_{n-2}^\prime,\,1\}\,,
\end{equation}
and the following $b = 0$ RMEs with $U_1 = 0$ 
\begin{equation}\label{eq:XII-b1}
    X_\alpha^{II} = \mel{\overline{U}^\prime_{II}}{0}{0}\,,\qquad \overline{U}_{II} = \{0,\,\overline{U}_{n-2}^\prime,\,0\}\,,
\end{equation}
where the sets $\overline{U}_{n-2}^\prime$ are defined as all sets $\overline{U}_{n-2}$ for which $U_{n-2} = 0$. The above are the only non-zero matrix elements. There are $N^{n-2}_0$ of sets $\overline{U}_{n-2}^\prime$ and thus there are $N^{n-2}_0$ of RMEs of each type $X_\alpha^I$ and $X_\alpha^{II}$. Note that $X^{II}_\alpha$ are all $b = 0$ RMEs with $U_1 = 0$, while $X^{I}_\alpha$ are only some of the $b = 1$ RMEs with $U_1 = 1$ as in the latter case $U_{n-2}\neq 0$ is also possible.

As we discussed above, the rank of the matrix $\left[C_{j\alpha}\right]$ is determined by coefficients $C^*(\overline{m}^d,\,\overline{U})$, see the discussion below Eq.~\eqref{eq:C_jalpha-def}. Using Eq.~\eqref{eq:C*-def} we find the following for the coefficients $C^*(\overline{m}^d,\,\overline{U}_I^\prime)$ and $C^*(\overline{m}^d,\,\overline{U}_{II}^\prime)$ that correspond to the RMEs in Eqs.~\eqref{eq:XI-b1} and~\eqref{eq:XII-b1} respectively 
\begin{equation}
    C^*(\overline{m}^d,\,\overline{U}_I^\prime) = 
    C^*(\overline{m}^d,\,\overline{U}_{II}^\prime) = 
    \frac{1}{\sqrt{2}} \times C^*(\overline{m}_{n-2},\,\overline{U}_{n-2}^\prime)\times \delta_{M,\,0}\,,
\end{equation}
where $\delta_{M,\,0}$ is Kronecker delta, $M = m_{n-1} + m_n$, and we exclude the amplitudes of the system of doublets for which $m_{n-1} = -1/2$, $m_n = +1/2$. Thus we learn that the RMEs of the form in Eq.~\eqref{eq:XI-b1} that enter the decomposition at order $b = 1$ are linearly dependent with the $b = 0$ RMEs in Eq.~\eqref{eq:XII-b1}. The number of these RMEs is equal to $N^{n-2}_0$. Since the addition of these $b = 1$ RMEs does not affect the counting of sum rules we exclude them from the decomposition and conclude that
\begin{equation}
    \text{rank\,} \widetilde{C}_{j\alpha}^{(b = 1)} \leq \text{rank\,} {C}_{j\alpha}^{(b = 1)} - N^{n-2}_{0}  = \sum_{b^\prime = 0}^{1} \left(N^n_{b^\prime} - N^{n-2}_{b^\prime - 1}\right)\,.
\end{equation}

Finally, we are adding to the decomposition RMEs with $b\geq2$. Consider the following $U_1 = 1$ RMEs at order $b$
\begin{equation}\label{eq:XI}
    X_\alpha^{I} = \mel{\overline{U}^\prime_{III}}{b}{0}\,,\qquad \overline{U}_{III}^\prime = \{1,\,\overline{U}_{n-2}^\prime,\,b\}\,,
\end{equation}
$U_1 = 0$ RMEs at order $b - 1$
\begin{equation}\label{eq:XII}
    X_\alpha^{II} = \mel{\overline{U}^\prime_{II}}{b-1}{0}\,,\qquad \overline{U}_{II}^\prime = \{0,\,\overline{U}_{n-2}^\prime,\,b-1\}\,,
\end{equation}
and $U_1 = 1$ RMEs at order $b - 2$
\begin{equation}\label{eq:XIII}
    X_\alpha^{III} = \mel{\overline{U}^\prime_I}{b-2}{0}\,,\qquad \overline{U}_{III}^\prime = \{1,\,\overline{U}_{n-2}^\prime,\,b-2\}\,,
\end{equation}
where the sets $\overline{U}^\prime_{n-2}$ are defined as all sets $\overline{U}_{n-2}$ for which $U_{n-2} = b-1$. There are $N^{n-2}_{b-1}$ of sets $\overline{U}_{n-2}^\prime$ and thus there are $N^{n-2}_{b-1}$ of RMEs of each type in Eqs.~\eqref{eq:XI}--~\eqref{eq:XIII}. 

Using Eq.~\eqref{eq:C*-def} we find the following for the coefficients $C^*(\overline{m}^d,\,\overline{U}_I^\prime)$, $C^*(\overline{m}^d,\,\overline{U}_{II}^\prime)$ and $C^*(\overline{m}^d,\,\overline{U}_{III}^\prime)$ that correspond to the RMEs in Eqs.~\eqref{eq:XI}--~\eqref{eq:XIII},~respectively,
\begin{align}
    C^*(\overline{m}^d,\,\overline{U}_I^\prime) & = \frac{1}{\sqrt{2}}\times C^*(\overline{m}_{n-2},\,\overline{U}_{n-2}^\prime)\times  \hspace{-16pt}\mathop{C_{1, M}}_{\hspace{28pt} b-1, -M}^{\hspace{6pt} b, 0}\,,\label{eq:C*-XI}\\
    C^*(\overline{m}^d,\,\overline{U}_{II}^\prime) & = \frac{1}{\sqrt{2}}\times C^*(\overline{m}_{n-2},\,\overline{U}_{n-2}^\prime)\times \delta_{M,\,0}\,,\label{eq:C*-XII}\\
    C^*(\overline{m}^d,\,\overline{U}_{III}^\prime) & = \frac{1}{\sqrt{2}}\times C^*(\overline{m}_{n-2},\,\overline{U}_{n-2}^\prime)\times \hspace{-16pt}\mathop{C_{1, M}}_{\hspace{28pt} b-1, -M}^{\hspace{16pt} b-2, 0}\,,\label{eq:C*-XIII}
\end{align}
where as in the case $b=1$, we exclude the amplitudes of the system of doublets for which $m_{n-1} = -1/2$, $m_n = +1/2$. Next, we show that the RMEs $X_\alpha^{I}$ that we add at order $b$ are linearly dependent with the lower order RMEs $X_\alpha^{II}$ and $X_\alpha^{III}$. In order to do this, we find non-zero coefficients $\omega_{I}$, $\omega_{II}$, $\omega_{III}$, such that
\begin{equation}\label{eq:LD-cond}
    \omega_I\, C^*(\overline{m}^d,\,\overline{U}_I^\prime) + \omega_{II}\, C^*(\overline{m}^d,\,\overline{U}_{II}^\prime) + \omega_{III}\,C^*(\overline{m}^d,\,\overline{U}_{III}^\prime) = 0\,,
\end{equation}
for all $\overline{m}^d$ except the ones with $m_{n-1} = -1/2$, $m_n = +1/2$. This condition for the linear dependence can be rewritten using Eqs.~\eqref{eq:C*-XI}--\eqref{eq:C*-XIII} as a system of two equations
\begin{align}
    \omega_I \hspace{-12pt}\mathop{C_{1, +1}}_{\hspace{20pt} b-1, -1}^{\hspace{4pt} b, 0} + \,\omega_{III} \hspace{-10pt}\mathop{C_{1, +1}}_{\hspace{20pt} b-1, -1}^{\hspace{12pt} b-2, 0} &=0\,,\label{eq:system-eq-1}\\
    \omega_I \hspace{-12pt}\mathop{C_{1, 0}}_{\hspace{20pt} b-1, 0}^{\hspace{8pt} b, 0} + \,\omega_{II} + \omega_{III} \hspace{-10pt}\mathop{C_{1, 0}}_{\hspace{20pt} b-1, 0}^{\hspace{18pt} b-2, 0} &=0\,,\label{eq:system-eq-2}
\end{align}
where the first equation is written for $M = +1$ and the second equation for $M = 0$. The case $M = -1$ results in the same equation as for $M = +1$ due to the following property of CG coefficients 
\begin{equation}\label{eq:CG_sym}
\mathop{C_{u_1, m_1}}_{\hspace{8pt} u_2, m_2}^{\hspace{25pt} u_3, m_1 + m_2} = (-1)^{u_1 + u_2 - u_3} \times \!\!\!\!\!\!\!\mathop{C_{u_1, -m_1}}_{\hspace{8pt} u_2, -m_2}^{\hspace{25pt} u_3, -m_1 - m_2}.
\end{equation}
Setting $\omega_{II} = 1$, we can solve the system in Eqs.~\eqref{eq:system-eq-1}--\eqref{eq:system-eq-2} for $\omega_I$ and $\omega_{III}$, we find
\begin{align}
    &\omega_I  = -\hspace{-10pt}\mathop{C_{1, +1}}_{\hspace{20pt} b-1, -1}^{\hspace{12pt} b-2, 0}\left(\hspace{-10pt}\mathop{C_{1, +1}}_{\hspace{20pt} b-1, -1}^{\hspace{12pt} b-2, 0}\hspace{-12pt}\mathop{C_{1, 0}}_{\hspace{20pt} b-1, 0}^{\hspace{8pt} b, 0} -  \hspace{-8pt}\mathop{C_{1, +1}}_{\hspace{20pt} b-1, -1}^{\hspace{4pt} b, 0} \hspace{-10pt}\mathop{C_{1, 0}}_{\hspace{20pt} b-1, 0}^{\hspace{18pt} b-2, 0}\right)^{-1}\,,\nonumber\\
    &\omega_{III}  = \hspace{-10pt}\mathop{C_{1, +1}}_{\hspace{20pt} b-1, -1}^{\hspace{2pt} b,0} \left(\hspace{-10pt}\mathop{C_{1, +1}}_{\hspace{20pt} b-1, -1}^{\hspace{12pt} b-2, 0}\hspace{-12pt}\mathop{C_{1, 0}}_{\hspace{20pt} b-1, 0}^{\hspace{8pt} b, 0} -  \hspace{-8pt}\mathop{C_{1, +1}}_{\hspace{20pt} b-1, -1}^{\hspace{4pt} b, 0} \hspace{-10pt}\mathop{C_{1, 0}}_{\hspace{20pt} b-1, 0}^{\hspace{18pt} b-2, 0}\right)^{-1}\,.
\end{align}
One can check explicitly that these expressions for $\omega_I$ and $\omega_{III}$ are finite and are non-zero for any $b \geq 2$. Thus we have found $\omega_I$, $\omega_{II}$ and $\omega_{III}$ that are non-zero for any $b\geq2$ and are such that the condition of the linear dependence in Eq.~\eqref{eq:LD-cond} is satisfied. Thus we conclude that there are $N^{n-2}_{b-1}$ RMEs of the form in Eq.~\eqref{eq:XI} that enter the decomposition in Eq.~\eqref{eq:A-decomp-0+1} at order $b$ and that are linearly dependent with RMEs at lower orders $b-1$ and $b-2$. The addition of these order $b$ RMEs does not affect the rank of the coefficient matrix, thus the following inequality is satisfied
\begin{equation}
    \text{rank\,} \left[\widetilde{C}_{j\alpha}^{(b)}\right] \leq \sum_{b^\prime = 0}^{b} \left(N^n_{b^\prime} - N^{n-2}_{b^\prime-1}\right)\,,
\end{equation}
where we emphasize the order of breaking $b$ as a superscript in the RHS.

Let's assume that there are additional linear dependencies, that reduce the rank of the $\left[\widetilde{C}_{j\alpha}^{(b)}\right]$ by some number that we denote as $N_{LD}(n,b)\geq 0$ such that
\begin{equation}\label{eq:NLD}
    \text{rank\,} \left[\widetilde{C}_{j\alpha}^{(b)}\right] = \sum_{b^\prime = 0}^{b} \left(N^n_{b^\prime} - N^{n-2}_{b^\prime-1}\right) - \sum_{b^\prime = 0}^{b}N_{LD}(n,b^\prime)\,.
\end{equation}
The expansion for the system of $n$ doublets saturates at the order of breaking $b = n/2$. That is at this order of breaking $\text{rank }\left[C_{j\alpha}\right] = n_A^d(n)$. Since the $0\oplus1$ system is constructed from the system of doublets, its expansion should saturate at order $b \leq n/2$, and thus
\begin{equation}
    \text{rank\,} \left[\widetilde{C}_{j\alpha}^{(b = n/2)}\right] = n_A^{0\oplus 1}(n)\,.
\end{equation}
On the other hand, evaluating Eq.~\eqref{eq:NLD} for $b = n/2$, we find
\begin{equation}
    \text{rank\,} \left[\widetilde{C}_{j\alpha}^{(b)}\right] = n_A^{0\oplus 1}(n) - \sum_{b^\prime = 0}^{n/2}N_{LD}(n,b^\prime)\,,
\end{equation}
see Eq.~(E19) from Ref.~\cite{Gavrilova:2022hbx} and the first line of Eq.~\eqref{eq:nA-0+1-final}. It follows that
$N_{LD}(n, b^\prime) = 0$ for all $n$ and $b^\prime$. Thus the only linear dependencies that arise in matrix $\left[C_{j\alpha}\right]$ due to the removal of rows corresponding to the amplitudes with $m_{n-1} = -1/2$, $m_n = +1/2$, are the ones that we find above and thus the rank of the matrix $\left[\widetilde{C}_{j\alpha}\right]$ is given by Eq.~\eqref{eq:rank-C-tilde}. \textbf{End of proof.}

\subsection{The number of sum rules $n_{SR}^{0\oplus1}(n,\,b)$}

Using Eqs.~\eqref{eq:nSR-0+1-def},~\eqref{eq:n-disc},~\eqref{eq:nA-0+1-final} and~\eqref{eq:rank-C-tilde}, we find the following for the number of sum rules in the $0\oplus1$ system at the order of breaking $b$
\begin{equation}
    n_{SR}^{0\oplus1}(n,b) = \binom{n}{n/2} - \binom{n-2}{n/2-1} - \sum_{U = 0}^b N_U^n + \sum_{U = 0}^{b-1} N_U^{n-2}\,.
\end{equation}
Interestingly, using Eq.~\eqref{eq:nSR_doublets}, $n_{SR}^{(0\oplus1)}(n,b)$ can be rewritten as
\begin{equation}\label{eq:nSR-0+1-app}
    n_{SR}^{0\oplus1}(n,b) = n^{d}_{SR}(n, b) - n^{d}_{SR}(n-2,b-1)\,,
\end{equation}
where $n_{SR}^{d}(n,b)$ is the number of sum rules for a system of $n$ doublets at order $b$. There is a subtle point to note regarding this expression for $n_{SR}^{0\oplus 1}(n, b)$ when $b = 0$. For $b = 0$, on the RHS, we have $n_{SR}^d(n - 2, -1)$. While the number of sum rules at order ``$-1$'' is not meaningful, and therefore in this case the result for the RHS of Eq.~\eqref{eq:nSR_doublets} does not correspond to a number of sum rules, Eq.~\eqref{eq:nSR-0+1-app} remains valid.

\section{Proof of the sum rule theorem for a $0\oplus 1$ system \label{sec:proof} }

In this appendix, we prove the theorem that establishes the algorithm for deriving amplitude sum rules for a $0 \oplus 1$ system at an order of breaking $b \geq 1$. We study systems with the group-theoretical structure described by Eq.~\eqref{eq:system-0+1}. That is, we consider, without loss of generality, a coefficient-free system (see the discussion in Section~\ref{sec:u0-coeff} and the introduction to Section~\ref{sec:0+1}), which is described by a direct sum of a singlet and triplet, and $n-2$ doublets, all in the final state. Note that the complete set of amplitude sum rules at order $b = 0$ was derived in Section~\ref{sec:case-no-coeff} by analyzing the general structure of the RME decomposition for systems with different parities. In this appendix, we focus on the case where $b \geq 1$.

As in Section~\ref{sec:0+1-from-n-doublets}, in this appendix we will use the two auxiliary systems that are obtained from the system of $n$ doublets via symmetrization (see Eq.~\eqref{eq:system-n-sym}) and anti-symmetrization (see Eq.~\eqref{eq:system-n-antisym}) of two doublets. For brevity we refer to this systems as $u = 1$ and $u = 0$ systems respectively. Throughout this appendix we use the notation for the amplitudes of the $u = 0$ and $u = 1$ systems introduced in Section~\ref{sec:0+1-from-n-doublets}, and for the RMEs we use the notation introduced in Section~\ref{sec:gen-case}.

Before we proceed to the formulation and proof of the sum rule theorem, we first formulate the following important lemma.\\

\textbf{Lemma.} All the sum rules of the $u = 1$ system that hold at order of breaking $b + 1$ also hold for the $0 \oplus 1$ system at order $b$.

\textbf{Proof.} This lemma follows directly from the proof of the formula for the rank of the matrix $\left[\widetilde{C}_{j\alpha}\right]$ in Appendix~\ref{app:rank-theorem}. The key insight necessary for deriving this formula is the realization that, as we have shown in Appendix~\ref{app:rank-theorem}, for any order $b$, RMEs of the $u = 0$ system (corresponding to $U_1 = 0$ in Appendix~\ref{app:rank-theorem} and denoted by $X^{II}_\alpha$ in Eq.~(\ref{eq:XII})) are linearly dependent on the RMEs of the $u = 1$ system at one higher order, $b + 1$, and one lower order, $b - 1$ (corresponding to $U_1 = 1$ and denoted by $X^{I}_\alpha$ and $X^{III}_\alpha$, respectively, see Eqs.~(\ref{eq:XI}) and (\ref{eq:XIII}).).

If there is a sum rule that holds for the $u = 1$ system at order $b + 1$, this means that there exist coefficients $\omega_i$ such that
\begin{equation}
    \sum_i \omega_i\, A^{(1)}_i = 0\,,
\end{equation}
where $A^{(1)}_i$ are amplitudes of the $u = 1$ system and at least one of the $\omega_i$ is non-zero. The existence of a $b + 1$ order sum rule implies that for all RMEs $X_\alpha^{(b^\prime)}(1)$ of the $u = 1$ system with the order of breaking $b^\prime \leq b + 1$, the sum of the CG coefficients $C_{i\alpha}^{(b^\prime)}$ weighted by coefficients $\omega_i$ is equal to zero:
\begin{equation}\label{eq:sum-rule-ex}
    \sum_i \omega_i\, C_{i\alpha}^{(b^\prime)} = 0\,, \qquad \forall \alpha,\,b^\prime \leq b + 1 \,.
\end{equation}
In particular, the cancellation holds for the coefficients in front of RMEs $X_\alpha^{I}$ and $X_\alpha^{III}$, which we denote for brevity as $C^{(b^\prime),\,I}_{i\alpha}$ and $C^{(b^\prime),\,III}_{i\alpha}$ respectively:
\begin{equation}\label{eq:sum-rule-cons}
    \sum_i \omega_i\,C^{(b^\prime),\,I}_{i\alpha} = 0\,,\quad \sum_i \omega_i\,C^{(b^\prime),\,III}_{i\alpha} = 0\,, \qquad \forall \alpha,\, b^\prime \leq b + 1\,.
\end{equation}
The linear dependence between RMEs $X_\alpha^{I}$, $X_\alpha^{II}$, and $X_\alpha^{III}$, given by Eq.~\eqref{eq:LD-cond}, has the form
\begin{equation}\label{eq:LD-lemma}
    \omega_I\, C_{i\alpha}^{(b^\prime),\,I} + C_{i\alpha}^{(b^\prime-1),\,II} + \omega_{III}\, C_{i\alpha}^{(b^\prime-2),\,III} = 0\,, \quad \forall \alpha,\, b^\prime\,,
\end{equation}
where, for brevity, we use $C_{i\alpha}^{(b),\,II}$ to denote the CG coefficient in front of $u = 0$ RME $X_\alpha^{II}$ and set $\omega_{II} = 1$. The linear dependence in Eq.~\eqref{eq:LD-lemma} and Eq.~\eqref{eq:sum-rule-cons} imply that
\begin{equation}\label{eq:sum-rule-cons-2}
    \sum_i \omega_i\, C_{i\alpha}^{(b^\prime - 1),\,II} = 0\,, \quad \forall \alpha, b^\prime \leq b + 1\,.
\end{equation}
Since the amplitudes of the $0 \oplus 1$ system are simply a sum of the amplitudes of the $u = 1$ and $u = 0$ systems (see the discussion in Section~\ref{sec:0+1-from-n-doublets} and Eqs.~\eqref{eq:A-0+1-def-1}--\eqref{eq:A-0+1-def-2}), Eq.~\eqref{eq:sum-rule-cons-2} means that the coefficients $\omega_i$ ensure the cancellation of the CG coefficients in front of all RMEs of the $u = 1$ system up to order $b + 1$ and the cancellation between CG coefficients for all RMEs of the $u = 0$ system up to order $b$. This means that the sum rule
\begin{equation}
    \sum \omega_i A^{0 \oplus 1}_i = 0\,,
\end{equation}
holds for the $0 \oplus 1$ system up to order $b$, which proves the statement of the Lemma.

\subsection{Formulation of the sum rule theorem for $0\oplus 1$ systems}

Since the amplitudes of the $0 \oplus 1$ system are given by the sums of the amplitudes of the $u = 0$ and $u = 1$ systems, any sum rules that hold for the $0 \oplus 1$ system should also hold for the $u = 0$ and $u = 1$ systems separately. We choose to formulate the sum rules for the $0 \oplus 1$ system in terms of the sum rules of the $u = 1$ system. We do this using the lattice approach, which we overview in Section~\ref{sec:lattice-review}. The logic is as follows: We start with the $u = 1$ system and think of the singlet in the direct sum $0 \oplus 1$ as a correction. The addition of this correction to the $u = 1$ system only affects the amplitudes with $M = 0$ and results in the breaking of some of the sum rules that hold for the $u = 1$ system. The sum rule theorem that we formulate below provides guidance on selecting the sum rules of the $u = 1$ system that are preserved in the $0 \oplus 1$ system.

In our formulation of the sum rule theorem below, we assume that we have constructed a $d = n/2 - 1$ dimensional lattice for the $u = 1$ system. Each node of the lattice represents an amplitude pair and is labeled by $d$ numbers that we call coordinates. The coordinates can take values from $1$ to $n-2$, such that:
\begin{itemize}
    \item Out of the $d$ coordinates, no more than one can be equal to $1,\,\dots,\,n-3$.
    \item No more than two can be equal to $n - 2$.
\end{itemize}
The reason is the structure of the $n$-tuple, which is given for example as 
\begin{align}
(\underset{0}{-}, 
\underset{1}{+},
\underset{2}{-},\text{\dots}, 
\underset{n-3}{-}, 
\underset{n-2}{--}) = (2, \text{\dots}, n-3, n-2, n-2)\,,
\end{align}
where on the LHS we show the $n$-tuple notation together with a numbering scheme of the entries of the $n$-tuple, and on the RHS the $d=n/2-1$ coordinates of the lattice. The leading minus sign is not reflected in the lattice (hence the \lq\lq{}$-1$\rq\rq{} in \lq\lq{}$d=n/2-1$\rq\rq{}). The values of the coordinates reflect the position of the minus signs. So there can be at most 2 minus signs at position $n-2$, which corresponds to the triplet, and not more then one minus sign between position 1 to $n-3$, because these correspond to the doublet representations. 

The lattice nodes that have one or two of their coordinates equal to $n - 2$ are accompanied by $\mu$-factors, that are explained in Section~\ref{sec:lattice-review}. In what follows, when we refer to a sum over nodes in a given subspace of the lattice, we always imply that the nodes in the sum are weighted by the appropriate $\mu$-factors.\\

\textbf{Theorem.} The full set of amplitude sum rules for the $0 \oplus 1$ system is read off the lattice for the $u = 1$ system, and at even (odd) order of breaking $b \geq 1$ is given by:
\begin{enumerate}
    \item[(1)] $s$($a$)-type sum rules, which are given by sums of $s$($a$)-type amplitudes corresponding to the nodes in all $(b + 1)$-dimensional subspaces of the lattice. The $(b + 1)$-dimensional subspaces are obtained by fixing $d - (b + 1)$ coordinates and allowing the remaining $(b + 1)$ coordinates to take any of the allowed values. The number of these sum rules is given by the following sum of binomial coefficients:
    \begin{equation}\label{eq:nSR-1}
        \binom{n-3}{n/2-b-2} + \binom{n-3}{n/2 - b - 3} + \binom{n-3}{n/2 - b - 4}\,,
    \end{equation}
    and they form the full set of $s$($a$)-type sum rules for the $0 \oplus 1$ system at order $b$.
    
    \item[(2)] $a$($s$)-type sum rules, which are given by:
    \begin{enumerate}
        \item[(2a)] The sum of all $a$($s$)-type amplitudes corresponding to the nodes in all $b$-dimensional subspaces obtained by fixing $d - b$ coordinates, such that two of the $d - b$ fixed coordinates are equal to $n - 2$ and the rest can take any of the allowed values. There are
        \begin{equation}\label{eq:nSR-2a}
            \binom{n - 3}{n/2 - b - 3} 
        \end{equation}
        of these sum rules.
        
        \item[(2b)] For all $(b + 1)$-dimensional subspaces obtained by fixing $d - b - 1$ coordinates such that none of the fixed coordinates is equal to $n - 2$, the sum of all nodes in the subspace without any coordinates being equal to $n - 2$ minus the sum of all nodes in the subspace for which two of the coordinates are equal to $n - 2$. There are
        \begin{equation}\label{eq:nSR-2b}
            \binom{n-3}{n/2 - b - 2}
        \end{equation}
        of these sum rules.
        
        \item[(2c)] The sum of all nodes in all $(b + 2)$-dimensional subspaces obtained by fixing $d - (b + 2)$ coordinates such that none of the fixed coordinates is equal to $n - 2$. There are
        \begin{equation}\label{eq:nSR-2c}
            \binom{n-3}{n/2 - b - 3}
        \end{equation}
        of these sum rules.
    \end{enumerate}
    The sum rules described by (2a)--(2c) give the full set of $a$($s$)-type sum rules.
\end{enumerate}

\textbf{Outline of the proof:}
\begin{itemize}[leftmargin=2.3cm]
    \item[\textbf{Step 1:}] We show that the sum rules described by (1) and (2a)--(2b) in the formulation of the theorem are sum rules of the $0 \oplus 1$ system that hold at order $b$.
    \item[\textbf{Step 2:}] We show that the described sum rules are all linearly independent.
    \item[\textbf{Step 3:}] We show that these sum rules form the full set of sum rules of the $0 \oplus 1$ system at order $b$.
\end{itemize}

\subsection{Step 1}

\subsubsection{(1)}
 First, let us consider the sum rules defined by (1). We know that \lq\lq{}for even (odd) $b$, the $b$ dimensional subspaces of the lattice  correspond to $a$$(s)$-type sum rules\rq\rq{} (Sec. III~F of Ref.~\cite{Gavrilova:2022hbx}). Therefore,
at even (odd) $b$, the $b+1$ dimensional subspaces correspond to $s$($a$)-type sum rules of the $u = 1$ system that hold at order $b+1$. Thus, by the Lemma, these sum rules also hold for the $0 \oplus 1$ system at order $b$.  To count the number of these sum rules, we need to count the number of ways to fix the values of $d - (b + 1)$ coordinates. This counting is equivalent to counting the number of ways to choose $d - (b + 1)$ numbers from $1,\,\dots,\, n-2$ such that each number can only be chosen once, except $n-2$, which can be chosen twice. There are three options when fixing $d - (b + 1)$ coordinates: (\emph{i}) none of the fixed coordinates is equal to $n - 2$, (\emph{ii}) one of the fixed coordinates is equal to $n - 2$, and (\emph{iii}) two of the fixed coordinates are equal to $n - 2$. For each of these three options, the number of ways to fix $d - (b + 1)$ coordinates is given by the binomial coefficients in Eq.~\eqref{eq:nSR-1}, corresponding to (\emph{i}), (\emph{ii}), and (\emph{iii}), respectively. The sum of the three binomial coefficients gives the total number of sum rules described by (1). Note that in Eq.~\eqref{eq:nSR-1} we used $d = n/2-1$.

\subsubsection{(2a)}
Next, let us consider the sum rules described by (2a). For even (odd) $b$, these sum rules are $a$($s$)-type sum rules of the $u = 1$ system that hold at order $b$. A special feature of these sum rules is that all amplitudes in each sum rule have two coordinates equal to $n-2$. This means that all of the amplitudes that enter these sum rules have $M = \pm 1$, and thus they are unaffected by the singlet in the direct sum. Recall that only the amplitudes with $M = 0$ get contributions from the singlet in the direct sum. In the language of Sec.~\ref{sec:case-no-coeff}, these sum rules only involve amplitudes $A^{0\oplus 1}(1)$ and are therefore preserved in the $0\oplus 1$ system. As two out of $d - b$ fixed coordinates are chosen to be equal to $n-2$, the number of these sum rules is given by the number of ways to choose $d-b-2$ numbers from $1,\,\dots,\,n-3$, which is given by the binomial coefficient in Eq.~\eqref{eq:nSR-2a}.

\subsubsection{(2b)}\label{app:2b-proof}

Consider a $(b+1)$-dimensional subspace of the lattice obtained by fixing $d-b-1$ coordinates such that none of the fixed coordinates is equal to $n-2$. Without loss of generality, we choose the fixed coordinates to be $1,\,\dots,\,d-b-1$. For concreteness, let us consider the case when $b$ is even. In this case, we focus on $a$-type amplitudes within the subspace.  The case for odd $b$ is exactly the same, except for a relabeling of amplitudes, $a\,\rightarrow s$. Since all nodes in this subspace can be labeled by $b+1$ coordinates, we denote the $a$-type amplitudes corresponding to the nodes of this subspace using coordinate notation as
\begin{equation}\label{eq:amp-coord-not}
    a(x_1,\,\dots,\,x_{b+1})\,,\qquad x_i = d-b,\,\dots,\,n-2\,.
\end{equation}
In this notation, we omit the $d-b-1$ fixed coordinates that define the subspace since they are shared by all the amplitudes within the subspace. Recall that for any node in the subspace, there exists a node obtained by any permutation of the coordinates in Eq.~\eqref{eq:amp-coord-not}, which also lies in the same subspace. Also, note that none of the values from $d-b,\,\dots,\,n-3$ can appear in the coordinate notation twice, while $n-2$ can appear at most two times. When we write amplitudes using the notation in Eq.~\eqref{eq:amp-coord-not}, we always assume that the values of the coordinates are such that the corresponding node is allowed.

First, in addition to the $d - b - 1$ coordinates fixed above, let us also consider all the possible ways to fix one extra coordinate. Without loss of generality, we choose to fix the value of $x_{b+1}$. The following $a$-type sum rules, corresponding to fixing $d - b$ coordinates ($d - b - 1$ coordinates fixed above and $x_{b+1}$), hold for the $u = 1$ system at order $b$:
\begin{equation}\label{eq:sr-b-dim-subspace}
    \sum_{x_1,\,\dots,\, x_{b}} \mu(x_1,\,\dots,\,x_{b+1})\, a(x_1,\,\dots,\,x_{b+1}) = 0\,, \qquad x_{b+1} = d-b,\, \dots, n-2\,,
\end{equation}
where $\mu(x_1,\,\dots,\,x_{b+1})$ are $\mu$-factors that appear in the lattice for the $u = 1$ system due to the symmetrization of two doublets. For a discussion of $\mu$-factors, see Section IV.C.2 and Eq.(4.34) of Ref.~\cite{Gavrilova:2022hbx}.
In the case of the $u = 1$ system, the $\mu$-factors are given by
\begin{equation}\label{eq:mu-u1-def}
    \mu(x_1,\,\dots,\,x_{b+1}) = \begin{cases}
        1, \quad \text{if } x_i \neq n-2 \text{ for all } i,\\
        \sqrt{2}, \quad \text{if only one of the } x_i \text{ is equal to } n-2,\\
        2, \quad \text{if two of the } x_i \text{ are equal to } n-2.
    \end{cases}
\end{equation}
Note that we use here for the purpose of the proof the same conventions for the coordinate notation of the $\mu$-factor as for the amplitudes introduced in Eq.~(\ref{eq:amp-coord-not}). The \lq\lq{}standard notation\rq\rq{} includes all indices and is shown in Eq.~(\ref{eq:mu-u1-def-main}).

There are $(n-2) - (d-b) + 1$ sum rules in Eq.~\eqref{eq:sr-b-dim-subspace}, each sum rule corresponds to a particular value of $x_{b + 1}$. Note that in Eq.~\eqref{eq:sr-b-dim-subspace}, all permutations of coordinates $x_1,\,\dots,\,x_{b}$ appear in the sums; we do not yet identify the amplitudes. Also note that by construction, all nodes of the $(b + 1)$-dimensional subspace that we consider appear in the sums in Eq.~\eqref{eq:sr-b-dim-subspace} exactly once. 

We construct the following linear combination of the sum rules in Eq.~\eqref{eq:sr-b-dim-subspace}:
\begin{equation}\label{eq:SR-set-up}
   \sum_{\substack{x_1,\,\dots,\, x_{b}\\ x_{b+1}\neq n-2}} \mu(x_1,\,\dots,\,x_{b+1})\, a(x_1,\,\dots,\,x_{b+1}) - b \sum_{x_1,\,\dots,\, x_{b}} \mu(x_1,\,\dots,n-2)\, a(x_1,\,\dots,\,n-2) = 0\,.
\end{equation}
In this linear combination we take the sum of the order $b$ sum rules of the $u = 1$ system in Eq.~\eqref{eq:sr-b-dim-subspace} that correspond to $x_{b+1} \neq n-2$ and subtract the sum rule with $x_{b+1} = n-2$ multiplied by the order of breaking $b$. Since any linear combination of sum rules is also a sum rule, the relation in Eq.~\eqref{eq:SR-set-up} is an order $b$ sum rule of the $u = 1$ system. In what follows, we show that in this sum rule, all the amplitudes that have only one coordinate equal to $n-2$ cancel out. Thus, the only remaining amplitudes are those unaffected by the singlet in the direct sum, meaning this sum rule also holds for the $0\oplus 1$ system at order $b$. Recall that the nodes for which two coordinates are equal to $n - 2$ correspond to $M = \pm 1$ and thus they do not get a contribution from the singlet in the direct sum.

To demonstrate the advertised cancellation, we rewrite the first sum in Eq.~\eqref{eq:SR-set-up} as
\begin{align}\label{eq:1st-sum}
    \sum_{\substack{x_1,\,\dots,\, x_{b}\\ x_{b+1}\neq n-2}} &\mu(x_1,\,\dots,\,x_{b+1})\, a(x_1,\,\dots,\,x_{b+1}) = \nonumber\\
    & = \sum_{x_1,\,\dots,\, x_{b+1}\neq n-2} a(x_1,\,\dots,\,x_{b+1}) \nonumber\\
    & + \sqrt{2}\,b \sum_{\substack{x_1,\,\dots,\, x_{b-1} \neq n-2\\ x_{b+1}\neq n-2}} a(x_1,\,\dots,\, x_{b-1},\, n-2,\,x_{b+1}) \nonumber\\
    & + b(b-1) \sum_{\substack{x_1,\,\dots,\, x_{b-2} \neq n-2\\ x_{b+1}\neq n-2}} a(x_1,\,\dots,\, x_{b-2},\, n-2,\, n-2,\, x_{b+1})\,.
\end{align}
In this equation, we explicitly separate the contributions from nodes for which none of the coordinates equals $n-2$ (second line), one coordinate equals $n-2$ (third line), and two coordinates equal $n-2$ (fourth line). We substitute the explicit values of the $\mu$-factors from Eq.~\eqref{eq:mu-u1-def}. The $\mu$-factors are equal to $1$ for all amplitudes in the first line, $\sqrt{2}$ for amplitudes in the second line, and $2$ for amplitudes in the fourth line. When writing Eq.~\eqref{eq:1st-sum}, we also identify some of the nodes. In the third line, we only keep the amplitudes for which $x_{b} = n-2$; however, the sum in the first line includes all possible permutations of coordinates where one of the $x_i$, for $i = 1,\,\dots, b$, equals $n-2$. Thus, for any particular combination of values of $x_1,\,\dots,\,x_{b-1}$ in the third line, there are $b$ nodes, all of which we identify with the node where $x_{b} = n-2$. Hence the multiplication by a factor of $b$. Similarly, in the fourth line, we only keep the nodes for which $x_{b-1} = x_{b} = n-2$. There are
\begin{align}
\binom{b}{2} &=  \frac{b!}{2!(b-2)!} = \frac{b (b-1)}{2}
\end{align}
nodes in the first line that are identical to each of the nodes in the fourth line, hence the additional overall factor of $b(b-1)/2$ in front of the sum in the fourth line. The factor of $1/2$ is cancelled with the $\mu$-factor and we obtain the final result.

The second sum in Eq.~\eqref{eq:SR-set-up} can be written as
\begin{align}\label{eq:2nd-sum}
    b \sum_{x_1,\,\dots,\, x_{b}} & \mu(x_1,\,\dots,\,x_{b+1})\, a(x_1,\,\dots,\,n-2)  = \nonumber\\
    & = \sqrt{2}\, b\sum_{x_1,\,\dots,\, x_{b}\neq n-2} a(x_1,\,\dots,\,n-2) \nonumber\\
    & + 2 b^2 \sum_{x_1,\,\dots,\, x_{b-1}\neq n-2} a(x_1,\,\dots,\, x_{b-1},\, n-2,\, n-2)\,.
\end{align}
In this equation, we use the explicit values of the $\mu$-factors from Eq.~\eqref{eq:mu-u1-def}: $\sqrt{2}$ for the amplitudes in the second line and $2$ for the amplitudes in the third line. In the third line, we also identify $b$ nodes for which one of the $x_i$ coordinates, $i = 1,\,\dots,\, b$, is equal to $n-2$ with the node for which $x_b = n-2$, hence the overall factor of $b$.

Using Eqs.~\eqref{eq:1st-sum} and~\eqref{eq:2nd-sum}, the sum rule in Eq.~\eqref{eq:SR-set-up} becomes
\begin{equation}\label{eq:SR-almost-final}
    \sum_{x_1,\,\dots,\, x_{b+1}\neq n-2} \hspace{-20pt} a(x_1,\,\dots,\,x_{b+1}) - b (b + 1) \hspace{-20pt} \sum_{x_1,\,\dots,\, x_{b-1}\neq n - 2} \hspace{-20pt} a(x_1,\,\dots,\, x_{b-1},\, n - 2,\, n - 2) = 0\,.
\end{equation}
The cancellation between the third line in Eq.~\eqref{eq:1st-sum} and the second line in Eq.~\eqref{eq:2nd-sum} can be seen explicitly from
\begin{align}
    \sum_{\substack{x_1,\,\dots,\, x_{b-1} \neq n-2\\ x_{b+1}\neq n-2}} a(x_1,\,\dots,\, x_{b-1},\, n-2,\,x_{b+1}) &= \nonumber\\
    \sum_{\substack{x_1,\,\dots,\, x_{b-1} \neq n-2\\ x_{b+1}\neq n-2}} a(x_1,\,\dots,\, x_{b-1},\, x_{b+1},\, n-2) &= \nonumber\\
    \sum_{\substack{x_1,\,\dots,\, x_{b-1},\,x_b \neq n-2}} a(x_1,\,\dots,\, x_{b-1},\, x_{b},\, n-2)\,,
\end{align}
where the first equality is the consequence of the permutation invariance of the coordinate notation, and the second equality is obtained by the relabeling of the summation index from $x_{b+1}$ to $x_b$. Similarly, the permutation invariance and the relabeling of the summation index in the fourth line of Eq.~\eqref{eq:1st-sum} allow us to combine the terms in the fourth line of Eq.~\eqref{eq:1st-sum} with the third line of Eq.~\eqref{eq:2nd-sum}.

Let us take a closer look at Eq.~\eqref{eq:SR-almost-final}. The first sum contains many lattice nodes that correspond to a single amplitude of the $U$-spin system. In particular, for any set of values of $x_i$, $i = 1,\, \dots,\, b+1$, there are $(b+1)!$ nodes that are related by permutations of coordinates. All these nodes map to a single amplitude. Without loss of generality, we choose to identify all nodes that differ only by permutations of coordinates with a single node for which $x_1 < \dots < x_{b+1}$. Thus, the first sum in Eq.~\eqref{eq:SR-almost-final} can be rewritten as
\begin{equation}\label{eq:1st-sum-ident}
    \sum_{x_1,\,\dots,\, x_{b+1}\neq n-2} \hspace{-20pt} a(x_1,\,\dots,\,x_{b+1}) = (b + 1)! \hspace{-20pt} \sum_{x_1 < \dots < x_{b+1}\neq n-2} \hspace{-20pt}  a(x_1,\,\dots,\,x_{b+1})\,.
\end{equation}
Similarly, in the second sum in Eq.~\eqref{eq:SR-almost-final}, for any set of values of $x_i$, $i = 1,\, \dots,\, b-1$, there are $(b-1)!$ nodes that are related by permutations of coordinates. Thus, we have
\begin{equation}\label{eq:2nd-sum-ident}
    \sum_{x_1,\,\dots,\, x_{b-1}\neq n - 2} \hspace{-20pt} a(x_1,\,\dots,\, x_{b-1},\, n - 2,\, n - 2) = (b-1)! \hspace{-20pt} \sum_{x_1 < \dots < x_{b-1}\neq n - 2} \hspace{-20pt} a(x_1,\,\dots,\, x_{b-1},\, n - 2,\, n - 2)\,.
\end{equation}
Plugging in Eqs.~\eqref{eq:1st-sum-ident} and~\eqref{eq:2nd-sum-ident} into Eq.~\eqref{eq:SR-almost-final} and canceling the overall factor $(b+1)!$, we find the following sum rule for the $u = 1$ system that holds at order $b$:
\begin{equation}\label{eq:sr-final}
    \sum_{x_1 < \dots < x_{b+1}\neq n-2} \hspace{-20pt}  a(x_1,\,\dots,\,x_{b+1}) \hspace{10pt} - \hspace{-20pt}\sum_{x_1 < \dots < x_{b-1}\neq n - 2} \hspace{-20pt} a(x_1,\,\dots,\, x_{b-1},\, n - 2,\, n - 2) = 0\,.
\end{equation}
Since this sum rule does not contain amplitudes for which exactly one of the coordinates is equal to $n - 2$, it also holds for the $0\oplus 1$ system at order $b$.

All amplitudes in the sum rule in Eq.~\eqref{eq:sr-final} belong to a subspace of the lattice for which $d - b - 1$ coordinates are fixed to some values from $1,\,\dots,\, n-3$. The sum rule is given by the difference of two sums: the sum of all unique nodes of the subspace for which none of the coordinates is equal to $n - 2$, and the sum of all unique nodes of the subspace for which two of the coordinates are equal to $n - 2$. This is the sum rule described by (2b).

Since for any $(b+1)$-dimensional subspace in which none of the $d - b - 1$ fixed coordinates is equal to $n - 2$, there exists a sum rule of the form in Eq.~\eqref{eq:sr-final}, the number of sum rules in (2b) can be found by counting the number of these $(b+1)$-dimensional subspaces. This number is equal to the number of ways we can choose $d - b - 1$ numbers from $1,\,\dots,\, n - 3$ and is given by the binomial coefficient in Eq.~\eqref{eq:nSR-2b}.

\subsubsection{(2c)}

The case of (2c) is very similar to (1). For even (odd) $b$, the sum rules in (2c) are $a$($s$)-type sum rules that hold for the $u = 1$ system at order $b + 2$. Since these sum rules hold at order $b + 2$, they also hold at order $b + 1$ for the $u = 1$ system and thus, by the Lemma, they are also sum rules of the $0 \oplus 1$ system at order $b$, as stated by the theorem. Since the sum rules in (2c) are obtained from the subspaces of the lattice where none of the $d - (b + 2)$ fixed coordinates is equal to $n - 2$, counting the subspaces (and thus the sum rules) is equivalent to counting the number of ways to choose $d - (b + 2)$ numbers from $1,\,\dots,\,n-3$. This number is given by the binomial coefficient in Eq.~\eqref{eq:nSR-2c}.

\subsection{Step 2}

In order to show that all the sum rules described by (1), (2a)--(2c) are linearly independent, we first demonstrate that the sum rules within each of the four groups are linearly independent and then show that there are no linear dependencies between the sum rules of different groups.

The sum rules in each of (1), (2a), and (2c) are sum rules of the $u = 1$ system that were studied in Ref.~\cite{Gavrilova:2022hbx} and were shown to be linearly independent. The different sum rules in (2b) belong to different subspaces, and thus each sum rule contains a non-overlapping set of amplitudes, making all the sum rules in (2b) linearly independent.

Next, we note that for a given even (odd) $b$, the sum rules in (1) are $s$($a$)-type sum rules, while the sum rules in (2a)--(2c) are $a$($s$)-type sum rules. Due to the decoupling of $a$- and $s$-type amplitudes that always occurs for a system with a singlet in the direct sum, see Section~\ref{sec:u0-coeff}, the sum rules in (1) are linearly independent from the sum rules in (2a)--(2c). Note that the decoupling of $a$- and $s$-type sum rules is obvious from Table~\ref{fig:CG-u1u2} as $a$- and $s$-type amplitudes always depend on different RMEs.
We conclude that if there are linear dependencies among the sum rules described in the formulation of the theorem, the only possibility is that there are linear dependencies between the sum rules in (2a)--(2c). Let us explore this possibility.

Let us consider the sum rules in (2a) and (2b). Recall that the sum rules in (2a) only contain amplitudes for which two of the coordinates are equal to $n - 2$, while each sum rule in (2b) contains both the amplitudes for which two of the coordinates are equal to $n - 2$ and the amplitudes for which none of the coordinates are equal to $n - 2$. The key point here is that each sum rule in (2b) contains amplitudes that do not enter the sum rules in (2a). This, combined with the fact that the sum rules in both (2a) and (2b) are linearly independent among themselves, respectively, ensures that there are no linear dependencies between the sum rules of the two groups.

Finally, the sum rules in (2c) are obtained by summing over a subspace where none of the fixed coordinates is equal to $n - 2$. Since this subspace has dimension $b + 2 > 0$ and the sum covers all nodes in the subspace, each sum rule in (2c) contains at least one amplitude with exactly one of its coordinates equal to $n - 2$. These amplitudes do not appear in the sum rules in (2a) and (2b). Due to this, and because the sum rules in (2c) are linearly independent, and there are no linear dependencies between (2a) and (2b), we conclude that all the sum rules described by (2a)--(2c) are linearly independent.

Thus, we have shown that the sum rules in (1), (2a)--(2c) form a set of linearly independent sum rules for the $0\oplus 1$ system.

\subsection{Step 3}

It is straightforward to check that the sum of the binomial coefficients in Eqs.~\eqref{eq:nSR-1}--\eqref{eq:nSR-2c} equals the total number of sum rules for the $0\oplus 1$ system, $n^{0\oplus 1}_{SR}(n,\, b)$, given in Eq.~\eqref{eq:nSR-0+1-app} for any $n$ and $b \geq 1$. This, together with the fact that all the sum rules in (1)--(2c) are linearly independent, implies that we have found the complete set of linearly independent sum rules for the $0\oplus 1$ system.

%%%%%%%%%%%%%%%%%%%%%%%%%%%%%%
\bibliography{draft.bib}

\bibliographystyle{JHEP}

\end{document}